\documentclass[format=acmsmall,screen=true,nonacm,pdftex,svgnames]{acmart}

\acmJournal{PACMPL}
\acmVolume{1}
\acmNumber{OOPSLA} 
\acmArticle{1}
\acmYear{2022}
\acmMonth{1}
\acmDOI{} 
\startPage{1}

\setcopyright{none}

\usepackage{booktabs}   
\usepackage{subcaption} 
\usepackage{bussproofs}
\usepackage[cal=boondoxo]{mathalfa}
\DeclareMathAlphabet{\mathpzc}{OT1}{pzc}{m}{it}
\usepackage{amsmath}
\newtheorem{theorem}{Theorem}[section]

\usepackage{color}
\usepackage{xspace}

\newcommand{\ignore}[1]{{}}

\newcommand{\syntaxDef}[3]{\rulebox{%
\syntaxKeyword$#1\mathrel{::=}{#2}$ \ifthenelse{\equal{#3}{}}{}{[#3]}%
}%
}

\makeatletter
\newcommand{\shorteq}{%
  \settowidth{\@tempdima}{-}
  \resizebox{\@tempdima}{\height}{=}%
}
\makeatother

\usepackage{xcolor}
\usepackage{colortbl}

\usepackage[utf8]{inputenc}

\usepackage{amsmath}
\usepackage{hyperref}

\usepackage{subcaption}
\usepackage{wrapfig}

\usepackage{xspace}
\usepackage{float}
\usepackage{balance}

\usepackage{textcomp} 

\usepackage{changepage}
\usepackage{array}

\usepackage{cleveref} 

\usepackage{wrapfig}
\usepackage{caption}

\usepackage{longtable}
\usepackage{tabu}

\usepackage{mathpartir}

\usepackage{mathtools} 
\usepackage{xfrac} 

\usepackage[qm,braket]{qcircuit}

\newcommand{\name}{\textsc{vqo}\xspace}
\newcommand{\qvm}{\textsc{qvm}\xspace}
\newcommand{\sourcelang}{\ensuremath{\mathcal{O}\textsc{qimp}}\xspace}
\newcommand{\oqasm}{\ensuremath{\mathcal{O}\textsc{qasm}}\xspace}
\newcommand{\pqasm}{\ensuremath{\mathcal{O}\textsc{qasm+}}\xspace}
\newcommand{\intlang}{\oqasm}
\newcommand{\vqimp}{\sourcelang}
\newcommand{\vqir}{\intlang}
\newcommand{\sqir}{\textsc{sqir}\xspace}
\newcommand{\qwire}{\ensuremath{\mathcal{Q}\textsc{wire}}\xspace}
\newcommand{\qbricks}{\ensuremath{\mathcal{Q}\textsc{bricks}}\xspace}

\newcommand{\voqc}{\textsc{voqc}\xspace}

\newcommand{\myparagraph}[1]{\paragraph{\textbf{#1}}}
\newcommand{\qket}[1]{\ket{\Phi(#1)}}
\usepackage{tikz}

\newcommand{\CC}{\ensuremath{\mathbb{C}}\xspace}

\usepackage{pgf}

\usepackage{tikz} 
\usetikzlibrary{%
  arrows,%
  shapes.misc,
  shapes.geometric, 
  shapes.arrows,%
  shapes.callouts,
  shapes.gates.logic.US,
  chains,%
  matrix,%
  positioning,
  scopes,%
  decorations.pathmorphing,
  decorations.text,
  decorations.pathreplacing, 
  shadows,%
  automata
}
\tikzset{ machine/.style={
    rectangle,
    minimum width=25mm,
    minimum height=18mm,
    text width=24mm,
    align=center,
    very thick,
    draw=black,
    color=black,
    fill=white,
  }
}

\DeclarePairedDelimiter\abs{\lvert}{\rvert}
\DeclarePairedDelimiter\norm{\lVert}{\rVert}

\makeatletter
\let\oldabs\abs
\def\abs{\@ifstar{\oldabs}{\oldabs*}}
\let\oldnorm\norm
\def\norm{\@ifstar{\oldnorm}{\oldnorm*}}
\makeatother

\makeatletter
\DeclareRobustCommand{\vardivision}{%
  \mathbin{\mathpalette\@vardivision\relax}%
}
\newcommand{\@vardivision}[2]{%
  \reflectbox{$\m@th\smallsetminus$}%
}
\makeatother

\usepackage{booktabs}

\usepackage{alltt}
\usepackage{listings,lstcoq}
\definecolor{ltblue}{rgb}{0,0.4,0.4}
\definecolor{dkblue}{rgb}{0,0.1,0.6}
\definecolor{dkgreen}{rgb}{0,0.35,0}
\definecolor{dkviolet}{rgb}{0.3,0,0.5}
\definecolor{dkred}{rgb}{0.5,0,0}
\newcommand{\code}[1]{{\small\texttt{#1}}}

 \newcommand{\rulelab}[1]{{\small \textsc{#1}}}
\newcommand{\steps}{\ensuremath{\longrightarrow}}
\newcommand{\tbool}{\texttt{bool}}
\newcommand{\tsizeof}{\texttt{sizeof}}

\newcommand{\verror}{\texttt{Error}}

\newcommand{\tfixed}{\texttt{fixedp}}
\newcommand{\tnat}{\texttt{nat}}
\newcommand{\tarr}[2]{\texttt{array}~{#1}~{#2}}
\newcommand{\econst}[2]{({#1}){#2}}
\newcommand{\eindex}[2]{{#1}\texttt{[}{#2}\texttt{]}}
\newcommand{\sassign}[4]{{#1} \leftarrow {#3}~{#2}~{#4}}
\newcommand{\ssassign}[3]{{#1} \xleftarrow{#2} {#3}}
\newcommand{\sif}[3]{\texttt{if}~{#1}~{#2}~{#3}}
\newcommand{\sfor}[3]{\texttt{for}~{#1}~{#2}~{#3}}
\newcommand{\scall}[3]{{#1}\leftarrow {#2}~{#3}}
\newcommand{\sseq}[2]{{#1}\,\texttt{;}\,{#2}}
\newcommand{\sinv}[1]{\texttt{inv}~{#1}}
\newcommand{\inst}[3][ ]{\texttt{#2}^{#1}~{#3}}
\newcommand{\insttwo}[4][ ]{\texttt{#2}^{#1}~{#3}~{#4}}

\newcommand{\iskip}[1]{\inst{ID}{#1}}

\newcommand{\inot}[1]{\inst{X}{#1}}
\newcommand{\ictrl}[2]{\insttwo{CU}{#1}{#2}}
\newcommand{\irz}[3][ ]{\insttwo[#1]{RZ}{#2}{#3}}
\newcommand{\isr}[3][ ]{\insttwo[#1]{SR}{#2}{#3}}

\newcommand{\ilshift}[1]{\inst{Lshift}{#1}}
\newcommand{\irshift}[1]{\inst{Rshift}{#1}}
\newcommand{\irev}[1]{\inst{Rev}{#1}}
\newcommand{\iqft}[3][ ]{\insttwo[#1]{QFT}{#2}{#3}}
\newcommand{\tphi}[1]{\texttt{Phi}~{#1}}
\newcommand{\thad}[1]{\texttt{Had}~{#1}}
\newcommand{\ihad}[1]{\inst{H}{#1}}

\newcommand{\hsp}[1]{\mathcal{#1}}
\newcommand{\iseq}[2]{{#1}\,\texttt{;}\,{#2}}
\newcommand{\inval}[2]{\insttwo{Nval}{#1}{#2}}

\newcommand{\itext}[1]{\texttt{#1}}

\newcommand{\instr}{\iota}

\newcommand{\app}[3]{#2\texttt{[}{#3}\mapsto{#1}\texttt{]}}
\newcommand{\xsem}{\texttt{xg}}

\newcommand{\qsem}{\texttt{qt}}
\newcommand{\psem}{\texttt{pm}}
\newcommand{\rsem}{\texttt{rz}}
\newcommand{\rrsem}{\texttt{rrz}}

\newcommand{\csem}{\texttt{cu}}

\newcommand{\Omegasz}{\Sigma}
\newcommand{\Omegaty}{\Omega}

\newcommand{\dabs}[1]{|\!| #1 |\!|}

\usepackage{newunicodechar}
\let\Alpha=A
\let\Beta=B
\let\Epsilon=E
\let\Zeta=Z
\let\Eta=H
\let\Iota=I
\let\Kappa=K
\let\Mu=M
\let\Nu=N
\let\Omicron=O
\let\omicron=o
\let\Rho=P
\let\Tau=T
\let\Chi=X

\newunicodechar{Α}{\ensuremath{\Alpha}}
\newunicodechar{α}{\ensuremath{\alpha}}
\newunicodechar{Β}{\ensuremath{\Beta}}
\newunicodechar{β}{\ensuremath{\beta}}
\newunicodechar{Γ}{\ensuremath{\Gamma}}
\newunicodechar{γ}{\ensuremath{\gamma}}
\newunicodechar{Δ}{\ensuremath{\Delta}}
\newunicodechar{δ}{\ensuremath{\delta}}
\newunicodechar{Ε}{\ensuremath{\Epsilon}}
\newunicodechar{ε}{\ensuremath{\epsilon}}
\newunicodechar{ϵ}{\ensuremath{\varepsilon}}
\newunicodechar{Ζ}{\ensuremath{\Zeta}}
\newunicodechar{ζ}{\ensuremath{\zeta}}
\newunicodechar{Η}{\ensuremath{\Eta}}
\newunicodechar{η}{\ensuremath{\eta}}
\newunicodechar{Θ}{\ensuremath{\Theta}}
\newunicodechar{θ}{\ensuremath{\theta}}
\newunicodechar{ϑ}{\ensuremath{\vartheta}}
\newunicodechar{Ι}{\ensuremath{\Iota}}
\newunicodechar{ι}{\ensuremath{\iota}}
\newunicodechar{Κ}{\ensuremath{\Kappa}}
\newunicodechar{κ}{\ensuremath{\kappa}}
\newunicodechar{Λ}{\ensuremath{\Lambda}}
\newunicodechar{λ}{\ensuremath{\lambda}}
\newunicodechar{Μ}{\ensuremath{\Mu}}
\newunicodechar{μ}{\ensuremath{\mu}}
\newunicodechar{Ν}{\ensuremath{\Nu}}
\newunicodechar{ν}{\ensuremath{\nu}}
\newunicodechar{Ξ}{\ensuremath{\Xi}}
\newunicodechar{ξ}{\ensuremath{\xi}}
\newunicodechar{Ο}{\ensuremath{\Omicron}}
\newunicodechar{ο}{\ensuremath{\omicron}}
\newunicodechar{Π}{\ensuremath{\Pi}}
\newunicodechar{π}{\ensuremath{\pi}}
\newunicodechar{ϖ}{\ensuremath{\varpi}}
\newunicodechar{Ρ}{\ensuremath{\Rho}}
\newunicodechar{ρ}{\ensuremath{\rho}}
\newunicodechar{ϱ}{\ensuremath{\varrho}}
\newunicodechar{Σ}{\ensuremath{\Sigma}}
\newunicodechar{σ}{\ensuremath{\sigma}}
\newunicodechar{ς}{\ensuremath{\varsigma}}
\newunicodechar{Τ}{\ensuremath{\Tau}}
\newunicodechar{τ}{\ensuremath{\tau}}
\newunicodechar{Υ}{\ensuremath{\Upsilon}}
\newunicodechar{υ}{\ensuremath{\upsilon}}
\newunicodechar{Φ}{\ensuremath{\Phi}}
\newunicodechar{φ}{\ensuremath{\phi}}
\newunicodechar{ϕ}{\ensuremath{\varphi}}
\newunicodechar{Χ}{\ensuremath{\Chi}}
\newunicodechar{χ}{\ensuremath{\chi}}
\newunicodechar{Ψ}{\ensuremath{\Psi}}
\newunicodechar{ψ}{\ensuremath{\psi}}
\newunicodechar{Ω}{\ensuremath{\Omega}}
\newunicodechar{ω}{\ensuremath{\omega}}

\newunicodechar{ℕ}{\ensuremath{\mathbb{N}}}
\newunicodechar{∅}{\ensuremath{\emptyset}}

\newunicodechar{∙}{\ensuremath{\bullet}}
\newunicodechar{≈}{\ensuremath{\approx}}
\newunicodechar{≅}{\ensuremath{\cong}}
\newunicodechar{≡}{\ensuremath{\equiv}}
\newunicodechar{≤}{\ensuremath{\le}}
\newunicodechar{≥}{\ensuremath{\ge}}
\newunicodechar{≠}{\ensuremath{\neq}}
\newunicodechar{∀}{\ensuremath{\forall}}
\newunicodechar{∃}{\ensuremath{\exists}}
\newunicodechar{±}{\ensuremath{\pm}}
\newunicodechar{∓}{\ensuremath{\pm}}
\newunicodechar{·}{\ensuremath{\cdot}}
\newunicodechar{⋯}{\ensuremath{\cdots}}
\newunicodechar{…}{\ensuremath{\ldots}}
\newunicodechar{∷}{~\mathrel{:\!\!\!:}~}
\newunicodechar{×}{\ensuremath{\times}}
\newunicodechar{∞}{\ensuremath{\infty}}
\newunicodechar{→}{\ensuremath{\to}}
\newunicodechar{←}{\ensuremath{\leftarrow}}
\newunicodechar{⇒}{\ensuremath{\Rightarrow}}
\newunicodechar{↦}{\ensuremath{\mapsto}}
\newunicodechar{↝}{\ensuremath{\leadsto}}
\newunicodechar{∨}{\ensuremath{\vee}}
\newunicodechar{∧}{\ensuremath{\wedge}}
\newunicodechar{⊢}{\ensuremath{\vdash}}
\newunicodechar{⊣}{\ensuremath{\dashv}}
\newunicodechar{∣}{\ensuremath{\mid}}
\newunicodechar{∈}{\ensuremath{\in}}
\newunicodechar{⊆}{\ensuremath{\subseteq}}
\newunicodechar{⊂}{\ensuremath{\subset}}
\newunicodechar{∪}{\ensuremath{\cup}}
\newunicodechar{⋓}{\ensuremath{\Cup}}
\newunicodechar{∉}{\ensuremath{\not\in}}
\newunicodechar{√}{\ensuremath{\sqrt}}

\newunicodechar{⊸}{\ensuremath{\multimap}}
\newunicodechar{⊗}{\ensuremath{\otimes}}
\newunicodechar{⨂}{\ensuremath{\bigotimes}}
\newunicodechar{⊕}{\ensuremath{\oplus}}
\newunicodechar{〈}{\ensuremath{\langle}}
\newunicodechar{⟨}{\ensuremath{\langle}}
\newunicodechar{⟩}{\ensuremath{\rangle}}
\newunicodechar{〉}{\ensuremath{\rangle}}
\newunicodechar{¡}{\ensuremath{\upsidedownbang}}
\newunicodechar{∘}{\ensuremath{\circ}}
\newunicodechar{†}{\ensuremath{\dagger}}
\newunicodechar{⊤}{\ensuremath{\top}}
\newunicodechar{⊥}{\ensuremath{\bot}}

\newunicodechar{〚}{\ensuremath{\llbracket}}
\newunicodechar{〛}{\ensuremath{\rrbracket}}

\usepackage{etoolbox} 
\newtoggle{comments}
\toggletrue{comments}

\iftoggle{comments}{
  \newcommand{\fixme}[1]{\textbf{\textcolor{red}{[ Fixme: #1]}}}
  \newcommand{\todo}[1]{\textbf{\textcolor{green}{[ TODO: #1 ]}}}
  \newcommand{\mwh}[1]{\textbf{\textcolor{red}{[ Mike: #1 ]}}}
  
  \newcommand{\khh}[1]{\textbf{\textcolor{orange}{[ Kesha: #1 ]}}}
  \newcommand{\shh}[1]{\textbf{\textcolor{purple}{[ Shih-Han: #1 ]}}}
  \newcommand{\liyi}[1]{\textbf{\textcolor{blue}{[ Liyi: #1 ]}}}
  \newcommand{\oth}[2]{\textbf{\textcolor{red}{[ #1: #2 ]}}}
  \newcommand{\xwu}[1]{\textbf{\textcolor{purple}{[ Xiaodi: #1 ]}}}
  
}{
  \newcommand{\fixme}[1]{}
  \newcommand{\todo}[1]{}
  \newcommand{\rnr}[1]{}
  \newcommand{\mwh}[1]{}  
  \newcommand{\khh}[1]{}
  \newcommand{\liyi}[1]{}
  \newcommand{\shh}[1]{}
  \newcommand{\xwu}[1]{}
  \newcommand{\oth}[2]{}
}

\newtoggle{submission}
\toggletrue{submission}

\iftoggle{submission}{
  
}{
  
}

\usepackage{pifont}

\newcommand{\cmark}{\text{\ding{51}}}
\newcommand{\xmark}{\text{\ding{55}}}
\begin{document}

\title{Verified Compilation of Quantum Oracles}                      



\def\titlerunning{A Verified Optimizer for Quantum Oracles}
\def\authorrunning{L. Li, F. Voichick, K. Hietala, Y. Peng, X. Wu \& M. Hicks}

\author{Liyi Li}
\affiliation{
  \institution{University of Maryland}
  \country{USA}
}
\email{liyili2@umd.edu}

\author{Finn Voichick}
\affiliation{
  \institution{University of Maryland}
  \country{USA}
}
\email{finn@umd.edu}

\author{Kesha Hietala}
\affiliation{
  \institution{University of Maryland}
  \country{USA}
}
\email{kesha@cs.umd.edu}

\author{Yuxiang Peng}
\affiliation{
 \institution{University of Maryland}
 \country{USA} 
}
\email{ypeng15@umd.edu}

\author{Xiaodi Wu}
\affiliation{
 \institution{University of Maryland}
 \country{USA} 
}
\email{xwu@cs.umd.edu}

\author{Michael Hicks}
\affiliation{
 \institution{University of Maryland}
 \country{USA} 
}
\email{mwh@cs.umd.edu}

\begin{abstract}
  Quantum algorithms often apply classical operations, such as
  arithmetic or predicate checks, over a quantum superposition of
  classical data; these so-called \emph{oracles} are often the largest
  components of a quantum program. To ease the construction of
  efficient, correct oracle functions, this paper presents \name, a
  high-assurance framework implemented with the Coq proof
  assistant. The core of \name is \oqasm, the \emph{oracle
  quantum assembly language}. \oqasm operations move qubits between
  two different bases via the quantum
  Fourier transform, thus admitting important optimizations, but
  without inducing \emph{entanglement} and the exponential blowup that
  comes with it. \vqir's design enabled us to prove
  correct \name's compilers---from a simple imperative language called
  \vqimp to \vqir, and from \vqir to \sqir, a general-purpose quantum
  assembly language---and allowed us to efficiently test properties of
  \vqir programs using the QuickChick property-based testing
  framework. 
  We have used \name to implement a variety of arithmetic and geometric operators
  that are building blocks for important oracles, including those used in Shor's and Grover's algorithms. 
  We found that \name's QFT-based arithmetic oracles require fewer qubits, sometimes substantially fewer, than those constructed using ``classical'' gates; \name's versions of the latter were nevertheless on par with or better than (in terms of both qubit and gate counts) oracles produced by Quipper, a state-of-the-art but unverified quantum programming platform.

\end{abstract}




\maketitle

\section{Introduction}
\label{sec:intro}


Quantum computers offer unique capabilities that can be used to
program substantially faster algorithms compared to those written for
classical computers. For example, Grover's search algorithm \cite{grover1996,grover1997}
can query unstructured data in sub-linear time (compared to linear
time on a classical computer), and Shor's algorithm \cite{shors} can factorize a
number in polynomial time (compared to the sub-exponential time for the
best known classical algorithm). An important source of speedups in these
algorithms are the quantum computer's ability to apply an \emph{oracle
function} coherently, i.e., to a \emph{superposition} of
classical queries, thus carrying out in one step a function that would
potentially take many steps on a classical computer. For Grover's, the
oracle is a predicate function that determines 
when the searched-for data is found. For Shor's, it is a classical
modular exponentiation function; the algorithm finds the period of this function where the modulus is the number being factored.


While the classical oracle function is perhaps the least interesting
part of a quantum algorithm, it contributes a significant fraction of
the final program's compiled quantum circuit.  For example,
\citet{Gidney2021howtofactorbit} estimated that Shor's modular
exponentiation function constitutes 90\% of the final code. In our own
experiments with Grover's, our oracle makes up over 99\% of the total gate count (the oracle has 3.3 million gates). Because quantum
computers will be resource-limited for the foreseeable
future~\cite{quantumcomputercurrent1,quantumcomputercurrent2}, 
programmers and programming tools will be expected to heavily optimize their
quantum circuits, especially the oracles. Such optimizations,
including ones that involve approximation, risk bugs that can be hard
to detect. This is because quantum programs are inherently difficult to
simulate, test, and debug---qubits on real quantum computers are noisy
and unreliable; observing a quantum program state mid-execution
may change that state; and simulating a
general quantum program on a classical computer is intractable because
quantum states can require resources exponential in the
number of qubits.


\begin{figure}[t]
  \includegraphics[width=.85\textwidth]{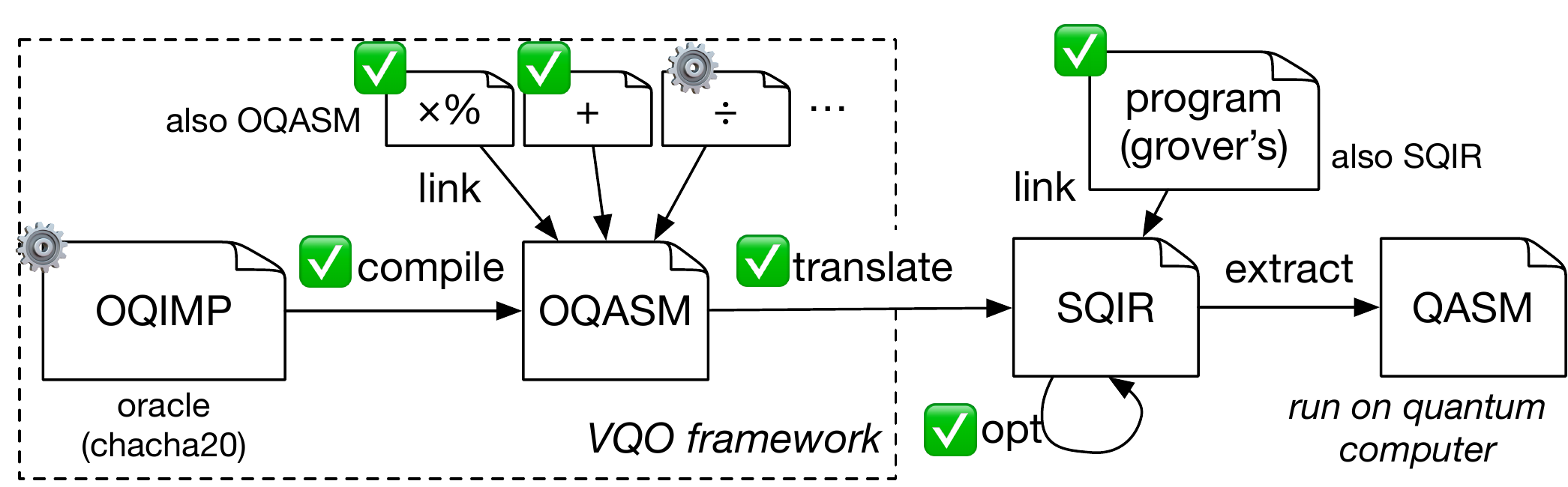}
  \caption{The \qvm high-assurance compiler stack. Checkbox means
    verified; gear means property-tested.}
\label{fig:arch}
\end{figure}

In this paper, we report on a framework we have been developing
called \name, the \emph{Verified Quantum Oracle} framework, whose goal
is to help programmers write quantum oracles that are \emph{correct}
and \emph{efficient}. \name is part of \qvm, for \emph{Quantum
  Verified Machine}, which has several elements, as shown in \Cref{fig:arch}.
\begin{itemize}
\item Using \name, an oracle can be specified in a simple, high-level programming
language we call \vqimp, which has standard imperative features and
can express arbitrary classical programs. It distinguishes quantum
variables from classical parameters, allowing the latter to be
\emph{partially evaluated}~\cite{partialeval}, thereby saving qubits
during compilation.

\item The resulting \vqimp program is compiled to \oqasm (pronounced
``O-chasm''), the 
\emph{oracle quantum assembly language}. \oqasm was
designed to be efficiently simulatable while nevertheless admitting
important optimizations; it is our core technical contribution and we
say more about it below. 
The generated \oqasm code links against implementations of standard
operators (addition, multiplication, sine, cosine, etc.) also written
in \oqasm.

\item The \oqasm oracle is then translated to \sqir, the
\emph{Simple Quantum Intermediate Representation}, which is a circuit
language embedded in the Coq proof assistant. \sqir has been used to
prove correct both quantum algorithms~\cite{PQPC} and
optimizations~\cite{VOQC}, the latter as part of \voqc, the
\emph{Verified Optimizer for Quantum Circuits}. After linking the
oracle with the quantum program that uses it, the complete \sqir program
can be optimized and extracted to OpenQASM 2.0 \cite{Cross2017} to run
on a real quantum machine. Both \name's compilation from
\vqimp to \oqasm and translation from 
\oqasm to \sqir have been proved correct in Coq.
\end{itemize}


\name helps programmers ensure their oracles are correct by supporting
both testing and verification, and ensures they are efficient by
supporting several kinds of optimization. Both aspects
are captured in the design of \oqasm, a quantum assembly language
specifically designed for oracles. 

Because oracles are classical functions, a reasonable approach would
have been to design \oqasm to be a circuit language comprised of
``classical'' gates; e.g., prior work has targeted gates \texttt{X}
(``not''), \texttt{CNOT} (``controlled not''), and \texttt{CCNOT}
(``controlled controlled not,'' aka \emph{Toffoli}). Doing so would
simplify proofs of correctness and support efficient testing by
simulation because an oracle's behavior could be completely
characterized by its behavior on computational basis states
(essentially, classical bitstrings).  ReverC~\cite{reverC} and
ReQWIRE~\cite{Rand2018ReQWIRERA} take this approach. However, doing so
cannot support optimized oracle implementations that use fundamentally
quantum functionality, e.g., as in \emph{quantum Fourier transform}
(QFT)-based arithmetic circuits~\cite{qft-adder,2000quant}.  These
circuits employ quantum-native operations (e.g., controlled-phase
operations) in the \emph{QFT basis}.  Our key insight is that
expressing such optimizations does not require expressing all quantum
programs, as is possible in a language like \sqir. Instead, \oqasm's
type system restricts programs to those that admit important
optimizations while keeping simulation tractable. \oqasm also supports
\emph{virtual qubits}; its type system ensures that position shifting
operations, commonly used when compiling arithmetic functions, require
no extra \texttt{SWAP} gates when compiled to \sqir, so there is no
added run-time cost.

Leveraging \oqasm's efficient simulatability, we implemented a
\emph{property-based random testing} (PBT) framework for \oqasm
programs in QuickChick~\cite{quickchick}, a variant of Haskell's
QuickCheck~\cite{10.1145/351240.351266} for Coq programs. This
framework affords two benefits. First, we can test that an \oqasm
operator or \vqimp program is correct according to its
specification. Formal proof in Coq can be labor-intensive, so PBT
provides an easy-to-use confidence boost, especially prior to
attempting formal proof. Second, we can use testing to assess the
effect of \emph{approximations} when developing oracles. For example,
we might like to use approximate QFT, rather than full-precision QFT,
in an arithmetic oracle in order to save gates. PBT can be used to test the
effect of this approximation within the overall oracle by measuring
the \emph{distance} between the fully-precise result and the
approximate one.


To assess \name's effectiveness we have used it to build several
efficient oracles and oracle 
components, and have either tested or proved their correctness.
\begin{itemize}
\item Using \vqimp we implemented sine, cosine, and other geometric
  functions used in Hamiltonian
  simulation~\cite{feynman1982simulating}, leveraging the arithmetic
  circuits described below. Compared to a sine function implemented in
  Quipper~\cite{Green2013}, a state-of-the-art quantum programming
  framework, \name's uses far fewer qubits thanks to \vqimp's partial
  evaluation.
  
\item We have implemented a variety of arithmetic operators in \oqasm,
including QFT-, approximate QFT- and Toffoli-based multiplication, addition, modular
multiplication, and modular division. Overall,
circuit sizes are competitive with, and oftentimes better than, those
produced by Quipper.  Qubit
counts for the final QFT-based circuits are always lower, sometimes
significantly so (up to 53\%), compared to the Toffoli-based circuits. 

\item We have proved correct both QFT and Toffoli-based adders, and QFT
and Toffoli-based modular multipliers (which are used in Shor's
algorithm). These constitute the first proved-correct implementations
of these functions, as far as we are aware.

\item We used PBT to test the correctness of various \oqasm operators.
Running 10,000 generated tests on 8- or 16-bit versions
of the operators takes just a few seconds. Testing 60-bit versions of the
adders and multipliers takes just a few minutes, whereas running a general
quantum simulator on the final circuits fails.
We found several interesting bugs in the process
of doing PBT and proof, including in the original algorithmic
description of the QFT-based modular multiplier~\cite{qft-adder}.  

\item We used PBT to analyze the precision difference between QFT and 
  approximate QFT (AQFT) circuits, and the suitability of AQFT in
  different algorithms.
We found that the AQFT adder (which uses AQFT in place of QFT) is not an accurate implementation of addition, but that it can be used as a subcomponent of division/modulo with no loss of precision, reducing gate count by 4.5--79.3\%.

\item Finally, to put all of the pieces together, we implemented the
  ChaCha20 stream cipher \cite{chacha} in \vqimp and used it as an oracle for
  Grover's search, previously implemented and proved correct in \sqir~\cite{PQPC}. We used PBT to test
  the oracle's correctness. Combining its tested
  property with Grover's correctness property, we demonstrate that Grover's is able to
  invert the ChaCha20 function and find collisions.
\end{itemize}

The rest of the paper is organized as follows. We begin with some
background on quantum computing (\Cref{sec:background}) and then 
present \oqasm's syntax, typing, and semantics (\Cref{sec:vqir}). Then we
discuss \name's implementation: \oqasm's translator and property-based
tester, and \vqimp (\Cref{sec:implementation}). Finally, we present our results
(\Cref{sec:arith-oqasm,sec:partial-eval,sec:grovers}), compare against related work (\Cref{sec:related}), and conclude. All
code presented in this paper is freely available at \url{https://github.com/inQWIRE/VQO}.

\section{Background}
\label{sec:background}

We begin with some background on quantum computing and quantum algorithms. 

\myparagraph{Quantum States} A quantum state consists of one or more quantum bits (\emph{qubits}). A qubit can be expressed as a two dimensional vector $\begin{psmallmatrix} \alpha \\ \beta \end{psmallmatrix}$ where $\alpha,\beta$ are complex numbers such that $|\alpha|^2 + |\beta|^2 = 1$.  The $\alpha$ and $\beta$ are called \emph{amplitudes}. 
We frequently write the qubit vector as $\alpha\ket{0} + \beta\ket{1}$ where $\ket{0} = \begin{psmallmatrix} 1 \\ 0 \end{psmallmatrix}$ and $\ket{1} = \begin{psmallmatrix} 0 \\ 1 \end{psmallmatrix}$ are \emph{computational basis states}. When both $\alpha$ and $\beta$ are non-zero, we can think of the qubit as being ``both 0 and 1 at once,'' a.k.a. a \emph{superposition}. For example, $\frac{1}{\sqrt{2}}(\ket{0} + \ket{1})$ is an equal superposition of $\ket{0}$ and $\ket{1}$. 

We can join multiple qubits together to form a larger quantum state with the \emph{tensor product} ($\otimes$) from linear algebra. For example, the two-qubit state $\ket{0} \otimes \ket{1}$ (also written as $\ket{01}$) corresponds to vector $[~0~1~0~0~]^T$. 
Sometimes a multi-qubit state cannot be expressed as the tensor of individual states; such states are called \emph{entangled}. One example is the state $\frac{1}{\sqrt{2}}(\ket{00} + \ket{11})$, known as a \emph{Bell pair}.
Entangled states lead to exponential blowup: A general $n$-qubit state must be described with a $2^n$-length vector, rather than $n$ vectors of length two. The latter is possible for unentangled states like $\ket{0} \otimes \ket{1}$; \vqir's type system guarantees that qubits remain unentangled.

\begin{figure}[t]
  \centering
\begin{tabular}{c@{$\quad\qquad$}c}
\begin{minipage}[b]{\textwidth}
  \Small
  \Qcircuit @C=0.5em @R=0.5em {
    \lstick{} & \gate{H} & \gate{R_2} & \gate{R_3} & \gate{R_4} & \qw        & \qw            & \qw     &\qw   & \qw &\qw &\qw\\
    \lstick{} & \qw        & \ctrl{-1}       & \qw   &\qw         & \gate{H} & \gate{R_2} & \gate{R_3}      & \qw  & \qw &\qw&\qw\\
    \lstick{} & \qw        & \qw            & \ctrl{-2}       & \qw      &\qw   & \ctrl{-1}      &\qw & \gate{H} & \gate{R_2} &\qw&\qw\\
    \lstick{} & \qw        & \qw           &\qw & \ctrl{-3}       & \qw    &\qw     & \ctrl{-2}      & \qw & \ctrl{-1} & \gate{H} & \qw 
    }
\end{minipage}
&
\begin{minipage}[b]{\textwidth}
  \Small
  \Qcircuit @C=0.5em @R=0.5em {
    \lstick{} & \gate{H} & \gate{R_2} & \gate{R_3} & \qw & \qw        & \qw            & \qw     &\qw   & \qw &\qw &\qw\\
    \lstick{} & \qw        & \ctrl{-1}       & \qw   &\qw         & \gate{H} & \gate{R_2} & \qw      & \qw  & \qw &\qw&\qw\\
    \lstick{} & \qw        & \qw            & \ctrl{-2}       & \qw      &\qw   & \ctrl{-1}      &\qw & \gate{H} & \qw &\qw&\qw\\
    \lstick{} & \qw        & \qw           &\qw & \qw       & \qw    &\qw     & \qw      & \qw & \qw & \gate{H} & \qw 
    }
\end{minipage}
\end{tabular}
\caption{Example quantum circuits: QFT over 4 qubits (left) and approximate QFT with 3 qubit precision (right). $R_m$ is a $z$-axis rotation by $2\pi / 2^m$.}
\label{fig:background-circuit-example}
\end{figure}
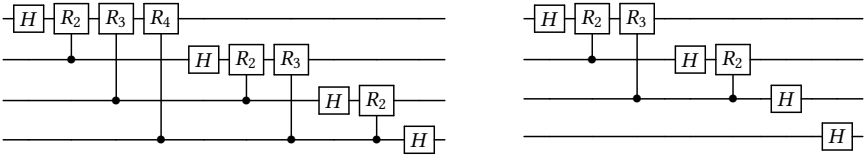

\myparagraph{Quantum Circuits} 
Quantum programs are commonly expressed as \emph{circuits}, like those shown in \Cref{fig:background-circuit-example}. In these circuits, each horizontal wire represents a qubit, and boxes on these wires indicate quantum operations, or \emph{gates}. Gates may be \emph{controlled} by a particular qubit, as indicated by a filled circle and connecting vertical line. The circuits in  \Cref{fig:background-circuit-example} use four qubits and apply 10 (left) or 7 (right) gates: four \emph{Hadamard} ($H$) gates and several controlled $z$-axis rotation (``phase'') gates.
When programming, circuits are often built by meta-programs embedded in a host language, e.g., Python (for Qiskit~\cite{Qiskit}, Cirq~\cite{cirq}, PyQuil~\cite{PyQuil}, and others), Haskell (for Quipper~\cite{Green2013}), or Coq (for \sqir and our work).

\myparagraph{Quantum Fourier Transform}
The quantum Fourier transform (QFT) is the quantum analogue of the discrete Fourier transform.
It is used in many quantum algorithms, including the phase estimation portion of Shor's factoring algorithm~\cite{shors}.
The standard implementation of a QFT circuit (for 4 qubits) is shown on the left of \Cref{fig:background-circuit-example}; an \emph{approximate QFT} (AQFT) circuit can be constructed by removing select controlled phase gates~\cite{ApproximateQFT,appox-qft2,appox-qft1}.
This produces a cheaper circuit that implements an operation mathematically similar to the QFT.\@
The AQFT circuit we use in \name (for 4 qubits) is shown on the right of \Cref{fig:background-circuit-example}. 
When it is appropriate to use AQFT in place of QFT is an open research problem, and one that is partially addressed by our work on \oqasm, which allows efficient testing of the effect of AQFT inside of oracles.

\myparagraph{Computational and QFT Bases}
The computational basis is just one possible basis for the underlying vector space.
Another basis is the \emph{Hadamard basis},  written as a tensor product of $\{\ket{+}, \ket{-}\}$, obtained by applying a \emph{Hadamard transform} to elements of the computational basis, where $\ket{+}=\frac{1}{\sqrt{2}}(\ket{0}+\ket{1})$ and  $\ket{-}=\frac{1}{\sqrt{2}}(\ket{0}-\ket{1})$.
A third useful basis is the \emph{Fourier (or QFT) basis}, obtained by applying a \emph{quantum Fourier transform} (QFT) to elements of the computational basis.


\ignore{
Applying a gate to a state \emph{evolves} the state. The semantics of doing so is expressed by multiplying the state vector by the gate's corresponding matrix representation; single-qubit gates are 2-by-2 matrices, and two-qubit gates are 4-by-4 matrices. A gate's matrix must be \emph{unitary}, ensuring that it preserves the unitarity invariant of quantum states' amplitudes. An entire circuit can be expressed as a matrix by composing its constituent gates.
}

\myparagraph{Measurement} A special, non-unitary \emph{measurement} operation extracts classical information from a quantum state, typically when a computation completes. Measurement collapses the state to a basis states with a probability related to the state's amplitudes. For example, measuring $\frac{1}{\sqrt{2}}(\ket{0} + \ket{1})$ in the computational basis will collapse the state to $\ket{0}$ with probability $\frac{1}{2}$ and likewise for $\ket{1}$, returning classical value 0 or 1, respectively. In all the programs discussed in this paper, we leave the final measurement operation implicit.

\myparagraph{Quantum Algorithms and Oracles}

Quantum algorithms manipulate input information encoded in ``oracles,'' which are callable black box circuits. For example, Grover's algorithm for unstructured quantum search \cite{grover1996,grover1997} is a general approach for searching a quantum ``database,''  which is encoded in an oracle for a function $f : \{0, 1\}^n \to \{0, 1\}$. Grover's finds an element $x \in \{0, 1\}^n$ such that $f(x) = 1$ using $O(2^{n/2})$ queries, a quadratic speedup over the best possible classical algorithm, which requires $\Omega(2^n)$ queries.
An oracle can be constructed for an arbitrary function $f$ simply by constructing a reversible classical logic circuit implementing $f$ and then replacing classical logic gates with corresponding quantum gates, e.g.,
\texttt{X} for ``not,'' \texttt{CNOT} for ``xor,'' and \texttt{CCNOT} (aka \emph{Toffoli}) for ``and.'' However, this approach does not always produce the most efficient circuits; for example, quantum circuits for arithmetic can be made more space-efficient using the quantum Fourier transform \cite{2000quant}.


Transforming an irreversible computation into a quantum circuit often requires introducing ancillary qubits, or \emph{ancillae}, to store intermediate information \cite[Chapter 3.2]{mike-and-ike}.
Oracle algorithms typically assume that the oracle circuit is reversible, so any data in ancillae must be \emph{uncomputed} by inverting the circuit that produced it.
Failing to uncompute this information leaves it entangled with the rest of the state, potentially leading to incorrect program behavior.
To make this uncomputation more efficient and less error-prone, recent programming languages such as Silq \cite{sliqlanguage} have developed notions of \emph{implicit} uncomputation.
We have similar motivations in developing \name: we aim to make it easier for programmers to write efficient quantum oracles, and to assure, through verification and randomized testing, that they are correct.


\section{\oqasm: An Assembly Language for Quantum Oracles}
\label{sec:vqir}

We designed \oqasm to be able to express efficient quantum
oracles that can be easily tested and, if desired, proved
correct.
\oqasm operations leverage both the standard
computational basis and an alternative basis connected by the quantum
Fourier transform (QFT). 
\oqasm's type system tracks the bases of variables in
\oqasm programs, forbidding operations that would introduce
entanglement. \oqasm states are therefore efficiently
represented, so programs can be effectively tested and are simpler to
verify and analyze. In addition, \oqasm uses \emph{virtual qubits}
to support \emph{position shifting operations}, which support
arithmetic operations without introducing extra gates during
translation. All of these features are novel to quantum assembly
languages. 

\begin{figure*}[t]
  \centering
  \begin{tabular}{c @{\quad} c}
  \begin{minipage}[b]{.55\textwidth}
    \Small
    \Qcircuit @C=0.5em @R=0.75em {
      \lstick{\ket{a_{n-1}}} & \qw & \ctrl{5} & \qw & \qw & \qw & \qw & \qw & \qw & \qw & \rstick{\ket{a_{n-1}}} \\
      \lstick{\ket{a_{n-2}}} & \qw & \qw & \ctrl{4} & \qw & \qw & \qw & \qw & \qw & \qw & \rstick{\ket{a_{n-2}}}\\
      \lstick{\vdots} & & & & & & & & & & \rstick{\vdots} \\
      \lstick{} & & & & & & & & & & \\
      \lstick{\ket{a_0}} & \qw & \qw & \qw & \qw & \qw & \qw & \ctrl{1} & \qw & \qw & \rstick{\ket{a_0}} \\
      \lstick{\ket{b_{n-1}}} & \multigate{5}{\texttt{QFT}} & \gate{\texttt{SR 0}} & \multigate{3}{\texttt{SR 1}} & \qw & \qw & \qw & \multigate{5}{\texttt{SR (n-1)}} & \multigate{5}{\texttt{QFT}^{-1}} & \qw & \rstick{\ket{a_{n-1} + b_{n-1}}} \\
      \lstick{} & & & & & \dots & & & & \\
      \lstick{\ket{b_{n-2}}} & \ghost{\texttt{QFT}} & \qw  &  \ghost{\texttt{SR 1}} & \qw & \qw & \qw & \ghost{\texttt{SR (n-1)}} & \ghost{\texttt{QFT}^{-1}} & \qw & \rstick{\ket{a_{n-2} + b_{n-2}}} \\
      \lstick{\vdots} & & & & & & & & & & \rstick{\vdots} \\
      \lstick{} & & & & & & & & & & \\
      \lstick{\ket{b_0}} & \ghost{\texttt{QFT}} & \qw & \qw & \qw & \qw & \qw & \ghost{\texttt{SR (n-1)}} & \ghost{\texttt{QFT}^{-1}}  & \qw & \rstick{\ket{a_0 + b_0}} 
      }
  \subcaption{Quantum circuit}
  \end{minipage} &
  \begin{minipage}[b]{.35\textwidth}
  \begin{coq}
  Fixpoint rz_adder' (a b:var) (n:nat) 
    := match n with 
       | 0 => ID (a,0)
       | S m => CU (a,m) (SR m b); 
                rz_adder' a b m
       end.
  Definition rz_adder (a b:var) (n:nat) 
    := Rev a ; Rev b ; $\texttt{QFT}$ b ;
       rz_adder' a b n;
       $\texttt{QFT}^{-1}$ b; Rev b ; Rev a.
  \end{coq}
  \subcaption{\oqasm metaprogram (in Coq)}
  \end{minipage}
  \end{tabular}
  \vspace{-0.5em}
  \caption{Example \oqasm program: QFT-based adder}
  \label{fig:circuit-example}
  \end{figure*}
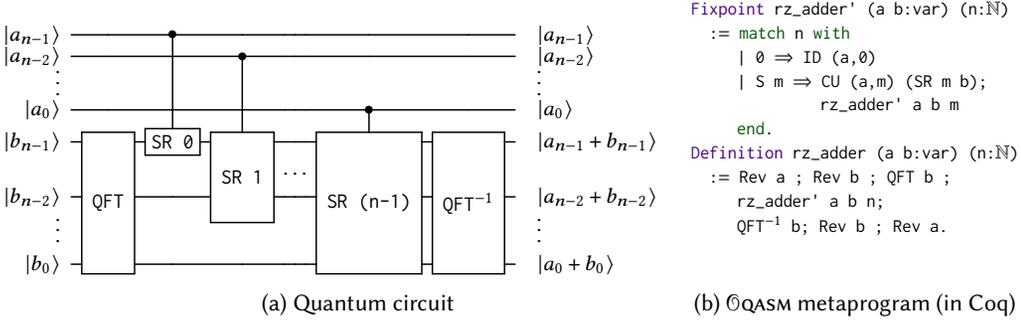

This section presents \oqasm states and the language's syntax,
semantics, typing, and soundness results.  As a running example, we use the QFT
adder~\cite{qft-adder} shown in \Cref{fig:circuit-example}. The Coq
function \coqe{rz_adder} generates an \oqasm program that adds two
natural numbers \coqe{a} and \coqe{b}, each of length \coqe{n} qubits.

\subsection{\oqasm States} \label{sec:pqasm-states}

\begin{figure}[t]
  \small
  \[\hspace*{-0.5em}
\begin{array}{l>{$} p{1.2cm} <{$} c l}
      \text{Bit} & b & ::= & 0 \mid 1 \\
      \text{Natural number} & n & \in & \mathbb{N} \\
      \text{Real} & r & \in & \mathbb{R}\\
      \text{Phase} & \alpha(r) & ::= & e^{2\pi i r} \\
      \text{Basis} & \tau & ::= & \texttt{Nor} \mid \texttt{Phi}\;n \\
      \text{Unphased qubit} & \overline{q} & ::= & \ket{b} ~~\mid~~ \qket{r} \\
      \text{Qubit} & q & ::= &\alpha(r) \overline{q}\\
      \text{State (length $d$)} & \varphi & ::= & q_1 \otimes q_2 \otimes \cdots \otimes q_d
    \end{array}
  \]
  \caption{\oqasm state syntax}
  \label{fig:vqir-state}
\end{figure}

An \oqasm program state is represented according to the grammar in
\Cref{fig:vqir-state}. A state $\varphi$ of $d$ qubits is 
a length-$d$ tuple of qubit values $q$; the state models the tensor
product of those values. This means that the size of $\varphi$ is
$O(d)$ where $d$ is the number of qubits. A $d$-qubit state in a
language like \sqir is represented as a length $2^d$ vector of complex
numbers, which is $O(2^d)$ in the number of qubits.  Our linear state
representation is possible because applying any well-typed \oqasm
program on any well-formed \oqasm state never causes qubits to be
entangled.

A qubit value $q$ has one of two forms $\overline{q}$, scaled by a
global phase $\alpha(r)$. The two forms depend on the \emph{basis}
$\tau$ that the qubit is in---it could be either \texttt{Nor} or \texttt{Phi}. A \texttt{Nor} qubit has form
$\ket{b}$ (where $b \in \{ 0, 1 \}$), which is a
computational basis value. 
A \texttt{Phi} qubit has form $\qket{r} = \frac{1}{\sqrt{2}}(\ket{0}+\alpha(r)\ket{1})$, which is a value of the (A)QFT basis.
The number $n$ in \texttt{Phi}$\;n$ indicates the precision of the state $\varphi$.
As shown by~\citet{qft-adder}, arithmetic on the computational basis can sometimes be more efficiently carried out on the QFT basis, which leads to the use of quantum operations (like QFT) when implementing circuits with classical input/output behavior.
 
\subsection{\oqasm Syntax, Typing, and Semantics}\label{sec:oqasm-syn}

\begin{figure}[t]
\begin{minipage}[t]{0.5\textwidth}
{\small \centering

  $ \hspace*{-0.8em}
\begin{array}{llcl}
      \text{Position} & p & ::= & (x,n) \qquad   \text{Nat. Num}~n
                                  \qquad   \text{Variable}~x\\
      \text{Instruction} & \instr & ::= & \iskip{p} \mid \inot{p}
                                          \mid \iseq{\instr}{\instr}\\
                & & \mid &  \isr[\lbrack -1 \rbrack]{n}{x} \mid \iqft[\lbrack -1 \rbrack]{n}{x} \mid \ictrl{p}{\instr}  \\
                      & & \mid & \ilshift{x} \mid \irshift{x} \mid \irev{x} 
    \end{array}
  $
}
  \caption{\oqasm syntax. For an operator \texttt{OP}, $\texttt{OP}^{\lbrack -1 \rbrack}$ indicates that the operator has a built-in inverse available.}
  \label{fig:vqir}
\end{minipage}
\hfill
\begin{minipage}[t]{0.45\textwidth}
\centering
\begin{tabular}{c@{$\quad=\quad$}c}
  \begin{minipage}{0.3\textwidth}
  \Small
  \Qcircuit @C=0.5em @R=0.5em {
    \lstick{} & \qw     & \multigate{4}{\texttt{SR m}} & \qw & \qw \\
    \lstick{} & \qw     & \ghost{\texttt{SR m}}           & \qw & \qw \\
    \lstick{} & \vdots & & \vdots & \\
    \lstick{} & & & & \\
    \lstick{} & \qw     & \ghost{\texttt{SR m}}           & \qw  & \qw
    }
  \end{minipage} & 
  \begin{minipage}{0.3\textwidth}
  \Small
  \Qcircuit @C=0.5em @R=0.5em {
    \lstick{} & \qw     & \gate{\texttt{RZ (m+1)}} & \qw & \qw \\
    \lstick{} & \qw     & \gate{\texttt{RZ m}}          & \qw & \qw \\
    \lstick{} & & \vdots & & \\
    \lstick{} & & & & \\
    \lstick{} & & & & \\
    \lstick{} & \qw     & \gate{\texttt{RZ 1}}           & \qw  & \qw
    }
  \end{minipage} 
\end{tabular}
\caption{\texttt{SR} unfolds to a series of \texttt{RZ} instructions}
\label{fig:sr-meaning}
\end{minipage}
\end{figure}

\Cref{fig:vqir} presents \oqasm's syntax. An \oqasm program consists of
a sequence of instructions $\instr$. Each instruction applies an
operator to either a variable $x$, which represents a group of qubits,
or a \emph{position} $p$, which identifies a particular offset into a variable $x$. 

The instructions in the first row correspond to simple single-qubit
quantum gates---$\iskip{p}$ and $\inot{p}$---and
instruction sequencing.
The instructions in the next row apply to whole variables: $\iqft{n}{x}$
applies the AQFT to variable $x$ with $n$-bit precision and
$\iqft[-1]{n}{x}$ applies its inverse.
If $n$ is equal to the size of $x$, then the AQFT operation is exact.
$\isr[\lbrack -1 \rbrack]{n}{x}$
applies a series of \texttt{RZ} gates (\Cref{fig:sr-meaning}). 
Operation $\ictrl{p}{\instr}$
applies instruction $\instr$ \emph{controlled} on qubit position
$p$. All of the operations in this row---\texttt{SR}, \texttt{QFT}, and \texttt{CU}---will be translated to multiple \sqir
gates. Function \coqe{rz_adder} in \Cref{fig:circuit-example}(b) uses
many of these instructions; e.g., it uses \texttt{QFT} and \texttt{QFT}$^{-1}$ and applies
\texttt{CU} to the $m$th position of variable \texttt{a} to control
instruction \texttt{SR m b}.

In the last row of \Cref{fig:vqir}, instructions $\ilshift{x}$,
$\irshift{x}$, and $\irev{x}$ are \emph{position shifting operations}.
Assuming that $x$ has $d$ qubits and $x_k$ represents the $k$-th qubit
state in $x$, $\texttt{Lshift}\;x$ changes the $k$-th qubit state to
$x_{(k + 1)\% d}$, $\texttt{Rshift}\;x$ changes it to
$x_{(k + d - 1)\% d}$, and \texttt{Rev} changes it to $x_{d-1-k}$. In
our implementation, shifting is \emph{virtual} not physical. The \oqasm
translator maintains a logical map of variables/positions to concrete
qubits and ensures that shifting operations are no-ops, introducing no extra gates.

Other quantum operations could be added to \oqasm to
allow reasoning about a larger class of quantum programs, while still
guaranteeing a lack of entanglement. In \Cref{sec:extended-oqasm}, we
show how \oqasm can be extended to include the Hadamard gate
\texttt{H}, $z$-axis rotations \texttt{Rz}, and a new basis
\texttt{Had} to reason directly about implementations of QFT and AQFT\@.
However, this extension compromises the property of type reversibility
(\Cref{thm:reversibility}, \Cref{sec:metatheory}), and we have not found it necessary in
oracles we have developed.

\begin{figure}[t]
\begin{minipage}[t]{0.6\textwidth}
{\Small
  \begin{mathpar}
    \inferrule[X]{\Omegaty(x)=\texttt{Nor} \\ n < \Omegasz(x)}{\Sigma;\Omega \vdash \inot{(x,n)}\triangleright \Omega}
  
    \inferrule[SR]{\Omegaty(x)=\tphi{n} \\ m < n}{\Sigma;\Omega \vdash \texttt{SR}\;m\;x\triangleright \Omega}   

    \inferrule[QFT]{\Omegaty(x)=\texttt{Nor}\\n \le \Omegasz(x)}{\Sigma; \Omega \vdash \iqft{n}{x}\triangleright \Omega[x\mapsto \tphi{n}]}    
     
    \inferrule[RQFT]{\Omegaty(x)=\tphi{n}\\n \le \Omegasz(x)}{\Sigma; \Omega \vdash \iqft[-1]{n}{x}\triangleright \Omega[x\mapsto \texttt{Nor}]}             
    
    \inferrule[CU]{\Omegaty(x)=\texttt{Nor} \\ \texttt{fresh}~(x,n)~\instr \\\\ \Sigma; \Omega\vdash \instr\triangleright \Omega \\ \texttt{neutral}(\instr)}{\Sigma; \Omega \vdash \texttt{CU}\;(x,n)\;\instr \triangleright \Omega} 
     
    \inferrule[LSH]{\Omegaty(x)=\texttt{Nor}}{\Sigma; \Omega \vdash \texttt{Lshift}\;x\triangleright \Omega}

     \inferrule[SEQ]{\Sigma; \Omega\vdash \instr_1\triangleright \Omega' \\ \Sigma; \Omega'\vdash \instr_2\triangleright \Omega''}{\Sigma; \Omega \vdash \instr_1\;;\;\instr_2\triangleright \Omega''} 
    
  \end{mathpar}
}
  \caption{Select \oqasm typing rules}
  \label{fig:exp-well-typed}
\end{minipage}
\hfill
\begin{minipage}[t]{0.35\textwidth}
{\footnotesize
\begin{center}\hspace*{-1em}
\begin{tikzpicture}[->,>=stealth',shorten >=1pt,auto,node distance=3.2cm,
                    semithick]
  \tikzstyle{every state}=[fill=black,draw=none,text=white]

  \node[state] (A)              {$\texttt{Nor}$};
  \node[state]         (C) [left of=A] {$\tphi{n}$};

  \path (A) edge [loop above]            node {$\Big\{\begin{array}{l}\texttt{ID},~\texttt{X},~\texttt{CU},~\texttt{Rev},\\
              \texttt{Lshift},\texttt{Rshift}\end{array}\Big\}$} (A)
            edge   node [above] {\{$\texttt{QFT}\;n$\}} (C);
  \path (C) edge [loop above]            node {$\{\texttt{ID},~\texttt{SR}^{\lbrack -1 \rbrack}\}$} (C)
            edge  [bend right]             node {$\{\texttt{QFT}^{-1}\;n\}$} (A);
\end{tikzpicture}
\end{center}
}
\caption{Type rules' state machine}
\label{fig:state-machine}
\end{minipage}
\end{figure}

\myparagraph{Typing}
\label{sec:vqir-typing}

In \oqasm, typing is with respect to a \emph{type environment}
$\Omega$ and a \emph{size
  environment} $\Sigma$, which map \oqasm
variables to their basis and size (number of qubits), respectively.
The typing judgment is written $\Sigma; \Omega\vdash \instr \triangleright \Omega'$ which
states that $\instr$ is well-typed under $\Omega$ and $\Sigma$, and
transforms the variables' bases to be as in $\Omega'$ 
($\Sigma$ is unchanged). Select type rules are given in
\Cref{fig:exp-well-typed}; the rules not shown (for \texttt{ID}, \texttt{Rshift}, \texttt{Rev}, and \texttt{SR}$^{-1}$) are similar.

The type system enforces three invariants.  First, it enforces that
instructions are well-formed, meaning that gates are applied to valid
qubit positions (the second premise in \rulelab{X}) and that any control qubit is distinct from the
target(s) (the \texttt{fresh} premise in
\rulelab{CU}).  This latter property enforces the quantum
\emph{no-cloning rule}.
For example, we can apply the \texttt{CU} in \code{rz\_adder'} (\Cref{fig:circuit-example})
because position \code{a,m} is distinct from variable \code{b}.

Second, the type system enforces that instructions leave affected
qubits in a proper basis (thereby avoiding entanglement). The
rules implement the state machine shown in
\Cref{fig:state-machine}. For example, $\texttt{QFT}\;n$ transforms a variable from \texttt{Nor} to
$\tphi{n}$ (rule \rulelab{QFT}), while $\texttt{QFT}^{-1}\;n$
transforms it from $\tphi{n}$ back to \texttt{Nor} (rule
\rulelab{RQFT}). Position shifting operations 
are disallowed on variables $x$ in
the \texttt{Phi} basis because the qubits that make up $x$ are
internally related (see \Cref{def:well-formed}) and cannot be rearranged. Indeed, applying a
\texttt{Lshift} and then a $\texttt{QFT}^{-1}$ on $x$ in \texttt{Phi}
would entangle $x$'s qubits.

%
%
%

Third, the type system enforces that the effect of position shifting
operations can be statically tracked. The \texttt{neutral} condition of
\rulelab{CU} requires that any shifting within $\instr$ is restored by the time it
completes. 
For example, $\sseq{\ictrl{p}{(\ilshift{x})}}{\inot{(x,0)}}$ is not well-typed, because knowing the final physical position of qubit $(x,0)$ would require statically knowing $p$. 
On the other hand, the program $\sseq{\ictrl{c}{(\sseq{\ilshift{x}}{\sseq{\inot{(x,0)}}{\irshift{x}}})}}{\inot{(x,0)}}$ is well-typed 
because the effect of the \texttt{Lshift} is ``undone'' by an \texttt{Rshift} inside the body of the \texttt{CU}.


\myparagraph{Semantics}\label{sec:pqasm-dsem}

\begin{figure}[t]
{\footnotesize
\[
\begin{array}{lll}
\llbracket \iskip{p} \rrbracket\varphi &= \varphi\\[0.2em]

\llbracket \inot{(x, i)} \rrbracket\varphi &= \app{\uparrow\xsem(\downarrow\varphi(x,i))}{\varphi}{(x,i)}
& \texttt{where  }\xsem(\ket{0})=\ket{1} \qquad\, \xsem(\ket{1})=\ket{0}
\\[0.5em]

\llbracket \ictrl{(x,i)}{\instr} \rrbracket\varphi &=  \csem(\downarrow\varphi(x,i),\instr,\varphi)
&
\texttt{where  }
\csem({\ket{0}},{\instr},\varphi)=\varphi\quad\;\,
\csem({\ket{1}},{\instr},\varphi)=\llbracket \instr \rrbracket\varphi
\\[0.4em]

\llbracket \isr{m}{x} \rrbracket\varphi &
                                            \multicolumn{2}{l}{= \app{\uparrow \qket{r_i+\frac{1}{2^{m-i+1}}}}{\varphi}{\forall i \le m.\;(x,i)}
\qquad \texttt{when  }
\downarrow\varphi(x,i) = \qket{r_i}}\\[0.5em]

\llbracket \isr[-1]{m}{x} \rrbracket\varphi&\multicolumn{2}{l}{= \app{\uparrow \qket{r_i-\frac{1}{2^{m-i+1}}}}{\varphi}{\forall i \le m.\;(x,i)}
\qquad \texttt{when  }
\downarrow\varphi(x,i) = \qket{r_i}}\\[0.5em]

\llbracket \iqft{n}{x} \rrbracket\varphi &= \app{\uparrow\qsem(\Sigma(x),\downarrow\varphi(x),n)}{\varphi}{x}
& \texttt{where  }\qsem(i,\ket{y},n)=\bigotimes_{k=0}^{i-1}(\qket{\frac{y}{2^{n-k}}})
\\[0.5em]

\llbracket \iqft[-1]{n}{x} \rrbracket\varphi &=  \app{\uparrow\qsem^{-1}(\Sigma(x),\downarrow\varphi(x),n)}{\varphi}{x}
\\[0.5em]

\llbracket \ilshift{x} \rrbracket\varphi &= \app{{\psem}_{l}(\varphi(x))}{\varphi}{x}
&
\texttt{where  }{\psem}_{l}(q_0\otimes q_1\otimes \cdots \otimes q_{n-1})=q_{n-1}\otimes q_0\otimes q_1 \otimes \cdots
\\[0.5em]

\llbracket \irshift{x} \rrbracket\varphi &= \app{{\psem}_{r}(\varphi(x))}{\varphi}{x}
&
\texttt{where  }{\psem}_{r}(q_0\otimes q_1\otimes \cdots \otimes q_{n-1})=q_1\otimes \cdots \otimes q_{n-1} \otimes q_0
\\[0.5em]

\llbracket \irev{x} \rrbracket\varphi &= \app{{\psem}_{a}(\varphi(x))}{\varphi}{x}
&
\texttt{where  }{\psem}_{a}(q_0\otimes \cdots \otimes q_{n-1})=q_{n-1}\otimes \cdots \otimes q_0
\\[0.5em]

\llbracket \iota_1; \iota_2 \rrbracket\varphi &= \llbracket \iota_2 \rrbracket (\llbracket \iota_1 \rrbracket\varphi)
\end{array}
\]
}
{\footnotesize
$
\begin{array}{l}
\\[0.2em]
\downarrow \alpha(b)\overline{q}=\overline{q}
\qquad
\downarrow (q_1\otimes \cdots \otimes q_n) = \downarrow q_1\otimes \cdots \otimes \downarrow q_n
\\[0.2em]
\app{\uparrow \overline{q}}{\varphi}{(x,i)}=\app{\alpha(b)\overline{q}}{\varphi}{(x,i)}
\qquad \texttt{where  }\varphi(x,i)=\alpha(b)\overline{q_i}
\\[0.2em]
\app{\uparrow \alpha(b_1)\overline{q}}{\varphi}{(x,i)}=\app{\alpha(b_1+b_2)\overline{q}}{\varphi}{(x,i)}
\qquad \texttt{where  }\varphi(x,i)=\alpha(b_2)\overline{q_i}
\\[0.2em]
\app{q_x}{\varphi}{x}=\app{q_{(x,i)}}{\varphi}{\forall i < \Sigma(x).\;(x,i)}
\\[0.2em]
\app{\uparrow q_x}{\varphi}{x}=\app{\uparrow q_{(x,i)}}{\varphi}{\forall i < \Sigma(x).\;(x,i)}
\end{array}
$
}
\vspace*{-0.5em}
\caption{\oqasm semantics}
  \label{fig:deno-sem}
\end{figure}

We define the semantics of an \oqasm program as a partial function
$\llbracket\rrbracket$ from
an instruction $\instr$ and input state $\varphi$ to an output state
$\varphi'$, written 
$\llbracket \instr \rrbracket\varphi=\varphi'$, shown in \Cref{fig:deno-sem}.



Recall that a state $\varphi$ is a tuple of $d$ qubit values,
modeling the tensor product $q_1\otimes \cdots \otimes q_d$. 
The rules implicitly map each variable $x$ to a
range of qubits in the state, e.g., 
$\varphi(x)$ corresponds to some sub-state $q_k\otimes \cdots \otimes q_{k+n-1}$
where $\Omegasz(x)=n$.
Many of the rules in \Cref{fig:deno-sem} update a \emph{portion} of a
state. We write $\app{q_{(x,i)}}{\varphi}{(x,i)}$ to update the $i$-th
qubit of variable $x$ to be the (single-qubit) state $q_{(x,i)}$, and
$\app{q_{x}}{\varphi}{x}$ to update variable $x$ according to
the qubit \emph{tuple} $q_x$.
$\app{\uparrow q_{(x,i)}}{\varphi}{(x,i)}$ and $\app{\uparrow q_{x}}{\varphi}{x}$ 
are similar, except that they also accumulate the previous global phase of $\varphi(x,i)$ (or $\varphi(x)$).
We use $\downarrow$ to convert a qubit $\alpha(b)\overline{q}$ to an unphased qubit $\overline{q}$.

Function $\xsem$ updates the state of a single
qubit according to the rules for the standard quantum gate $X$.  
\texttt{cu} is a conditional operation
depending on the \texttt{Nor}-basis qubit $(x,i)$. 
\texttt{SR} (or
$\texttt{SR}^{-1}$) applies an $m+1$ series of \texttt{RZ} (or
$\texttt{RZ}^{-1}$) rotations where the $i$-th rotation
applies a phase of $\alpha({\frac{1}{2^{m-i+1}}})$
(or $\alpha({-\frac{1}{2^{m-i+1}}})$).
$\qsem$ applies an approximate quantum Fourier transform; $\ket{y}$ is an abbreviation of
$\ket{b_1}\otimes \cdots \otimes \ket{b_i}$ (assuming $\Omegasz(y)=i$)
and $n$ is the degree of approximation.
If $n = i$, then the operation is the standard QFT\@.
Otherwise, each qubit in the state is mapped to $\qket{\frac{y}{2^{n-k}}}$, which is equal to $\frac{1}{\sqrt{2}}(\ket{0} + \alpha(\frac{y}{2^{n-k}})\ket{1})$ when $k < n$ and $\frac{1}{\sqrt{2}}(\ket{0} + \ket{1}) = \ket{+}$ when $n \leq k$ (since $\alpha(n) = 1$ for any natural number $n$).
$\qsem^{-1}$ is the inverse function of $\qsem$. 
Note that the input state to $\qsem^{-1}$ is guaranteed to have the form $\bigotimes_{k=0}^{i-1}(\qket{\frac{y}{2^{n-k}}})$ because it has type $\tphi{n}$.
$\psem_l$, $\psem_r$, and
$\psem_a$ are the semantics for \itext{Lshift}, 
\itext{Rshift}, and \itext{Rev}, respectively.   

\subsection{\oqasm Metatheory}\label{sec:metatheory}

\myparagraph{Soundness}
We prove that well-typed \oqasm programs are well defined; i.e., the
type system is sound with respect to the semantics. 
We begin by defining the well-formedness of an \oqasm state.

\begin{definition}[Well-formed \oqasm state]\label{def:well-formed}\rm 
  A state $\varphi$ is \emph{well-formed}, written
  $\Sigma;\Omega \vdash \varphi$, iff:
\begin{itemize}
\item For every $x \in \Omega$ such that $\Omegaty(x) = \texttt{Nor}$,
  for every $k <\Omegasz(x)$, $\varphi(x,k)$ has the form
  $\alpha(r)\ket{b}$.

\item For every $x \in \Omega$ such that $\Omegaty(x) = \tphi{n}$ and $n \le \Omegasz(x)$,
  there exists a value $\upsilon$ such that for
  every $k < \Omegasz(x)$, $\varphi(x,k)$ has the form
  $\alpha(r)\qket{\frac{\upsilon}{ 2^{n- k}}}$.\footnote{Note that $\Phi(x) = \Phi(x + n)$, where the integer $n$ refers to phase $2 \pi n$; so multiple choices of $\upsilon$ are possible.}
\end{itemize}
\end{definition}

\noindent
Type soundness is stated as follows; the proof is by induction on $\instr$, and is mechanized in Coq.

\begin{theorem}\label{thm:type-sound-oqasm}\rm[\oqasm type soundness]
If $\Sigma; \Omega \vdash \instr \triangleright \Omega'$ and $\Sigma;\Omega \vdash \varphi$ then there exists $\varphi'$ such that $\llbracket \instr \rrbracket\varphi=\varphi'$ and $\Sigma;\Omega' \vdash \varphi'$.
\end{theorem}

\myparagraph{Algebra}
Mathematically, the set of well-formed $d$-qubit \oqasm states for a given $\Omega$ can be interpreted as a subset $\hsp{S}^d$ of a $2^d$-dimensional Hilbert space $\hsp{H}^d$,\footnote{A \emph{Hilbert space} is a vector space with an inner product that is complete with respect to the norm defined by the inner product. $\hsp{S}^d$ is a sub\emph{set}, not a sub\emph{space} of $\hsp{H}^d$ because $\hsp{S}^d$ is not closed under addition: Adding two well-formed states can produce a state that is not well-formed.}
and the semantics function $\llbracket \rrbracket$ can be interpreted as a $2^d \times 2^d$ unitary matrix, as is standard when representing the semantics of programs without measurement~\cite{PQPC}.
Because \oqasm's semantics can be viewed as a unitary matrix, correctness properties extend by linearity from $\hsp{S}^d$ to $\hsp{H}^d$---an oracle that performs addition for classical \texttt{Nor} inputs will also perform addition over a superposition of \texttt{Nor} inputs.
We have proved that $\hsp{S}^d$ is closed under well-typed \oqasm programs.

Given a qubit size map $\Sigma$ and type environment $\Omega$, the set of \oqasm programs that are well-typed with respect to $\Sigma$ and $\Omega$ (i.e., $\Sigma;\Omega \vdash \instr \triangleright \Omega'$) form a groupoid $(\{\instr\},\Sigma, \Omega,\hsp{S}^d)$, where $\hsp{S}^d$ is the set of $d$-qubit states that are well-formed ($\Omega \vdash \varphi$) according to \Cref{def:well-formed}.

We can extend the groupoid to $(\{\instr\},\Sigma,\hsp{H}^d)$ by defining a general $2^d$ dimensional Hilbert space $\hsp{H}^d$, such that $\hsp{S}^d \subseteq \hsp{H}^d$, and removing the typing requirements on $\{\instr\}$. Clearly, $(\{\instr\},\Sigma,\hsp{H}^d)$ is still a groupoid because every \oqasm operation is valid in a traditional quantum language like \sqir. We then have the following two two theorems to connect \oqasm operations with operations in the general Hilbert space: 

 \begin{theorem}\label{thm:subgroupoid}\rm
   $(\{\instr\},\Sigma, \Omega,\hsp{S}^d) \subseteq (\{\instr\},\Sigma,\hsp{H}^d)$ is a subgroupoid.
 \end{theorem}

\begin{theorem}\label{thm:sem-same}\rm
Let $\ket{y}$ be an abbreviation of $\bigotimes_{m=0}^{d-1} \alpha(r_m) \ket{b_m}$ for $b_m \in \{0,1\}$.
If for every $i\in [0,2^d)$, $\llbracket \instr \rrbracket\ket{y_i}=\ket{y'_i}$, then $\llbracket \instr \rrbracket (\sum_{i=0}^{2^d-1} \ket{y_i})=\sum_{i=0}^{2^d-1} \ket{y'_i}$.
\end{theorem}

We prove these theorems as corollaries of the compilation correctness theorem from \oqasm to \sqir (\Cref{thm:vqir-compile}). 
\Cref{thm:subgroupoid} suggests that the space $\hsp{S}^d$ is closed under the application of any well-typed \oqasm operation.
\Cref{thm:sem-same} says that \oqasm oracles can be safely applied to superpositions over classical states.\footnote{Note that a superposition over classical states can describe \emph{any} quantum state, including entangled states.}

\begin{figure}[t]
  {\Small
    \begin{mathpar}
      \inferrule[ ]{}{\inot{(x,n)}\xrightarrow{\text{inv}} \inot{(x,n)}}
    
      \inferrule[  ]{}{\texttt{SR}\;m\;x\xrightarrow{\text{inv}} \texttt{SR}^{-1}\;m\;x}
  
      \inferrule[ ]{}{\iqft{n}{x} \xrightarrow{\text{inv}}  \iqft[-1]{n}{x}}   
  
      \inferrule[ ]{}{\texttt{Lshift}\;x\xrightarrow{\text{inv}} \texttt{Rshift}\;x} 
       
      \inferrule[ ]{\instr \xrightarrow{\text{inv}} \instr'}{\texttt{CU}\;(x,n)\;\instr \xrightarrow{\text{inv}} \texttt{CU}\;(x,n)\;\instr'} 
  
      \inferrule[ ]{\instr_1 \xrightarrow{\text{inv}} \instr'_1 \\ \instr_2 \xrightarrow{\text{inv}} \instr'_2}{\instr_1\;;\;\instr_2\xrightarrow{\text{inv}} \instr'_2\;;\;\instr'_1} 
      
    \end{mathpar}
  }
  \caption{Select \oqasm inversion rules}
  \label{fig:exp-reversed-fun}
\end{figure}

\begin{figure}[t]
\centering
\begin{tabular}{c@{$\quad=\quad$}c@{\qquad}c@{$\quad=\quad$}c}
  \begin{minipage}{0.25\textwidth}
  \footnotesize
  \Qcircuit @C=0.25em @R=0.35em {
    & \qw & \multigate{3}{(x+a)_n} & \qw \\
    & \vdots & & \\
    & & & \\
    & \qw & \ghost{(x+a)_n} & \qw \\
    }
  \end{minipage}
&
\begin{minipage}{.45\textwidth}
  \footnotesize
  \Qcircuit @C=0.35em @R=0.55em {
     & \qw & \gate{\texttt{SR}\;0} & \multigate{3}{\texttt{SR}\;1} & \qw & \qw & \qw & \multigate{5}{\texttt{SR}\;(n-1)} & \qw  \\
      & & & & & \dots & & &  \\
      & \qw & \qw  &  \ghost{\texttt{SR}\; 1} & \qw & \qw & \qw & \ghost{\texttt{SR}\;(n-1)} & \qw \\
      & & & & & & & &  \\
     & & & & & & & &  \\
   & \qw & \qw & \qw & \qw & \qw & \qw & \ghost{\texttt{SR}\;(n-1)}  & \qw 
    }
\end{minipage}
&  
\begin{minipage}{0.25\textwidth}
  \footnotesize
  \Qcircuit @C=0.25em @R=0.35em {
    & \qw & \multigate{3}{(x-a)_n} & \qw \\
    & \vdots & & \\
    & & & \\
    & \qw & \ghost{(x+a)_n} & \qw \\
    }
  \end{minipage}
&
\begin{minipage}{.45\textwidth}
  \footnotesize
  \Qcircuit @C=0.35em @R=0.55em {
    & \qw & \multigate{5}{\texttt{SR}^{-1} (n-1)} & \qw & \qw & \qw & \multigate{3}{\texttt{SR}^{-1} 1} & \gate{\texttt{SR}^{-1} 0} & \qw \\
    &     &                                  &     & \dots &   &                              &                      &   \\
    & \qw & \ghost{\texttt{SR}^{-1} (n-1)}        & \qw & \qw   & \qw & \ghost{\texttt{SR}^{-1} 1} & \qw & \qw  \\
      & & & & & & & &  \\
     & & & & & & & &  \\
    & \qw & \ghost{\texttt{SR}^{-1} (n-1)} & \qw & \qw & \qw & \qw & \qw & \qw 
    }
\end{minipage}
\end{tabular}
\caption{Addition/subtraction circuits are inverses}
\label{fig:circuit-add-sub}
\end{figure}
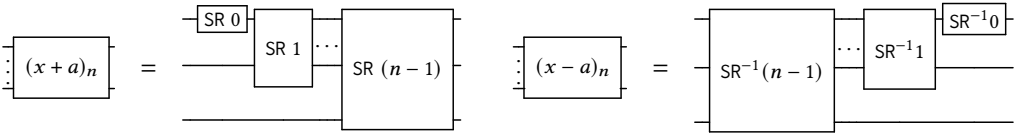

\oqasm programs are easily invertible, as shown by the rules in \Cref{fig:exp-reversed-fun}.
This inversion operation is useful for constructing quantum oracles; for example, the core logic in the QFT-based subtraction circuit is just the inverse of the core logic in the addition circuit (\Cref{fig:exp-reversed-fun}).
This allows us to reuse the proof of addition in the proof of subtraction.
The inversion function satisfies the following properties:

 \begin{theorem}\label{thm:reversibility}\rm[Type reversibility]
    For any well-typed program $\instr$, such that $\Sigma; \Omega \vdash \instr \triangleright \Omega'$, its inverse $\instr'$, where $\instr \xrightarrow{\text{inv}} \instr'$, is also well-typed and we have $\Sigma;\Omega' \vdash \instr' \triangleright \Omega$. Moreover, $\llbracket \instr ; \instr' \rrbracket \varphi=\varphi$.
 \end{theorem}

\section{\name Quantum Oracle Framework}
\label{sec:implementation}

This section presents \name, our framework for specifying, compiling,
testing, and verifying quantum oracles, whose architecture was
given in \Cref{fig:arch}. We start by considering translation
from \oqasm to \sqir and proof of its correctness. Next, we discuss \name's
property-based random testing framework for \oqasm
programs. Finally, we discuss \vqimp, a simple imperative language for
writing oracles, which compiles to \oqasm. We also present its proved-correct 
compiler and means to test the correctness of \vqimp oracles.

\subsection{Translation from \oqasm to \sqir}\label{sec:vqir-compilation}

\newcommand{\tget}{\texttt{get}}
\newcommand{\tstart}{\texttt{start}}
\newcommand{\tfst}{\texttt{fst}}
\newcommand{\tsnd}{\texttt{snd}}
\newcommand{\tucom}[1]{\texttt{ucom}~{#1}}
\newcommand{\tif}{\texttt{if}}
\newcommand{\tthen}{\texttt{then}}
\newcommand{\telse}{\texttt{else}}
\newcommand{\tlet}{\texttt{let}}
\newcommand{\tin}{\texttt{in}}

\name translates \oqasm to \sqir by mapping \oqasm positions to \sqir 
concrete qubit indices and expanding \oqasm instructions to sequences
of \sqir gates.
%
%
Translation is expressed as the judgment
$\Sigma\vdash (\gamma,\instr) \steps
(\gamma',\epsilon)$ where $\Sigma$ maps \oqasm variables to their sizes, 
$\epsilon$ is the output \sqir circuit, and $\gamma$ maps an \oqasm 
position $p$ to a \sqir concrete qubit index (i.e., offset into a 
global qubit register).  At the start of translation, for every
variable $x$ and $i < \Sigma(x)$, $\gamma$ maps $(x,i)$ to a unique
concrete index chosen from 0 to $\sum_{x}(\Sigma(x))$.

\begin{figure}[t]
{\Small

  \begin{mathpar}
    \inferrule{ }{\Sigma\vdash(\gamma,\inot{p}) \to (\gamma,\textcolor{blue}{\inot{\gamma(p)}})}
          
    \inferrule{\gamma'=\gamma[\forall i.\; i < \Sigma(x) \Rightarrow (x,i)\mapsto \gamma(x,(i+1)\%\Sigma(x))]}{\Sigma\vdash(\gamma,\ilshift{x}) \to (\gamma',\textcolor{blue}{\iskip{(\gamma'(x,0))}})}
    
    \inferrule{\Sigma\vdash(\gamma,\instr) \to (\gamma,\textcolor{blue}{\epsilon})\\
      \textcolor{blue}{\epsilon' = \texttt{ctrl}(\gamma(p),\epsilon)}}{\Sigma\vdash(\gamma,\ictrl{p}{\instr}) \to (\gamma,\textcolor{blue}{\epsilon')}}    
     
    \inferrule{ \Sigma\vdash (\gamma,\instr_1) \to (\gamma',\textcolor{blue}{\epsilon_1}) \\ \Sigma\vdash(\gamma',\instr_2) \to (\gamma'',\textcolor{blue}{\epsilon_2})}{\Sigma\vdash(\gamma,\sseq{\instr_1}{\instr_2}) \to (\gamma'',\textcolor{blue}{\sseq{\epsilon_1}{\epsilon_2}})}             
  
  \end{mathpar}
}
\vspace*{-1em}
\caption{Select \oqasm to \sqir translation rules (\sqir circuits are marked blue)}
\label{fig:compile-vqir}
\end{figure}

\Cref{fig:compile-vqir} depicts a selection of translation rules.\footnote{Translation in fact threads through the typing judgment, but we elide that for simplicity.}
The first rule shows how to translate
$\inot{p}$, which has a directly corresponding gate in \sqir.
The second rule left-shifts the qubits of the target variable in the
map $\gamma$, and produces an identity gate (which will be removed in a subsequent optimization pass).
For example, say we have variables $x$ and $y$ in the map $\gamma$ and variable $x$ has three qubits so $\gamma$ is $\{(x,0)\mapsto 0,(x,1)\mapsto 1, (x,2)\mapsto 2,(y,0)\mapsto 3,...\}$.
Then after $\ilshift{x}$ the $\gamma$ map becomes $\{(x,0)\mapsto 1,(x,1)\mapsto 2, (x,2)\mapsto 0,(y,0)\mapsto 3,...\}$. 
The last two rules translate the \texttt{CU} and sequencing instructions. In the \texttt{CU} translation, the rule assumes that $\instr$'s translation does not affect the $\gamma$ position map. This requirement is assured for well-typed programs per rule \rulelab{CU} in \Cref{fig:exp-well-typed}. \texttt{ctrl} generates the controlled version of an arbitrary \sqir program using standard decompositions \cite[Chapter 4.3]{mike-and-ike}.

\newcommand{\transs}[3]{[\!|{#1}|\!]^{#2}_{#3}}

We have proved \oqasm-to-\sqir translation correct. To formally state
the correctness property we relate $d$-qubit \oqasm states to \sqir states, which are vectors of $2^d$ complex numbers, via a function $\transs{-}{d}{\gamma}$, where $\gamma$ is the virtual-to-physical qubit map.
For example, say that our program uses two variables, $x$ and $y$, and both have two qubits.
The qubit states are $\ket{0}$ and $\ket{1}$ (meaning that $x$ has type \texttt{Nor}), and $\qket{r_1}$ and $\qket{r_2}$ (meaning that $y$ has type \texttt{Phi}).
Furthermore, say that $\gamma = \{(x,0)\mapsto 0,(x,1)\mapsto 1, (y,0)\mapsto 2, (y,1)\mapsto 3\}$. 
This \oqasm program state will be mapped to the $2^4$-element vector $\ket{0}\otimes \ket{1}\otimes (\ket{0}+e^{2\pi i r_1}\ket{1})\otimes (\ket{0}+e^{2\pi i r_2}\ket{1})$.

\begin{theorem}\label{thm:vqir-compile}\rm[\oqasm translation correctness]
  Suppose $\Sigma; \Omega \vdash \instr \triangleright \Omega'$ and
  $\Sigma \vdash(\gamma,\instr) \to (\gamma',\epsilon)$. 
Then for $\Sigma; \Omega \vdash \varphi$, $\llbracket \instr \rrbracket\varphi=\varphi'$, and we have 
$\llbracket \epsilon \rrbracket \times \transs{\varphi}{d}{\gamma} = \transs{\varphi'}{d}{\gamma'}$ where $\llbracket \epsilon \rrbracket$ is the matrix interpretation of $\epsilon$ per \sqir's semantics.
\end{theorem}

The proof of translation correctness is by induction on the \oqasm program $\instr$. 
Most of the proof simply shows the correspondence of operations in $\instr$ to their translated-to gates $\epsilon$ in \sqir, except for shifting operations, which update the virtual-to-physical map.

Note that to link a complete, translated oracle $\instr$ into a larger \sqir program may require that $\gamma = \gamma'$, i.e., $\texttt{neutral}(\instr)$, so that logical inputs match logical outputs. This requirement is naturally met for programs written to be reversible, as is the case for all arithmetic circuits in this paper, e.g., \coqe{rz_adder} from \Cref{fig:circuit-example}. 

\ignore{
\begin{lemma}\label{thm:subgroupoid-lemma}\rm
   For all $\epislon \in \{\epsilon^{(\Sigma,\Omega)}\}$, if $\epislon$ is valid operation in $\hsp{S}_n$, $n \le m$, and $\hsp{S}_m$ and it every qubit in $\hsp{S}_m$ satisfies $\Omega$'s restriction, then 
\end{lemma}

We view $(\mathcal{H}, \instr )$ as a groupoid over Hilbert space $\mathcal{H}$, we can then defined a subset of $\mathcal{H}$ as $\mathcal{H}^n_s$, where it has the following conditions:

\begin{itemize}
\item Each element in $\mathcal{H}^n_s$ has the form: $\ket{q_1}\otimes ... \otimes \ket{q_n}$, where $\ket{q_1}$,...,$\ket{q_n}$ are 1-dimensional qubit. 
\item For any element $\ket{q_1}$,...,$\ket{q_n}$ in $\mathcal{H}^n_s$, $\ket{q_i}$ has three possible forms:  $\alpha\ket{c}$, $\frac{1}{\sqrt{2}} \alpha( s_1 \ket{0}+ s_2 \ket{1})$, or $\frac{1}{\sqrt{2}}\alpha~(\ket{0}+\beta\ket{1})$.
\end{itemize}

We view $\Sigma;\Omega\vdash \iota \triangleright \Omega'$ as a predicate for each \oqasm operation $\iota$ on where a program $\iota$ is defined given a subspace $\mathcal{H}_{(x, p)}$, then $(\mathcal{H}^n_s, \instr )$ is a sub-groupoid of $(\mathcal{H}, \instr )$ for all $\instr$ that is type-checked in $\mathcal{H}^n_s$.

We then define a superoperator over $\instr$ as $\instr^*(\rho)= \llbracket \instr \rrbracket \rho \llbracket \instr \rrbracket^{\dag}$ where $\rho \in (\mathcal{H}^n_s)^*$. $(\mathcal{H}^n_s)^*$ is the collection of density matrices seen as linear transformations from $\mathcal{H}^n_s$ to $\mathcal{H}^n_s$.
The superoperator gives the density matrix interpretation of the \oqasm semantics. We define a $2^m$ dimensional database $D$ as $\mathcal{H}^n_s \otimes D$, and $D$ has the format $\ket{q_1}$,...,$\ket{q_{2^n}}$ where $q_i$ is a $k$ array bitstring, each of the bitstring position is either $0$ or $1$.
We define a new operation in $\instr$ as $\texttt{read}\;y\;x$, such that $y$ is a $k$-length qubit, and $x$ is a $m$ length qubit representing the position. The desity matrix semantics of the $\texttt{read}$ operation is given as:

{
\[\Sigma^{2^m-1}_{0}\ket{i}\bra{i}\otimes D_{i}\]
}

With finite bijection mapping $\tget(\rho)$ and $\varrho$, we develop the translation process as the function $(d * \Sigma * \rho * \instr * \varrho) \to (\tucom{d}* \rho * \varrho)$, where $d$ is the dimension number indicating the number of qubits in the system, $\Sigma$ maps variables to qubit numbers in \oqasm, $\rho$ is the position mapping database, $\varrho$ is the inverse function of $\tget(\rho)$, $\instr$ is an \oqasm program, and $\delta\in\tucom{d}$ is a \sqir circuit.

create a mapping database $\rho$ that maps positions $p$ to a data structure $\coqe{nat} * (\coqe{nat} \to \coqe{nat}) * (\coqe{nat} \to \coqe{nat})$. We assume that all qubit locations in \sqir are managed as a number line counting from $0$. The first \coqe{nat} value is the starting position for an \oqasm variable $x$ on the number line. We assume that $\texttt{start}(\rho,x)$ is a function to get the start position of $x$ in the map $\rho$. The second function ($\mu$, $\coqe{nat} \to \coqe{nat}$) is a mapping from position offset to the offset number of the physical position in \sqir. 
\khh{Moved from earlier text: Function $\tstart(\rho,x)$ is equivalent to $\rho(x,0)$. }
For example, a position $(x,i)$ is mapped to $\tstart(\rho,x)+\mu(i)$ in the number line. The third function ($\nu$, $\coqe{nat} \to \coqe{nat}$) is the inverse function mapping from an offset in \sqir back to the offset in \oqasm. For every offset $i$ for $x$ in \oqasm, if $\mu$ and $\nu$ are the two maps in $\rho(x)$, then $\mu(\nu(i)) = i$, and vice versa. We assume that the actual virtual to physical position mapping is $ \tget(\rho)$, which gets the physical position in \sqir for $p$.
$\tget(\rho,p)$ gives us the \sqir position for $p$ and its definition is $\texttt{start}(\rho,\texttt{fst}(p))+\texttt{get\_}\mu(\rho,\texttt{fst}(p))(\texttt{snd}(p))$.
On the other hand, since different virtual positions map to different physical positions, the function $\tget(\rho)$ is bijective; there is an inverse function $\varrho$ for $\tget(\rho)$, such that $\tget(\rho,p)=i \Rightarrow \varrho(i) = p$. The functions $\rho$, $\tget(\rho)$, and its inverse function $\varrho$ are also useful in the translation process, and we assume that they satisfy \textbf{finite bijection}, where for a set of positions $\overline{p}$, there exists a mapping database $\rho$, a dimension number $d$ and an inverse function $\varrho$, such that for all $p$ in $\overline{p}$, $\tget(\rho,p)=i$, $i<d$, $\varrho(\tget(\rho,p))=p$, and $\tget(\rho,\varrho(i))=i$.}
\ignore{
\begin{definition}\label{def:vars-def}\rm
(\textbf{finite bijection})
Given a virtual to physical position mapping database $\rho$, and the mapping function $\tget(\rho)$, its inverse function $\varrho$, a map from \oqasm variables to its qubit size $\Sigma$, and $d$ is the dimension of the qubits in \sqir and it is a maximum number that is larger than all physical position number in the image of $\tget{\rho}$, we say that $\rho$ and $\varrho$ is finitely bijective iff:

\begin{itemize}
  \item For all $p$, if $\tfst(p))$ is in the domain of $\rho$ and $\tsnd(p)< \Sigma(\tfst(p))$, then $\tget(\rho,p)<d$.
  \item For all $i$, if $i < d$, then $\tfst(\varrho(i))$ is in the domain of $\rho$ and $\tsnd(\varrho(i))< \Sigma(\tfst(\varrho(i)))$
\item For all $p$, if $\tfst(p))$ is in the domain of $\rho$ and $\tsnd(p)< \Sigma(\tfst(p))$, then $\varrho(\tget(\rho,p)) = p$.
\item For all $i$, if $i < d$, then $\tget(\rho, \varrho(i))=i$.
\item For all $p_1$ $p_2$, if $p_1 \neq p_2$, then $\tstart(\rho,p_1) \neq \tstart(\rho,p_1)$.
\item For all $x$ $y$, if $x \neq y$, then for all $i$ $j$, such that $i < \Sigma(x)$ and $j < \Sigma(y)$, $\tget(\rho,(x,i)) \neq \tget(\rho,(y, j))$.
\item For all $p$, if $\tsnd(p) < \Sigma(\tfst(p))$, then $\tget\_\mu(\rho,\tfst(p))(\tsnd(p))<\Sigma(\tfst(p))$.
\item For all $p$, if $i < \Sigma(x)$, then $\tget\_\nu(\rho,x)(i)<\Sigma(x)$.
\end{itemize}

\end{definition}
}

\subsection{Property-based Random Testing}\label{sec:rand-testing}

\oqasm's type system ensures that states can be efficiently
represented. We leverage this fact to implement a testing framework
for \oqasm programs using QuickChick \cite{quickchick}, which is
a property-based testing (PBT) framework for Coq in the style of
Haskell's QuickCheck~\cite{10.1145/351240.351266}. We use this
framework for two purposes: To test correctness
properties of \oqasm programs and to experiment with the consequences of
approximation for efficiency and correctness.

\myparagraph{Implementation}

PBT randomly generates inputs using a hand-crafted \emph{generator}, 
and confirms that a property holds for the given inputs. Since
arithmetic/oracle operations are defined over \texttt{Nor}-basis
inputs, we wrote a generator for these inputs. 
A single test of an \oqasm program involves five steps:
(1) generate (or specify) $n$, which is the number of qubits in the input;
(2) for each input variable $x$, generate uniformly random bitstrings $b_0b_1...b_{n-1}$ of length $n$, representing $x$'s initial qubit value $\bigotimes_{i=0}^{n-1} \alpha(0)\ket{b_i}$;
(3) prepare an \oqasm state $\varphi$ containing all input variables' encoded bitstrings;
(4) execute the \oqasm program with the prepared state; and
(5) check that the resulting state satisfies the desired property.

We took several steps to improve testing performance. 
First, we streamlined the representation of states:
Per the semantics in \Cref{fig:deno-sem}, in a state with $n$ qubits, the phase associated with each qubit can be written as $\alpha(\frac{\upsilon}{2^n})$ for some natural number $\upsilon$. 
Qubit values in both bases are thus pairs of natural numbers: the global phase $\upsilon$ (in range $[0,2^n)$) and $b$ (for $\ket{b}$) or $y$ (for $\qket{\frac{y}{2^n}}$). 
An \oqasm state $\varphi$ is a map from qubit positions $p$ to qubit values $q$; in our proofs, this map is implemented as a partial function, but for testing, we use an AVL tree implementation (proved equivalent to the functional map). 
To avoid excessive stack use, we implemented the \oqasm semantics function tail-recursively. 
To run the tests, QuickChick runs OCaml code that it \emph{extracts} from the Coq definitions; during extraction, we replace natural numbers and operations thereon with machine integers and operations. We present performance results in \Cref{sec:arith-oqasm}.

\myparagraph{Testing Correctness}

A full formal proof is the gold standard for correctness, but it is also
laborious. It is especially deflating to be well through a proof only
to discover that the intended property does not hold and, worse,
that nontrivial changes to the program are necessary.
Our PBT framework gives assurance that an \oqasm
program property is correct by attempting to falsify it using
thousands of randomly generated instances, with good coverage of the
program's input space. We have used PBT to test the correctness of a
variety of operators useful in oracle programs, as presented in
\Cref{sec:arith-oqasm}. When implementing a QFT-adder circuit, using
PBT revealed that we had encoded the wrong endianness. 
We have also used PBT with \vqimp programs by first
compiling them to \oqasm and then testing their correctness at that
level.

\myparagraph{Assessing the Effect of Approximation}

Because of the resource limitations of near-term machines, programmers
may wish to \emph{approximate} the implementation of an operation to
save qubits or gates, rather than implement it exactly. For example, a
programmer may prefer to substitute QFT with an approximate QFT, which
requires fewer gates. Of course, this substitution will affect the
circuit's semantics, and the programmer will want to understand the
\emph{maximum distance} (similarity) between the approximate and exact
implementations, to see if it is tolerable. To this end, we can test
a relational property between the outputs of an exact and approximate circuit, 
on the same inputs, to see if the difference is
acceptable. \Cref{sec:approx-circs} presents experiments comparing
the effect of approximation on circuits using QFT-based adders.

\subsection{\vqimp: A High-level Oracle Language}\label{sec:qimp}

\begin{figure}[t]
{\footnotesize
\centering
\[\hspace*{-1em}
\begin{array}{c}
\begin{array}{l}
\texttt{fixedp sin}(\textcolor{red}{Q~\texttt{fixedp }x_{/8}},\;\textcolor{red}{Q~\texttt{fixedp }x_r},\;C~\texttt{nat }n)\{
\\[0.2em]
\;\;\textcolor{red}{x_r\texttt{ = }x_{/8};} \;\;C~\texttt{fixedp }n_y;\;\;\textcolor{red}{Q~\texttt{fixedp }x_z;}\;\;
\textcolor{red}{Q~\texttt{fixedp }x_1;}
\\[0.2em]
\;\;
C~\texttt{nat }n_1;\;\;
C~\texttt{nat }n_2;\;\;
C~\texttt{nat }n_3;\;\;
C~\texttt{nat }n_4;\;\;
C~\texttt{nat }n_5;
\\[0.2em]
\;\;\texttt{for }(C~\texttt{nat }i\texttt{ = }0;\; i\texttt{ < }n;\;i\texttt{++})\{\\
\;\;\quad n_1\texttt{ = }i+1;\;\;n_2\texttt{ = }2*n_1;\;\;n_3\texttt{ = }\texttt{pow}(8,n_2);\;\;
          n_4\texttt{ = }n_2+1;
\\[0.2em]
\;\;\quad n_5\texttt{ = }n_4!;\;\;n_y\texttt{ = }n_3 / n_5;\;\;
\textcolor{red}{x_z\texttt{ = }\texttt{pow}(x_{/8},n_4);}\\[0.2em]
\;\;\quad \texttt{if }(\texttt{even}(n_1))\;\;{\{
\textcolor{red}{{x_1}\texttt{ = }{n_y}*{x_z};\;\;
x_r\texttt{ += }{x_1};}

\}}\\[0.2em]
\;\;\quad\texttt{else } {\{\textcolor{red}{{x_1}\texttt{ = }{n_y}*{x_z};\;\;{x_r}\texttt{ -= }{x_1};}\};}\\[0.2em]
\;\;\quad\textcolor{red}{\texttt{inv}(x_1);\;\;\texttt{inv}(x_z);}

\}\\
\;\;\texttt{return }\textcolor{red}{(8*x_r)};\\
\}
\end{array}\\[8em]
\sin{x}\approx 8*(\frac{x}{8}-\frac{8^2}{3!}(\frac{x}{8})^3+\frac{8^4}{5!}(\frac{x}{8})^5-\frac{8^6}{7!}(\frac{x}{8})^7+...+(-1)^{n-1}\frac{8^{2n-2}}{(2n-1)!}(\frac{x}{8})^{2n-1})
\end{array}
\]
}
\vspace*{-1em}
\caption{Implementing sine in \vqimp}
\label{fig:sine-impl}
\end{figure}

It is not uncommon for programmers to write oracles as metaprograms in
a quantum assembly's host language, e.g., as we did for \coqe{rz_adder} in
\Cref{fig:circuit-example}. But this process can be tedious and error-prone,
especially when trying to write optimized code.
To make writing efficient arithmetic-based quantum oracles easier,
we developed \vqimp, a higher-level imperative language that compiles
to \oqasm. Here we discuss \vqimp's basic features, describe how we 
optimize \vqimp programs during compilation using partial
evaluation, and provide correctness guarantees for \vqimp programs. 
Using \vqimp, we have defined operations for the ChaCha20 hash-function \cite{chacha}, exponentiation, sine, arcsine, and cosine, and tested program correctness by running inputs through \vqimp's semantics. 
More details about \vqimp are available in \Cref{sec:appendix}.

\myparagraph{Language Features}

An \vqimp program is a sequence of function definitions, with the last
acting as the ``main'' function. Each function definition is a series
of statements that concludes by returning a value $v$.  \vqimp statements contain
variable declarations, assignments (e.g., $x_r\texttt{ = }x_{/8}$),
arithmetic computations ($n_1\texttt{ = }i+1$), loops, conditionals,
and function calls.
Variables $x$ have types $\tau$, which are either primitive types
$\omega^m$ or arrays thereof, of size $n$. A primitive type pairs a
base type $\omega$ with a \emph{quantum mode} $m$. There are three base
types: type $\tnat$ indicates non-negative (natural) numbers; type
$\tfixed$ indicates fixed-precision real numbers in the range $(-1,1)$;
and type $\tbool$ represents booleans. The programmer specifies the
number of qubits to use to represent $\tnat$ and $\tfixed$ numbers
when invoking the \vqimp compiler.  
The mode $m \in\{C, Q\}$ on a primitive type indicates when a
type's value is expected to be known: $C$ indicates that the value is based
on a classical parameter of the oracle, and should be known at compile
time; $Q$ indicates that the value is a quantum input to the oracle, 
computed at run-time. 

\Cref{fig:sine-impl} shows the \vqimp implementation of the sine function,
which is used in quantum applications such as Hamiltonian
simulation~\cite{feynman1982simulating,Childs_2009}. 
Because $\tfixed$ types must be in the range $(-1,1)$, the function
takes $\frac{1}{8}$ times the input angle in variable $x_{/8}$ (the input 
angle $x$ is in $[0,2\pi)$). The result, stored in variable $x_r$, 
is computed by a Taylor expansion of $n$ terms.
The standard formula for the Taylor expansion is
$\sin{x}\approx
x-\frac{x^3}{3!}+\frac{x^5}{5!}-\frac{x^7}{7!}+...+(-1)^{n-1}\frac{x^{2n-1}}{(2n-1)!}$;
the loop in the algorithm computes an equivalent formula given input
$\frac{1}{8}x$, as shown at the bottom of the figure. 



\myparagraph{Reversible Computation}
\label{sec:revcomp}

Since programs that run on quantum computers must be
\emph{reversible}, \vqimp compiles functions to reverse their
effects upon returning. In \Cref{fig:sine-impl}, after the
\texttt{main} function returns, only the return value is copied and
stored to a target variable. For other values, like $x_{/8}$, the 
compiler will insert an \emph{inverse circuit} to revert all side effects.

When variables are reused within a function, they must be
\emph{uncomputed} using \vqimp's $\sinv{x}$ operation. For
example, in \Cref{fig:sine-impl}, the second \texttt{inv} operation
returns $x_z$ to its state prior to the execution of
$\textcolor{red}{x_z\texttt{=}\texttt{pow}(x_{/8},n_4)}$ so that $x_z$ 
can be reassigned in the next iteration.
We plan to incorporate automatic uncomputation techniques to insert $\sinv{x}$ calls automatically, but doing so requires care to avoid blowup in the generated circuit \cite{unqomp}. 

The \name compiler imposes three restrictions on the use of $\sinv{x}$, 
which aim to ensure that each use uncomputes just one assignment to $x$.
First, since the semantics of an \texttt{inv} operation reverses the
most recent assignment, we require that every \texttt{inv} operation
have a definite predecessor. Example \texttt{(1)} in \Cref{fig:inv-examples}
shows an \texttt{inv} operation on a variable that does not have a
predecessor; \texttt{(2)} shows a variable $z$ whose
predecessor is not always executed. Both are invalid in \vqimp.
Second, the statements between an \texttt{inv} operation and its
predecessor cannot write to any variables used in the body of the
predecessor. Example \texttt{(3)} presents an invalid case where $x$
is used in the predecessor of $z$, and is assigned between the
\texttt{inv} and the predecessor.  The third restriction is that, while
sequenced \texttt{inv} operations are allowed, the number of \texttt{inv}
operations must match the number of predecessors. Example \texttt{(4)}
is invalid, while \texttt{(5)} is valid, because the first
\texttt{inv} in \texttt{(5)} matches the multiplication assignment
and the second \texttt{inv} matches the addition assignment.

\begin{figure}[t]
\footnotesize
\[
\begin{array}{c}
\texttt{(1)}
\begin{array}{l}

a\texttt{=}x\texttt{ * }y;
\\
\texttt{inv}(z);
\textcolor{red}{\xmark}
\end{array}
\quad
\texttt{(2)}
\begin{array}{l}
\texttt{if}(x<y)\\
\;\;\;a\texttt{=}x\texttt{ * }y;\\
\texttt{else}\\
\;\;\;z\texttt{=}x\texttt{ * }y;
\\
\texttt{inv}(z);
\textcolor{red}{\xmark}
\end{array}

\quad
\texttt{(3)}
\begin{array}{l}

z\texttt{=}x\texttt{ * }y;
\\
x\texttt{=}x\texttt{ + }1;
\textcolor{red}{\xmark}
\\
\texttt{inv}(z);
\end{array}
\quad
\texttt{(4)}
\begin{array}{l}

z\texttt{=}x\texttt{ * }y;
\\
\texttt{inv}(z);
\\
\texttt{inv}(z);
\textcolor{red}{\xmark}
\end{array}
\quad
\texttt{(5)}
\begin{array}{l}

z\texttt{+=}x;
\\
z\texttt{=}x\texttt{ * }y;
\\
\texttt{inv}(z);
\\
\texttt{inv}(z);
\end{array}
\textcolor{green}{\cmark}
\end{array}
\]
\caption{Example (in)valid uses of \texttt{inv}}\label{fig:inv-examples}
\end{figure}

To implement these well-formedness checks, \name's \vqimp compiler maintains a 
stack of assignment statements. Once the compiler hits an \texttt{inv}
operation, it pops statements from the stack to find a match
for the variable being uncomputed. It also checks that none of the popped
statements contain an assignment of variables
used in the predecessor statement.

\myparagraph{Compilation from \vqimp to \oqasm}

The \vqimp compiler performs \emph{partial evaluation} \cite{partialeval} 
on the input program given
classical parameters; the residual program is compiled to a quantum
circuit.
In particular, we compile an \vqimp program by evaluating its $C$-mode
components, storing the results in a store $\sigma$, and then
using these results while translating its $Q$-mode components
into \oqasm code. For example, when compiling the \texttt{for} loop in
\Cref{fig:sine-impl}, the compiler will look up the value of
loop-bound variable $n$ in the store and update $i$'s value in the
store for each iteration. When compiling the loop-body statement
$n_1\texttt{=} i + 1$, variable $n_1$ will simply be updated in the
store, and no code generated. When compiling statement
$\textcolor{red}{x_z \texttt{=} \texttt{pow}({x_{/8}},n_4)}$, the fact that $x_z$ has
mode $Q$ means that \oqasm code must be generated. Thus, each iteration will
compile the non $C$-mode components of the body, essentially inlining the
loop. As an illustration, if we were to initialize $n$
to 3, the partially evaluated program would be equivalent to the
following (in \oqasm rather than \vqimp).

{
\begin{footnotesize}
\begin{center}
$
\begin{array}{l@{~}l}
\textcolor{red}{x_r\texttt{ = }x_{/8};}\; &
\textcolor{red}{{x_z}\texttt{ = }\texttt{pow}(x_{/8},3);}\;
\textcolor{red}{{x_1}\texttt{ = }{\frac{8^2}{3!}}*{x_z};\;{x_r}\texttt{ -= }{x_1};}\;
\textcolor{red}{\sinv{x_1};\;\sinv{x_z};}
\\
&\textcolor{red}{{x_z}\texttt{ = }\texttt{pow}(x_{/8},5);}\;
\textcolor{red}{{x_1}\texttt{ = }{\frac{8^4}{5!}}*{x_z};\;{x_r}\texttt{ += }{x_1};}\;
\textcolor{red}{\sinv{x_1};\;\sinv{x_z};}\;
\\
&\textcolor{red}{{x_z}\texttt{ = }\texttt{pow}(x_{/8},7);}\;
\textcolor{red}{{x_1}\texttt{ = }{\frac{8^6}{7!}}*{x_z};\;{x_r}\texttt{ -= }{x_1};}\;
\textcolor{red}{\sinv{x_1};\;\sinv{x_z};}
\end{array}
$
\end{center}
\end{footnotesize}
}

We have verified that compilation from \vqimp to \oqasm is correct, in
Coq, with a caveat: Proofs for assignment statements are
parameterized by correctness statements about the involved operators.
Each Coq operator function has a correctness statement associated with it; e.g., 
we state that the \oqasm code
produced by invoking \coqe{rz_adder} for addition corresponds to an addition at
the \vqimp level. In the case of \coqe{rz_adder} and a few others, we
have a proof of this in Coq; for the rest, we use PBT to provide some
assurance that the statement is true. Further details about \vqimp compilation and its correctness claims can be found in \Cref{sec:appendix}.

\section{Evaluation: Arithmetic Operators in \oqasm}
\label{sec:arith-oqasm}

We evaluate \name by (1) demonstrating how it can be used for validation, both by verification and random testing, and (2) by showing that it gets good performance in terms of resource usage compared to Quipper, a state-of-the-art quantum programming framework~\cite{Green2013}.
This section presents the arithmetic operators we have implemented in
\oqasm, while the next section discusses the geometric operators and
expressions implemented in \vqimp. The following section presents an
end-to-end case study applying Grover's search.

\subsection{Implemented Operators}

\Cref{fig:circ-evaluation,fig:op-table} summarize the operators we have implemented in \vqir. 

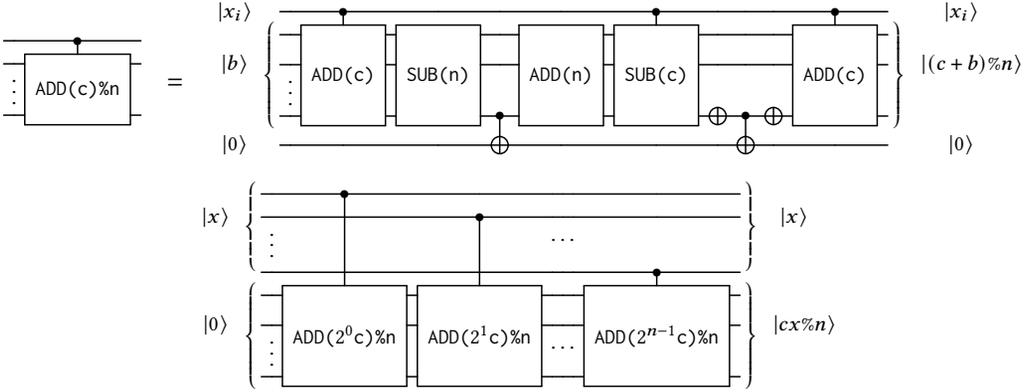
\begin{figure*}[t]
\centering
\centering
\begin{tabular}{c@{$\quad=\quad$}c}
  \begin{minipage}{0.2\textwidth}
  \Small
  \Qcircuit @C=0.5em @R=0.5em {
    & \qw & \ctrl{1} & \qw \\
    & \qw & \multigate{3}{\texttt{ADD(c)\%n}} & \qw \\
    & \vdots & & \\
    & & & \\
    & \qw & \ghost{\texttt{ADD(c)\%n}} & \qw \\
    }
  \end{minipage} &
  \begin{minipage}{0.75\textwidth}
  \Small
  \Qcircuit @C=0.5em @R=0.5em {
    & \ket{x_i}\quad & & \qw & \ctrl{1}  & \qw & \qw & \qw & \ctrl{1} & \qw & \qw & \qw & \ctrl{1} & \qw & \ket{x_i} \\
    & && \qw & \multigate{4}{\texttt{ADD(c)}} & \multigate{4}{\texttt{SUB(n)}} & \qw & \multigate{4}{\texttt{ADD(n)}} & \multigate{4}{\texttt{SUB(c)}} & \qw & \qw & \qw & \multigate{4}{\texttt{ADD(c)}} & \qw & \\
    & \push{\ket{b}\quad} & & \qw & \ghost{\texttt{ADD(c)}} & \ghost{\texttt{SUB(n)}} & \qw & \ghost{\texttt{ADD(n)}} & \ghost{\texttt{SUB(c)}} & \qw & \qw & \qw & \ghost{\texttt{ADD(c)}} & \qw & \push{\quad\ket{(c+b)\%n}} \\
    & & & \vdots & & & & & & & & & & & & \\
    & & & & & & & & & & & & & & & \\
    & & & \qw & \ghost{\texttt{ADD(c)}} & \ghost{\texttt{SUB(n)}} & \ctrl{1} & \ghost{\texttt{ADD(n)}} & \ghost{\texttt{SUB(c)}} & \targ & \ctrl{1} & \targ & \ghost{\texttt{ADD(c)}} & \qw & \\
    & \ket{0}\quad && \qw & \qw & \qw & \targ & \qw & \qw & \qw & \targ & \qw & \qw & \qw & \ket{0}
    \gategroup{2}{3}{6}{3}{1em}{\{}
    \gategroup{2}{14}{6}{14}{1em}{\}}
    }
  \end{minipage}
\end{tabular}

\vspace{1em}
\begin{tabular}{c}
  \begin{minipage}{0.5\textwidth}
  \Small
  \Qcircuit @C=0.5em @R=0.5em {
    & & & \qw & \ctrl{6} & \qw & \qw & \qw & \qw & \qw & \qw & \\
    & \push{\ket{x}\quad} & & \qw & \qw & \ctrl{5} & \qw & \qw & \qw & \qw & \qw & \ket{x} \\
    & & & \vdots & & & & \dots & & & & \\
    & & & & & & & & & & & \\
    & & & & & & & & & & & \\
    & & & \qw & \qw & \qw & \qw & \qw & \qw & \ctrl{1} & \qw & \\
    & & & \qw & \multigate{4}{\texttt{ADD($2^0$c)\%n}} & \multigate{4}{\texttt{ADD($2^1$c)\%n}} & \qw & \qw & \qw & \multigate{4}{\texttt{ADD($2^{n-1}$c)\%n}} & \qw & \\
    & \push{\ket{0}\quad} & & \qw & \ghost{\texttt{ADD($2^0$c)\%n}} & \ghost{\texttt{ADD($2^1$c)\%n}} & \qw & \qw & \qw & \ghost{\texttt{ADD($2^{n-1}$c)\%n}} & \qw & \push{\quad\ket{cx\%n}} \\
    & & & \vdots & & & & \dots & & & & \\
    & & & & & & & & & & & \\
    & & & \qw & \ghost{\texttt{ADD($2^0$c)\%n}} & \ghost{\texttt{ADD($2^1$c)\%n}} & \qw & \qw & \qw & \ghost{\texttt{ADD($2^{n-1}$c)\%n}} & \qw
    \gategroup{1}{3}{5}{3}{1em}{\{}
    \gategroup{7}{3}{11}{3}{1em}{\{}
    \gategroup{1}{11}{5}{11}{1em}{\}}
    \gategroup{7}{11}{11}{11}{1em}{\}}
    }
  \end{minipage}
\end{tabular}
\caption{Structure of modular multiplication circuits}
\label{fig:mod-mult}
\end{figure*}

The addition and modular multiplication circuits 
(parts (a) and (d) of \Cref{fig:circ-evaluation}) are components of the oracle used in Shor's factoring algorithm~\cite{shors}, which accounts for most of the algorithm's cost \cite{Gidney2021howtofactorbit}.
The oracle performs modular exponentiation on natural numbers via modular multiplication, which takes a quantum variable $x$ and two co-prime constants $M, N \in \mathbb{N}$ and produces $(x * M) \% N$. We have implemented two modular multipliers---inspired by \citet{qft-adder}
and \citet{ripple-carry-mod}---in \vqir. 
Both modular multipliers are constructed using controlled modular addition by a constant, which is implemented in terms of controlled addition and subtraction by a constant, as shown in \Cref{fig:mod-mult}.
The two implementations differ in their underlying adder and subtractor circuits: the first (QFT) uses a quantum Fourier transform-based circuit for addition and subtraction \cite{Draper2000AdditionOA}, while the second (TOFF) uses a ripple-carry adder \cite{ripple-carry-mod}, which uses classical controlled-controlled-not (Toffoli) gates.

Part (b) of \Cref{fig:circ-evaluation} shows results for \oqasm implementations of multiplication (without the modulo) and part (c) shows results for modular division by a constant, which is useful in Taylor series expansions used to implement operators like sine and cosine.
\Cref{fig:op-table} lists additional operations we have implemented in \vqir for arithmetic and Boolean comparison using natural and fixed-precision numbers.

\begin{figure*}[t]
{\footnotesize
\hspace*{-1em}
\begin{tabular}{c @{\quad} c}
\begin{minipage}[b]{0.48\textwidth}
\centering
\begin{tabular}{|l|c|c|c|}
\hline
		     & \# qubits  & \# gates & Verified \\
                     \hline
\oqasm TOFF & 33 & 423 & \cmark \\
\oqasm QFT & 32 & 1206 & \cmark \\
\oqasm QFT (const) & 16 & $756 \pm 42$ &  \cmark \\ \hline
Quipper TOFF & 47 & 768 &   \\
Quipper QFT & 33 & 6868 &   \\
Quipper TOFF (const) & 31 & $365 \pm 11$  &   \\
\hline                           
\end{tabular}
  \subcaption{Addition circuits (16 bits)}
\end{minipage}
&
\begin{minipage}[b]{0.52\textwidth}
\centering
\begin{tabular}{|l|c|c| >{\centering\arraybackslash} m{1cm} |}
\hline
                     & \# qubits  & \# gates & QC time (16 / 60 bits)\\
                     \hline
\oqasm TOFF & 49 & 11265  & 6 / 74 \\
\oqasm TOFF (const) & 33 & $1739 \pm 367$  & 3 / 31 \\
\oqasm QFT & 48 & 4339 & 4 / 138 \\
\oqasm QFT (const)  & 32 & $1372 \pm 26$  & 4 / 158 \\ \hline
Quipper TOFF & 63 & 8060  & \\
Quipper TOFF (const)  & 41 & $2870\pm 594$   & \\       
\hline                           
\end{tabular}
  \subcaption{Multiplication circuits (16 bits)}
\end{minipage}
\vspace{0.5mm}
\end{tabular}
\\
\hspace*{-1em}
\begin{tabular}{c @{\;\;} c}
\begin{minipage}[b]{0.52\textwidth}
\centering
\begin{tabular}{|l|c|c| >{\centering\arraybackslash} m{1cm} |}
\hline
		     & \# qubits  & \# gates &  QC time (16 / 60 bits) \\
                     \hline
\oqasm TOFF (const) & 49 & $28768 $ &  16 / 397 \\
\oqasm QFT (const) & 34 & 15288 & 5 / 412 \\
\oqasm AQFT (const) & 34 & 5948 & 4 / 323 \\\hline
Quipper TOFF & 98 & 37737 &  \\
\hline                           
\end{tabular}
  \subcaption{Division/modulo circuits (16 bits)}
\end{minipage}
&
\begin{minipage}[b]{0.52\textwidth}
\centering
\begin{tabular}{|l|c|c|c|}
\hline
      & \# qubits  & \# gates & Verified   \\
                     \hline
\oqasm TOFF (const) & 41 & 56160 & \cmark \\
\oqasm QFT (const) & 19 & 18503 &\cmark \\
\hline                           
\end{tabular}
  \subcaption{Modular multiplication circuits (8 bits)}
\end{minipage}
\end{tabular}
}

\caption{Comparison of \oqasm and Quipper arithmetic operators. In the ``const'' case, one argument is a classically-known constant parameter. 
For (a)-(b) we present the average ($\pm$ standard deviation) over 20 randomly selected constants $c$ with $0 < c < 2^{16}$.
For division/modulo, $x \textsf{ mod } n$, we only consider the case when $n=1$, which results in the maximum number of circuit iterations; the Quipper version assumes $n$ is a variable, but uses the same number of iterations as the constant case when $n=1$.
In (d), we use the constant 255 ($=2^8-1$) for the modulus and set the other constant to 173 (which is invertible mod 255). 
Quipper supports no QFT-based circuits aside from an adder. ``QC time'' is the time (in seconds) for QuickChick to run 10,000 tests.}
\label{fig:circ-evaluation}
\end{figure*}

\begin{figure}[t]
\centering
{\small
\hspace*{-1em}
\begin{tabular}{|l|>{\centering\arraybackslash}p{3.2cm}<{}|>{\centering\arraybackslash}p{3.3cm}<{}|}
\hline
type &Verified & Randomly Tested 
\\[0.5em]
\hline
Nat /Bool &
\vspace{-0.5em}
 \begin{adjustwidth}{-1em}{}
\begin{tabular}{l}
 $[x\texttt{-}N]_{q}\;$ $[N\texttt{-}x]_{q}\;$ $[x\texttt{-}y]_{q,t}\;$\\
  $[x\texttt{=}N]_{q,t}\;$
 $[x\texttt{<}N]_{q,t}\;$ \\$[x\texttt{=}y]_{q,t}\;$ $[x\texttt{<}y]_{q,t}$
\end{tabular}
\end{adjustwidth}
&
\vspace{-0.5em}
 \begin{adjustwidth}{-1em}{}
\begin{tabular}{l}
$[x\texttt{+}N]_{a}\;$
$[x\texttt{+}y]_{a}\;$
$[x\texttt{-}N]_{t}$ \\
$[N\texttt{-}x]_{t}\;$
 $[x\texttt{-}y]_{q}\;$
 $[x\texttt{\%}N]_{a,q,t}$\\
 $[x\texttt{/}N]_{a,q,t}\;$
\end{tabular}
\end{adjustwidth}
\\[0.5em]
\hline
FixedP &
\vspace{-0.5em}
 \begin{adjustwidth}{-1em}{}
\begin{tabular}{l}
$[x\texttt{+}N]_{q}\;$ $[x\texttt{+}y]_{t}\;$ $[x\texttt{-}N]_{q}$ \\
 $[N\texttt{-}x]_{q}\;$ $[x\texttt{-}y]_{t}\;$ $[x\texttt{=}N]_{q,t}$\\
 $[x\texttt{<}N]_{q,t}$ $[x\texttt{=}y]_{q,t}$ $[x\texttt{<}y]_{q,t}$
\end{tabular}
\end{adjustwidth}
&
\vspace{-0.5em}
 \begin{adjustwidth}{-1em}{}
\begin{tabular}{l}
$[x\texttt{+}N]_{t}\;$  $[x\texttt{+}y]_{q}\;$  $[x\texttt{-}N]_{t}$\\
 $[N\texttt{-}x]_{t}\;$  $[x\texttt{-}y]_{q}\;$  $[x\texttt{*}N]_{q,t}$\\
 $[x\texttt{*}y]_{q,t}\;$  $[x\texttt{/}N]_{q,t}$
\end{tabular}
\end{adjustwidth}
\\[0.5em]
\hline
\end{tabular}\\[1em]
}
{\scriptsize $x$,$y$ = variables, $N$ = constant,\\
 $[]_{a,q,t}$ = AQFT-based ($a$), QFT-based ($q$), or Toffoli-based ($t$)\\
All testing is done with 16-bit/60-bit circuits.
}
\caption{Other verified \& tested operations}
\label{fig:op-table}
\end{figure}

\subsection{Validating Operator Correctness}

As shown in \Cref{fig:circ-evaluation}, we have fully verified the
adders and modular multipliers used in Shor's
algorithm. These constitute the first proved-correct implementations
of these functions, as far as we are aware. 

All other operations in the figure were tested with Quick\-Chick. To
ensure these tests were efficacious, we confirmed they could find
hand-injected bugs; e.g., we reversed the input bitstrings for the QFT
adder (\Cref{fig:circuit-example}) and confirmed that testing found
the endianness bug.  The tables in \Cref{fig:circ-evaluation} give the
running times for the QuickChick tests---the times include the cost of
extracting the Coq code to OCaml, compiling it, and running it with
10,000 randomly generated inputs. We tested these operations both on
16-bit inputs (the number that's relevant to the reported qubit and
gate sizes) and 60-bit inputs. For the smaller sizes, tests complete
in a few seconds; for the larger sizes, in a few minutes. For
comparison, we translated the operators' \vqir programs to \sqir,
converted the \sqir programs to OpenQASM 2.0 \cite{Cross2017}, and
then attempted to simulate the resulting circuits on test inputs using
the DDSim~\cite{ddsim}, a state-of-the-art quantum simulator. Unsurprisingly, the simulation
of the 60-bit versions did not complete when running overnight. 

We also verified and property-tested several other operations, as
shown in \Cref{fig:op-table}.

During development, we found two bugs in the original presentation of the QFT-based modular multiplier \cite{qft-adder}. The first issue was discovered via random testing and relates to assumptions about the endianness of stored integers. The binary number in Figure 6 of the paper uses a little-endian format whereas the rest of the circuit assumes big-endian.
Quipper's implementation of this algorithm solves the problem by creating a function in their Haskell compiler to reverse the order of qubits. 
In \vqir, we can use the \texttt{Rev} operation (which does not insert \texttt{SWAP}s) to correct the format of the input binary number.

The second issue was discovered during verification. \citet{qft-adder} indicates that the input $x$ should be less than $2^n$ where $n$ is the number of bits. However, to avoid failure the input must \emph{actually} be less than $N$, where $N$ is the modulus defined in Shor's algorithm. To complete the proof of correctness, we needed to insert a preprocessing step to change the input to $x \% N$. 
The original on-paper implementation of the ripple-carry-based modular multiplier \cite{ripple-carry-mod} has the same issue. 

\ignore{
In addition to formal verification and random testing, we also ``spot-checked'' results by translating \vqir programs to \sqir, converting the \sqir programs to OpenQASM 2.0 \cite{Cross2017}, and simulating the resulting circuits on test inputs.
For this manual testing, we generated circuits for 8 bits and simulated each circuit with four manually generated inputs using DDSIM \cite{ddsim}.
This helped to prevent bugs in the parts of our toolchain not implemented in Coq (e.g., extraction from Coq to OCaml and OpenQASM file I/O).
}

\subsection{Operator Resource Usage}

\Cref{fig:circ-evaluation} compares the resources used by \vqir operators with counterparts in Quipper. In both cases, we compiled the operators to OpenQASM 2.0 circuits,\footnote{We converted the output Quipper files to OpenQASM 2.0 using a compiler produced at Dalhousie University \cite{quipper-qasm}.} and then ran the circuits through the \voqc optimizer~\cite{VOQC} to ensure that the outputs account for inefficiencies in automatically-generated circuit programs (e.g., no-op gates inserted in the base case of a recursive function). \voqc outputs the final result to use gates preferred by the Qiskit compiler~\cite{Qiskit}, which are the single-qubit gates $U_1, U_2, U_3$ and the two-qubit gate $CNOT$. 

We also provide resource counts (computed by the same procedure) for our implementations of 8-bit modular multiplication. Quipper does not have a built-in operation for modular multiplication (which is different from multiplication followed by a modulo operator in the presence of overflow). 

We define all of the arithmetic operations in \Cref{fig:circ-evaluation} for arbitrary input sizes; the limited sizes in our experiments (8 and 16 bits) are to account for inefficiencies in \voqc. For the largest circuits (the modular multipliers), running \voqc takes about 10 minutes.

\myparagraph{Comparing QFT and Toffoli-based operators}

The results show that the QFT-based implementations always use fewer
qubits. This is because they do not need ancillae
to implement reversibility. For both division/modulo and modular
multiplication (used in Shor's oracle), the savings are substantial
because those operators are not easily reversible using Toffoli-based
gates, and more ancillae are needed for uncomputation.

The QFT circuits also typically use fewer gates. 
This is partially due to algorithmic advantages of QFT-based arithmetic, 
partially due to \voqc (\voqc reduced QFT circuit gate
counts by 57\% and Toffoli circuit gate counts by 28\%) 
and partially due to the optimized decompositions we use to convert 
many-qubit gates to the one- and two-qubit gates supported by \voqc.%
\footnote{We use the decompositions for Toffoli and controlled-Toffoli at \url{https://qiskit.org/documentation/_modules/qiskit/circuit/library/standard_gates/x.html}; the decomposition for controlled-$Rz$ at \url{https://qiskit.org/documentation/_modules/qiskit/circuit/library/standard_gates/u1.html}; and the decomposition for controlled-controlled-$Rz$ at \url{https://quantumcomputing.stackexchange.com/questions/11573/controlled-u-gate-on-ibmq}. The decompositions we use are all proved correct in the \sqir development. All of the decompositions are ancilla free.}
We found during evaluation that gate counts are highly sensitive
to the decompositions used: Using a more
na\"{i}ve decomposition of the controlled-Toffoli gate (which simply
computes the controlled version of every gate in the standard Toffoli
decomposition) increased the size of our Toffoli-based modular multiplication circuit by 1.9x, and a similarly na\"{i}ve decomposition of the controlled-controlled-$Rz$ gate increased the size of our QFT-based modular multiplication circuit by 4.4x.
We also found that gate counts (especially for the Toffoli-based circuits) are sensitive to choice of constant parameter: The QFT-based constant multiplication circuits had between 1320 and 1412 gates, while the Toffoli-based circuits had between 988 and 2264.
Unlike gate counts, qubit counts
are more difficult to optimize because they require fundamentally
changing the structure of the circuit; this makes QFT's qubit savings
for modular multiplication even more impressive.

Overall, our results suggest that QFT-based arithmetic provides better and more consistent performance, so when compiling \vqimp programs (like the sine function in \Cref{fig:sine-impl}) to \oqasm, we should bias towards using the QFT-based operators.

\myparagraph{Comparing to Quipper}

Overall, \Cref{fig:circ-evaluation}(a)-(c) shows that operator
implementations in \vqir consume resources comparable to those
available in Quipper, often using fewer qubits and gates, both for
Toffoli- and QFT-based operations.
In the case of the QFT adder, the difference in results is because the Quipper-to-OpenQASM converter we use has a more expensive decomposition of controlled-$Rz$ gates.\footnote{\citet{quipper-qasm} decomposes a controlled-$Rz$ gate into a circuit that uses two Toffoli gates, an $Rz$ gate, and an ancilla qubit. In \voqc, each Toffoli gate is decomposed into 9 single-qubit gates and 6 two-qubit gates. In contrast, \name's decomposition for controlled-$Rz$ uses 3 single-qubit gates, 2 two-qubit gates, and no ancilla qubits.}
In the other cases (all Toffoli-based circuits), we made choices
when implementing the oracles that improved their
resource usage. Nothing fundamental stopped the Quipper
developers from having made the same choices, but we note they did
not have the benefit of the \oqasm type system and PBT
framework. Quipper has recently begun to develop a random testing
framework based on QuickCheck~\cite{10.1145/351240.351266},
but it only applies to Toffoli-based (i.e., classical) gates.

\subsection{Approximate Operators}
\label{sec:approx-circs}


\oqasm's efficiently-simulable semantics can be used to predict the effect of using approximate components, which enables a new workflow for optimizing quantum circuits:
Given an exact circuit implementation, replace a subcomponent with an
approximate implementation; use \name's PBT framework to compare the
outputs between the exact and approximate circuits; and finally
decide whether to accept or reject the approximation based on the results of these tests, iteratively improving performance.

In this section, we use \name's PBT framework to study the effect of replacing QFT circuits with AQFT circuits (\Cref{fig:circuit-example}) in addition and division/modulo circuits.

\myparagraph{Approximate Addition}

\Cref{fig:approx-results}(a) shows the results of replacing QFT with AQFT in the QFT adder from \Cref{fig:circ-evaluation}(a).
As expected, a decrease in precision leads to a decrease in gate count.
On the other hand, our testing framework demonstrates that this also increases error (measured as absolute difference accounting for overflow, maximized over randomly-generated inputs).
Random testing over a wider range of inputs suggests that dropping $b$ bits of precision from the exact QFT adder always induces an error of at most $\pm 2^b - 1$.
This exponential error suggests that the ``approximate adder'' is not particularly useful on its own, as it is effectively ignoring the least significant bits in the computation.
However, it computes the most significant bits correctly: if the inputs are both multiples of $2^b$ then an approximate adder that drops $b$ bits of precision will always produce the correct result.

\begin{figure*}[t]
  {\footnotesize
  \hspace*{-1em}
  \begin{tabular}{c @{\quad} c}
  \begin{minipage}[b]{0.4\textwidth}
  \centering
  \begin{tabular}{|l|c|c|}
    \hline
    Precision & \# gates & Error \\
    \hline
    16 bits (full) & 1206 & $\pm$ 0 \\
    15 bits & 1063 & $\pm$ 1 \\
    14 bits & 929 & $\pm$ 3 \\ \hline
  \end{tabular}
    \subcaption{Varying the precision in a 16-bit adder}
  \end{minipage}
  &
  \begin{minipage}[b]{0.55\textwidth}
  \centering
  \begin{tabular}{|l|c|c|c|c|}
  \hline
  \# iters. ($I+1$) & TOFF & QFT & AQFT & \% savings \\ \hline
  1 & 1798 & 1794 & 1717 & 4.5 / 4.5 \\
  $4$ & 7192 & 4432 & 3488 & 48.5 / 21.2 \\
  $8$ & 14384 & 8017 & 4994 & 65.2 / 37.7 \\
  ${12}$ & 21576 & 11637 & 5684 & 73.6 / 51.1 \\
  ${16}$ & 28768 & 15288 & 5948 & 79.3 / 61.1 \\
  \hline
  \end{tabular}
    \subcaption{Gate counts for TOFF vs. QFT vs. AQFT division/modulo circuits; the righthand column shows the savings for TOFF vs. AQFT and QFT vs. AQFT}
  \end{minipage}
  \vspace{0.5mm}
  \end{tabular}
  }
  
  \caption{Effects of approximation}
  \label{fig:approx-results}
  \end{figure*}

\myparagraph{Exact Division/Modulo using an Approximate Adder}
\label{sec:qft-moder}

\begin{figure*}[t]
{\hspace*{2.3em}
\begin{tabular}{c }
\begin{minipage}{.4\textwidth}
  \footnotesize
  \Qcircuit @C=0.25em @R=0.4em {
    \lstick{\qket{x_{n-1}}} & \multigate{4}{\texttt{$x-2^{I-i} n$}} & \multigate{4}{ \texttt{QFT}^{-1}\;N } & \ctrl{9} & \multigate{4}{ \texttt{QFT}\;N } & \multigate{4}{\texttt{$x+2^{I-i} n$}} & \qw & \qw & \qw \\
    \lstick{\qket{x_{n-2}}} &  \ghost{\texttt{$x-2^{I-i} n$}} & \ghost{ \texttt{QFT}^{-1}\;N } & \qw & \ghost{ \texttt{QFT}\;N } & \ghost{\texttt{$x+2^{I-i} n$}} & \qw & \qw & \qw \\
    \lstick{\vdots} & & & & & & & &  \rstick{\vdots} \\
    \lstick{} & &  & & & & & & \\
    \lstick{\qket{x_0}} & \ghost{\texttt{$x-2^{I-i} n$}}  & \ghost{ \texttt{QFT}^{-1}\;N } & \qw & \ghost{ \texttt{QFT}\;N } & \ghost{\texttt{$x+2^{I-i} n$}} & \qw & \qw & \qw  \\
\lstick{} & & & & & & & & &\\
    \lstick{\ket{b_{n-1}}} & \qw & \qw & \qw  & \qw & \qw & \qw & \qw & \qw   \\
    \lstick{\vdots} & & & & \dots &  & &   \\
    \lstick{} & & & & &  & & &   \\
    \lstick{\ket{b_{1}}} & \qw & \qw  &  \targ  & \qw & \ctrl{-5} & \qw & \targ & \qw   \\
    \lstick{\vdots} & & & & & & & &  \rstick{\vdots} \\
    \lstick{} & & & & & & & & &  \\
    \lstick{\ket{b_0}} & \qw & \qw & \qw & \qw & \qw & \qw & \qw & \qw  
    }
\subcaption{QFT-based}
\end{minipage} 
\\\\
\begin{minipage}{.7\textwidth}
  \footnotesize
  \Qcircuit @C=0.25em @R=0.4em {
    \lstick{\qket{x_{n-1}}} & \multigate{4}{\texttt{$x-2^{I-i} n$}} & \multigate{4}{ \texttt{QFT}^{-1}\;(N-i) } & \qw & \qswap & \qw & \multigate{4}{ \texttt{RSH} } & \multigate{4}{ \texttt{QFT}\;(N-i-1) } & \multigate{4}{x+(2^{I-i} n \!\mod 2^{I-i-1})} & \qw & \qw & \qw & \qw  \\
    \lstick{\qket{x_{n-2}}} &  \ghost{\texttt{$x-2^{I-i} n$}} & \ghost{ \texttt{QFT}^{-1}\;(N-i) } & \qw & \qw \qwx &\qw & \ghost{ \texttt{RSH} } & \ghost{ \texttt{QFT}\;(N-i-1) } & \ghost{x+(2^{I-i} n \!\mod 2^{I-i-1})} & \qw & \qw & \qw & \qw \\
    \lstick{\vdots} & & & & \qwx & & & & & & & &  \rstick{\vdots} \\
    \lstick{} & &  & & \qwx & & & & & & & &  \\
    \lstick{\qket{x_0}} & \ghost{\texttt{$x-2^{I-i} n$}}  & \ghost{ \texttt{QFT}^{-1}\;(N-i) } & \qw & \qw \qwx & \qw & \ghost{\texttt{RSH}} & \ghost{ \texttt{QFT}\;(N-i-1) } & \ghost{x+(2^{I-i} n \!\mod 2^{I-i-1})} & \qw & \qw & \qw & \qw  \\
    \lstick{} & & & & \qwx & & & & & & & & & \\
    \lstick{\ket{b_{n-1}}} & \qw & \qw & \qw & \qw \qwx & \qw & \qw & \qw & \qw & \qw & \qw & \qw & \qw \\
    \lstick{\vdots} & & & & \qwx & & & \dots &  & & & & \\
    \lstick{} & & & & \qwx & & & & & & & & & \\
    \lstick{\ket{b_{1}}} & \qw & \qw  & \qw &  \qswap \qwx & \qw & \qw & \qw & \ctrl{-5} & \qw & \targ & \qw & \qw \\
    \lstick{\vdots} & & & & & & & & &&  & &  \rstick{\vdots} \\
    \lstick{} & & & & & & & & && & &  \\
    \lstick{\ket{b_0}} & \qw & \qw & \qw & \qw & \qw & \qw & \qw & \qw & \qw & \qw & \qw  & \qw 
    }
\subcaption{AQFT-based (addition and subtraction are approximate)}
\end{minipage} 
\end{tabular}
}
\caption{One step of the QFT/AQFT division/modulo circuit}
\label{fig:qft-moder}
\end{figure*}
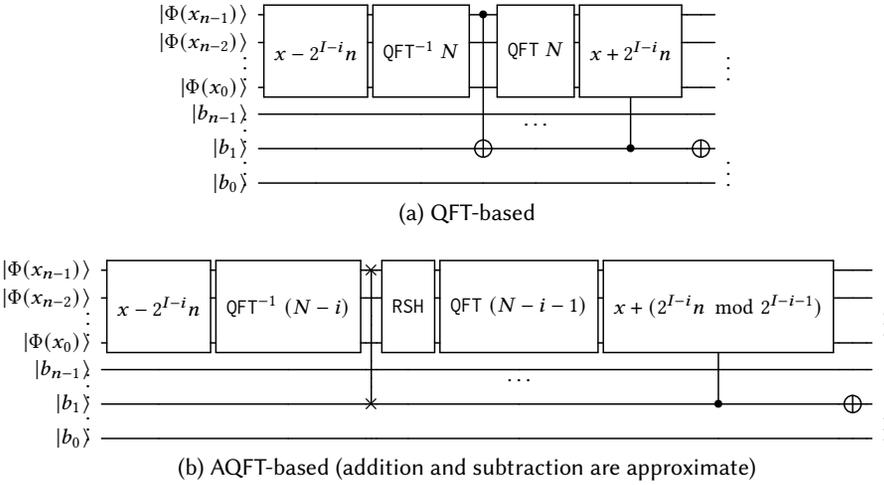

\ignore{
\begin{figure*}[t]
\centering

\begin{coq}
Fixpoint appx_moder' i (n:nat) (b:nat) (x ex:var) (M:nat -> bool) := 
     match i with 0 =>  (SKIP (x,0))
           | S j => appx_compare_half3 x n b (ex,j) M ;  Rshift x;
                     QFT x b; (CU (ex,j) ((appx_adder x n b M)));
                      (X (ex,j)); 
                       appx_moder' j n (b+1) x ex (cut_n (div_two_spec M) n)
     end.
\end{coq}
\caption{Approximate QFT Modulo Operation (Core of Addition/Subtraction in \Cref{fig:circuit-add-sub})}
\label{fig:qft-moder}
\end{figure*}
}

Even though the approximate adder is not particularly useful for addition, there are still cases where it can be useful as a subcomponent. 
For example, the modulo/division circuit relies on an addition subcomponent, but does not need every bit to be correctly added.

\Cref{fig:qft-moder}(a) shows one step of an $N$-bit QFT-based modulo circuit that computes $x\!\!\mod n$ for constant $n$.
The algorithm runs for $I+1$ iterations, where $2^{N-1} \le 2^I n <2^N$, with the iteration counter $i$ increasing from 0 to $I$ (inclusive).
In each iteration, the circuit in \Cref{fig:qft-moder}(a) computes $x-2^{I-i} n$ and uses the result's most significant bit (MSB) to check whether $x < 2^{N-1-i}$.
If the MSB is $0$, then $x \ge 2^{N-1-i}$ and the circuit continues to next iteration; otherwise, it adds $2^{I-i} n$ to the result and continues.

We can improve the resource usage of the circuit in \Cref{fig:qft-moder}(a) by replacing the addition, subtraction, and QFT components with approximate versions, as shown in \Cref{fig:qft-moder}(b).
At the start of each iteration, $x < 2^{N-i}$, so it is safe to replace components with versions that will perform the intended operation on the lowest $(N-i)$ bits.
The circuit in \Cref{fig:qft-moder}(b) begins by subtracting the top $(N-i)$ bits, and then converts $x$ back to the \texttt{Nor} basis using an $(N-i)$-bit precision QFT\@.
It then swaps the MSB with an ancilla, guaranteeing that the MSB is 0. 
Next, it uses a \texttt{Rshift} to move the cleaned MSB to become the lowest significant bit (effectively, multiplying $x$ by 2) and uses a $(N-i-1)$-bit precision QFT to convert back to the \texttt{Phi} basis.
Finally, it conditionally adds back the top $(N-i-1)$ bit of the value $(2^{I-i} n \!\mod 2^{I-i-1})$, ignoring the original MSB.

The result is a division/modulo circuit that uses approximate components, but, as our testing assures, is exactly correct.
\Cref{fig:approx-results}(b) shows the required resources for varying numbers of iterations.
Compared to the QFT-based circuit,
for a single iteration, the approximation provides a 4.5\% savings. 
And the saving increases with more iterations.
In the case of the maximum number of iterations (16 for $n=1$), the AQFT-based division/modulo circuit uses 61.1\% fewer gates than the QFT-based implementation and 79.3\% fewer gates than the Toffoli-based implementation.

\section{Evaluation: \vqimp Oracles and Partial Evaluation}
\label{sec:partial-eval}

The prior section considered arithmetic operators implemented in \oqasm, 
which are the building blocks for operators we have programmed using \vqimp, 
including sine, arcsine, cosine, arccosine, and exponentiation on fixed-precision
numbers. 
These operators are useful in near-term applications; for example, the
arcsine and sine functions are used in the quantum walk algorithm~\cite{Childs_2009}. 
We used \vqimp's source semantics to test each operator's correctness.

As discussed in \Cref{sec:qimp}, one of the key features of \vqimp is \emph{partial evaluation} during compilation to \vqir.
The simplest optimization similar to partial evaluation happens for a
binary operation $x := x\odot y$, where $y$ is a constant value. 
\Cref{fig:circ-evaluation} hints at the power of partial evaluation for this case---all constant operations (marked ``const'') generate circuits with significantly fewer qubits and gates.
Languages like Quipper take advantage of this by producing special circuits for operations that use classically-known constant parameters.

Partial evaluation takes this one step further, pre-evaluating as much of the circuit as possible.
For example, consider the fixed precision operation $\frac{x*y}{M}$ where
$M$ is constant and a natural number, and $x$ and $y$ are two fixed precision numbers that may be constants.
This is a common pattern, appearing in many quantum oracles (recall the $\frac{8^n*x}{n!}$ in the Taylor series decomposition of sine). 
In Quipper, this is expression compiled to ${r_1}\texttt{ = }{\frac{x}{M}}; {r_2}\texttt{ = }{r1*y}$.
The \vqimp compiler produces different outputs depending on whether $x$ and $y$ are constants. If they both are constant, \vqimp simply assigns the result of computing $\frac{x*y}{M}$ to a quantum variable. If $x$ is a constant, but $y$ is not, \vqimp evaluates $\frac{x}{M}$ classically, assigns the value to $r_1$, and evaluates $r_2$ using a constant multiplication circuit. If they are both quantum variables, \vqimp generates a circuit to evaluate the division first and then the multiplication.

In \Cref{fig:self-data}~(a) we show the size of the circuit generated for $\frac{x*y}{M}$ where zero, one, or both variables are classically known. 
It is clear that more classical variables in a program lead to a more efficient output circuit.
If $x$ and $y$ are both constants, then only a constant assignment circuit is needed, which is a series of \texttt{X} gates. 
Even if only one variable is constant, it may lead to substantial savings: In this example, if $x$ is constant, the compiler can avoid the division circuit and use a constant multiplier instead of a general multiplier.
These savings quickly add up: \Cref{fig:self-data}~(b) shows the qubit size difference between our implementation of sine and Quippers'. Both the TOFF and QFT-based circuits use fewer than $7\%$ of the qubits used by Quipper's sine implementation.
\footnote{\vqimp also benefits from its representation of fixed-precision numbers (\Cref{sec:qimp}), which is more restrictive than Quipper's. Our representation of fixed-precision numbers reduces the qubit usage of the sine function by half, so about half of the qubit savings can be attributed to this.}

\begin{figure}[t]
{\small
\begin{subfigure}[b]{.6\textwidth}
\centering
\begin{tabular}{| l | c | c |}
\hline
                     & \# qubits  & \# gates   \\
                     \hline
OQIMP ($x$, $y$ const) & 16 & 16\\
OQIMP TOFF ($x$ const) & 33 & $1739 \pm 376$ \\
OQIMP QFT ($x$ const) & 16 & $1372 \pm 26$ \\
OQIMP TOFF & 33 & 61470   \\
OQIMP QFT & 32  & 25609  \\
\hline                           
\end{tabular}
  \subcaption{
Fixed-precision circuits for $\frac{x*y}{M}$ with $M=5$ (16 bits)
}
\end{subfigure}
\hfill
\begin{subfigure}[b]{.35\textwidth}
\centering
\begin{tabular}{| l | c|}
\hline
                     & \# qubits   \\
                     \hline
OQIMP TOFF & 418   \\
OQIMP QFT & 384  \\
Quipper & 6142  \\
\hline                           
\end{tabular}
  \subcaption{
Sine circuits (64 bits)}
\end{subfigure}
}
\caption{Effects of partial evaluation}
\label{fig:self-data}
\end{figure}

\section{Case Study: Grover's Search}
\label{sec:grovers}

Here we present a case study of integrating an oracle implemented with \name into a full quantum algorithm, Grover's search algorithm, implemented and verified in \sqir.

Grover's search algorithm \cite{grover1996,grover1997}, described in \Cref{sec:background}, has implications for cryptography, in part because it can be used to find collisions in cryptographic hash functions \cite{grover-hash}. Thus, the emergence of quantum computers may require lengthening hash function outputs.

We have used \vqimp to implement the ChaCha20 stream cipher \cite{chacha} as an oracle for Grover's search algorithm. 
%
This cipher computes a hash of a 256-bit key, a 64-bit message number, and a 64-bit block number, and it is actively being used in the TLS protocol \cite{rfc7905,rfc8446}.
The procedure consists of twenty cipher rounds, most easily implemented when segmented into quarter-round and double-round subroutines. 
The only operations used are bitwise \textsc{xor}, bit rotations, and addition modulo $2^{32}$, all of which are included in \vqimp; the implementation is given in \Cref{fig:chacha-qr}.

To test our oracle implementation, we wrote our specification as a Coq function on bitstrings.
We then defined correspondence between these bitstrings and program states in \vqir semantics and conjectured that for any inputs, the semantics of our compiled oracle matches the corresponding outputs from our specification function.
Using random testing (\Cref{sec:rand-testing}),
we individually tested the quarter-round and double-round subroutines as well as the whole twenty-round cipher, performing a sort of unit testing.
We also tested the oracle for the boolean-valued function that checks whether the ChaCha20 output matches a known bitstring rather than producing the output directly.
This oracle can be compiled to \sqir using our verified compiler, and then the compiled oracle can be used by Grover's algorithm to invert the ChaCha20 function and find collisions.
Grover's algorithm was previously implemented and verified in \sqir \cite{PQPC}, and we have modified this implementation and proof to allow for oracles with ancillae like the ones generated by our compiler; thus, our successful QuickChick tests combined with the previously proved theorems for Grover's algorithm provide confidence that we can find Chacha20's hash collisions in a certain probability through Grover's algorithm.

\begin{figure}[t]
\[\footnotesize
\begin{array}{l}
Q\;\tnat[4]\;qr(Q\;\tnat\;x_1,Q\;\tnat\;x_2,Q\;\tnat\;x_3,Q\;\tnat\;x_4)\;\{
\\
\quad{x_1}\texttt{ += }{x_2};\; {x_4}{\; \oplus\texttt{= } }{x_1};\; x_4\,\texttt{<{}<{}<=}\,16;\\
\quad{x_3}\texttt{ += }{x_4};\; {x_2}{\; \oplus\texttt{= } }{x_3};\; x_4\,\texttt{<{}<{}<=}\,12;\\
\quad{x_1}\texttt{ += }{x_2};\; {x_2}{\; \oplus\texttt{= } }{x_1};\; x_4\,\texttt{<{}<{}<=}\,8;\\
\quad{x_3}\texttt{ += }{x_4};\; {x_2}{\; \oplus\texttt{= } }{x_3};\; x_4\,\texttt{<{}<{}<=}\,7\\
\quad\texttt{return}\;[x_1,x_2,x_3,x_4];\\
\}\\\\
\texttt{void}\;chacha20(Q\;\tnat[16]\;x)\;\{
\\
\quad\texttt{for}(C\;\tnat\;i=20;\;i>0;\;i \texttt{ -= } 2)\;\{\\
\qquad [x[0], x[4], x[8], x[12]] = qr(x[0], x[4], x[8], x[12]);\\
\qquad [x[1], x[5], x[9], x[13]] = qr(x[1], x[5], x[9], x[13]);\\
\qquad [x[2], x[6], x[10], x[14]] = qr(x[2], x[6], x[10], x[14]);\\
\qquad [x[3], x[7], x[11], x[15]] = qr(x[3], x[7], x[11], x[15]);\\
\qquad [x[0], x[5], x[10], x[15]] = qr(x[0], x[5], x[10], x[15]);\\
\qquad [x[1], x[6], x[11], x[12]] = qr(x[1], x[6], x[11], x[12]);\\
\qquad [x[2], x[7], x[8], x[13]] = qr(x[2], x[7], x[8], x[13]);\\
\qquad [x[3], x[4], x[9], x[14]] = qr(x[3], x[4], x[9], x[14]);\\
\quad\}\\
\}
\end{array}
\]
\caption{ChaCha20 implementation in \vqimp}
\label{fig:chacha-qr}
\end{figure}

\section{Related Work}
\label{sec:related}

\myparagraph{Oracles in Quantum Languages}

Quantum programming languages have proliferated in recent years. 
Many of these languages (e.g. Quil~\cite{quilc}, OpenQASM 2.0~\cite{Cross2017}, \sqir~\cite{VOQC}) describe low-level circuit programs and provide no abstractions for describing quantum oracles.
Higher-level languages may provide library functions for performing common oracle operations (e.g. Q\# \cite{qsharp}, Scaffold~\cite{scaffold,scaffCCnew}) or support compiling from classical programs to quantum circuits (e.g. Quipper~\cite{Green2013}), but still leave some important details (like uncomputation of ancilla qubits) to the programmer.

There has been some work on type systems to enforce that uncomputation happens correctly (e.g. Silq~\cite{sliqlanguage}), and on automated insertion of uncomputation circuits (e.g. Quipper~\cite{Green2013}, Unqomp~\cite{unqomp}), but while these approaches provide useful automation, they also lead to inefficiencies in compiled circuits.
For example, all of these tools force compilation into the classical gate set \texttt{X}, \texttt{CNOT},  and \texttt{CCNOT} (or ``Toffoli''), which precludes the use of QFT-based arithmetic, which uses fewer qubits than Toffoli-based approaches.
Of course, programmers are not obligated to use automation for constructing oracles---they can do it by hand for greater efficiency---but this risks mistakes.
\name allows programmers to produce oracles automatically from \vqimp using \texttt{inv} to uncompute, or to manually implement oracle functions in \vqir, in both cases supporting formal verification and testing.

\myparagraph{Verified Quantum Programming}

Recent work on formally verifying quantum programs includes \qwire~\cite{RandThesis}, \sqir~\cite{PQPC}, and \qbricks~\cite{qbricks}. These tools have been used to verify a range of quantum algorithms, from Grover's search to quantum phase estimation.
Like these tools, properties of \vqir programs are expressed and verified in a proof assistant.
But, unlike these tools, we focus on a quantum sub-language that, while not able to express any quantum program, is efficiently simulatable.
This allows us to reuse existing infrastructure (like QuickChick~\cite{quickchick}) for testing Coq properties.


\myparagraph{Verified Compilation of Quantum Programs}

Recent work has looked at verified optimization of quantum circuits (e.g., \voqc~\cite{VOQC}, CertiQ~\cite{Shi2019}), but the problem of verified \emph{compilation} from high-level languages to quantum circuits has received less attention.
The only examples of verified compilers for quantum circuits are ReVerC~\cite{reverC} and ReQWIRE~\cite{Rand2018ReQWIRERA}.
Both of these tools support verified translation from a low-level Boolean expression language to circuits consisting of \texttt{X}, \texttt{CNOT}, and \texttt{CCNOT} gates.
Compared to these tools, \name supports both a higher-level classical source language (\vqimp) and a more interesting quantum target language (\vqir).





\section{Conclusion}

We present \name, a framework for expressing, testing, and verifying quantum oracles. The key component of \name is \vqir, the oracle quantum assembly language, which can express a restricted class of quantum programs that are efficiently simulatable (and hence testable) and are useful for implementing quantum oracles. 
We have verified the translator from \vqir to \sqir and have verified (or randomly tested) many arithmetic circuits written in \vqir.
We also present \vqimp, a high-level imperative language, and compiler from \vqimp to \vqir (framework verified and arithmetic operations randomly tested). We have used \name to implement oracles and oracle components useful in quantum programming, like modular multiplication and sine, and showed that our performance is comparable to the state-of-the-art (unverified) framework Quipper. We also demonstrated the benefit of partial evaluation in \vqimp, showing that partial evaluation results in our implementation of sine using just 7\% of the qubits used in Quipper's implementation.


\begin{acks}                            
We thank Leonidas Lampropoulos for helping us with effective use of
QuickChick, and Aaron Green and Robert Rand for helpful comments and contributions
during the development of this work. This material is based upon work supported
by the U.S. Department of 
Energy, Office of Science, Office of Advanced Scientific Computing
Research, Quantum Testbed Pathfinder Program under Award Number
DE-SC0019040, and the Air Force Office of Scientific Research under Grant No.
FA95502110051.
\end{acks}

\bibliography{reference}
\newpage
\clearpage

\appendix

\section{\pqasm: Extending \oqasm with Additional Gates}\label{sec:extended-oqasm}

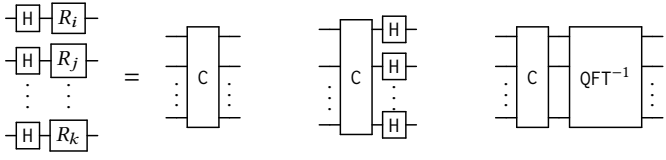
\begin{figure}[t]
  \centering
\begin{tabular}{c@{$\quad=\quad$}c @{$\quad\qquad$} c @{$\quad\qquad$} c}
  \begin{minipage}{0.3\textwidth}
  \Small
  \Qcircuit @C=0.5em @R=0.5em {
    \lstick{} & & \gate{\texttt{H}} & \gate{R_i} & \qw  \\
    \lstick{} & & \gate{\texttt{H}}        & \gate{R_j}       & \qw    \\
    \lstick{} & & \vdots & \vdots & \\
    \lstick{} & & & & \\
    \lstick{} & & & & \\
    \lstick{} & & \gate{\texttt{H}}        & \gate{R_k}           & \qw  
    }
  \end{minipage} 
&
  \begin{minipage}{0.3\textwidth}
  \Small
  \Qcircuit @C=0.5em @R=0.5em {
    \lstick{} & \qw     & \multigate{4}{\texttt{C}} & \qw & \qw \\
    \lstick{} & \qw     & \ghost{\texttt{C}}           & \qw & \qw \\
    \lstick{} & \vdots  &                                & \vdots & \\
    \lstick{} & & & & \\
    \lstick{} & \qw     & \ghost{\texttt{C}}           & \qw  & \qw
    }
  \end{minipage} & 
  \begin{minipage}{0.3\textwidth}
  \Small
  \Qcircuit @C=0.5em @R=0.5em {
    \lstick{} & \qw     & \multigate{4}{\texttt{C}} & \gate{\texttt{H}} & \qw \\
    \lstick{} & \qw     & \ghost{\texttt{C}}           & \gate{\texttt{H}} & \qw \\
    \lstick{} & \vdots  &                               & \vdots & \\
    \lstick{} & & & & \\
    \lstick{} & \qw     & \ghost{\texttt{C}}           & \gate{\texttt{H}}  & \qw
    }
  \end{minipage} 
&
  \begin{minipage}{0.3\textwidth}
  \Small
  \Qcircuit @C=0.5em @R=0.5em {
    \lstick{} & \qw     & \multigate{4}{\texttt{C}} & \qw& \multigate{4}{\texttt{QFT}^{-1}} & \qw & \qw \\
    \lstick{} & \qw     & \ghost{\texttt{C}}         & \qw & \ghost{\texttt{QFT}^{-1}} & \qw & \qw\\
    \lstick{} & \vdots &                              &    &                          & \vdots  \\
    \lstick{} & & & & & \\
    \lstick{} & \qw     & \ghost{\texttt{C}}           &\qw & \ghost{\texttt{QFT}^{-1}}  & \qw & \qw
    }
  \end{minipage} 
\end{tabular}
\caption{Example quantum pattern as part of a graph representation in the quantum walk algorithm \cite{Quantumcircuitdesign}. $R_i$ is a $z$-axis rotation by $2\pi / 2^i$ (written \texttt{RZ i} in \pqasm).}
\label{fig:quantum-walk-graph}
\end{figure}

 
\pqasm extends \oqasm by adding an $Rz$ gate, a Hadamard gate, and a $\texttt{Had}\;n$ basis, which uses the same qubit form as the \texttt{Phi} basis (i.e., $\alpha(r_1)\qket{r_2}$).
\pqasm is useful for defining common circuit patterns that appear as subcomponents of quantum algorithms.
For example, the circuit \texttt{C} in \Cref{fig:quantum-walk-graph} is used in several places in the encoding of the graph in a quantum walk algorithm \cite{Quantumcircuitdesign}. 
The behaviors of circuits like \texttt{C}, which consist only of Hadamard and $Rz$ gates, can be efficiently tested using \pqasm. 
Recall from \Cref{fig:background-circuit-example} that QFT and AQFT circuits also have this form, meaning that we can use \pqasm to evaluate different implementations of AQFT/QFT to use when compiling \oqasm to \sqir.

\begin{figure}[t]
{\footnotesize
\begin{center}\hspace*{-1em}
\begin{tikzpicture}[->,>=stealth',shorten >=1pt,auto,node distance=3.2cm,
                    semithick]
  \tikzstyle{every state}=[fill=black,draw=none,text=white]

  \node[state] (A)              {$\texttt{Nor}$};
  \node[state,fill=red]         (B) [right of=A] {$\thad{n}$};
  \node[state]         (C) [left of=A] {$\tphi{n}$};

  \path (A) edge [loop above]            node {$\Big\{\begin{array}{l}\texttt{ID},~\texttt{X},~\textcolor{red}{\texttt{RZ}^{\lbrack -1 \rbrack}},~\texttt{CU},\\
              \texttt{Lshift},\texttt{Rshift},\texttt{Rev}\end{array}\Big\}$} (A)
            edge [red]                node {$\{\texttt{H}\}^1$} (B)
            edge   node [above] {\{$\texttt{QFT}\;n$\}} (C);
  \path (B) edge [red,loop above] node {$\{\texttt{ID},\texttt{CU},\texttt{X},\texttt{RZ}^{\lbrack -1 \rbrack}\}_{\ge n}$} (B)
            edge [red,loop right] node {$\{\texttt{H}\}^+_{= n}$} (B)
            edge [red,loop below] node {$\{\texttt{ID},\texttt{RZ}^{\lbrack -1 \rbrack}\}_{< n}$} (B);
  \path (C) edge [loop above]            node {$\{\texttt{ID},~\texttt{SR}^{\lbrack -1 \rbrack}\}$} (C)
            edge  [bend right]             node {$\{\texttt{QFT}^{-1}\;n\}$} (A);
\end{tikzpicture}
\end{center}
}
\caption{Typing state machine. Red indicates new additions to \pqasm.}
\label{fig:state-machine-1}
\end{figure}
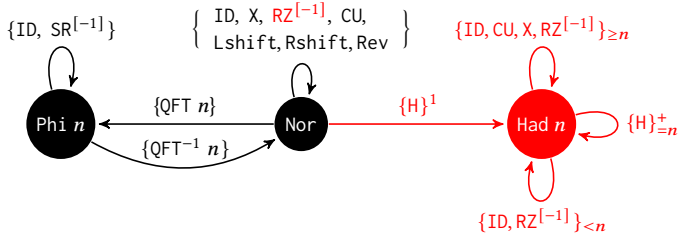

\subsection{\pqasm Syntax, Type System, and Semantics}\label{sec:pqasm-syn}

\begin{figure}[t]

{\small \centering

  $ \hspace*{-0.8em}
\begin{array}{llcl}
      \text{Position} & p & ::= & (x,n) \qquad   \text{Nat. Num}~n
                                  \qquad   \text{Variable}~x\\
      \text{Instruction} & \instr & ::= & \iskip{p} \mid \inot{p} \mid \iseq{\instr}{\instr}\\
                & & \mid &  \isr[\lbrack -1 \rbrack]{n}{x} \mid \iqft[\lbrack -1 \rbrack]{n}{x} \mid \ictrl{p}{\instr}  \\
                      & & \mid & \ilshift{x} \mid \irshift{x} \mid \irev{x} \\
                      &&  \textcolor{red}{\mid} & \textcolor{red}{\irz[\lbrack -1 \rbrack]{n}{p} \mid \ihad{p}}\\
    \end{array}
  $
}
  \caption{\pqasm syntax. For an operator \texttt{OP}, $\texttt{OP}^{\lbrack -1 \rbrack}$ indicates that the operator has a built-in inverse available.}
  \label{fig:pqasm-syn}

\end{figure}

\begin{figure}[t]
{\Small
\begin{tabular}{l}
\text{\textcolor{blue}{Typing:  }}

\\[0.5em]
  \begin{mathpar}
  
    \inferrule[RZ]{\Omegaty(x)=\texttt{Nor} \\ n < \Omegasz(x)}{\Sigma;\Omega \vdash \irz{q}{(x,n)} \triangleright \Omega}

    \inferrule[RRZ]{\Omegaty(x)=\texttt{Nor} \\ n < \Omegasz(x)}{\Sigma;\Omega \vdash \irz[-1]{q}{(x,n)} \triangleright \Omega}

    \inferrule[RZ-HAD]{\Omegaty(x)=\thad{n} \\ m < n}{\Sigma;\Omega \vdash \irz{q}{(x,m)} \triangleright \Omega}

    \inferrule[RRZ-HAD]{\Omegaty(x)=\thad{n} \\ m < n}{\Sigma;\Omega \vdash \irz[-1]{q}{(x,m)} \triangleright \Omega}
     
    \inferrule[HAD-1]{\Omegaty(x)=\texttt{Nor}\\ 0 < \Omegasz(x)}{\Sigma; \Omega \vdash \ihad{(x,0)}\triangleright \Omega[x\mapsto \thad{1}]}  
 
    \inferrule[HAD-2]{\Omegaty(x)=\thad{n}\\ n < \Omegasz(x)}{\Sigma; \Omega \vdash \ihad{(x,n)}\triangleright \Omega[x\mapsto \thad{n+1}]}  

    \inferrule[NOR-HAD]{\Sigma; \Omega[\texttt{var}(\instr)\mapsto \texttt{Nor}] \vdash \instr\triangleright \Omega'\\
                    \Omega'(\texttt{var}(\instr)) = \texttt{Nor}\\
\\\\\Omegaty(\texttt{var}(\instr))=\thad{n}\\ n \le \texttt{pos}(\instr) < \Omegasz(x)}{\Sigma; \Omega \vdash \instr\triangleright \Omega'[\mapsto \thad{n}]}  
    
  \end{mathpar}
\\[0.5em]
\text{\textcolor{blue}{Semantics:  }}
\\[0.5em]
$
\begin{array}{l l}
\llbracket \irz{m}{(x,i)} \rrbracket\varphi &= \app{\uparrow {\rsem}({m},\downarrow\varphi(x,i))}{\varphi}{(x,i)}\\
& \texttt{where  }{\rsem}(m,\ket{0})=\ket{0}\quad{\rsem}(m,\ket{1})=\alpha(\frac{1}{2^m})\ket{1}
\quad
{\rsem}(m,\qket{r})=\qket{r+\frac{1}{2^m}}
\\[0.5em]

\llbracket \irz[-1]{m}{(x,i)} \rrbracket\varphi &= \app{\uparrow {\rrsem}({m},\downarrow\varphi(x,i))}{\varphi}{(x,i)}\\
& \texttt{where  }{\rrsem}(m,\ket{0})=\ket{0}
\quad{\rrsem}(m,\ket{1})=\alpha(-\frac{1}{2^m})\ket{1}
\quad
{\rrsem}(m,\qket{r})=\qket{r-\frac{1}{2^m}}
\\[0.5em]

\llbracket \ihad{(x,i)} \rrbracket\varphi &= \app{\qket{\frac{b}{2}}}{\varphi}{(x,i)}
\qquad
\texttt{where  }\downarrow\varphi(x,i) = \ket{b}
\end{array}
$
\end{tabular}
}
  \caption{\pqasm additional typing and semantics rules}
  \label{fig:pqasm-sem}
\end{figure}

The new syntax in \pqasm is marked in red in \Cref{fig:pqasm-syn}. Every valid \oqasm program is also valid in \pqasm. 
All \oqasm typing rules are also valid in \pqasm; the new rules are summarized by the red parts of \Cref{fig:state-machine} and listed in the top of \Cref{fig:pqasm-sem}.
Rules~\rulelab{RZ}, \rulelab{RRZ}, \rulelab{RZ-HAD}, and \rulelab{RRZ-HAD} deal with $Rz$ gates.
Rules~\rulelab{HAD-1} and \rulelab{HAD-2} enforce that once a variable is in the $\thad{n}$ basis, it can never return to $\texttt{Nor}$.
These rules represent the gradual transition of a \texttt{Nor} basis variable to a \texttt{Had} basis variable.
Given a $\thad{i}$ basis variable $x$, $\texttt{H}$ can only be applied to $(x,i)$, and the result is in the $\thad{i+1}$ basis, as indicated by the $\{\texttt{H}\}^{+}_{=n}$ label in \Cref{fig:state-machine-1}.
For $j < i$, $(x,j)$ has the form $\alpha(b)\qket{b'}$; thus, only \texttt{ID} and $\texttt{RZ}^{-1}$ gates are allowed.
For $j \ge i$, $(x,j)$ has the form $\alpha(b)\qket{c}$; thus, all allowed \texttt{Nor} gates are permitted.
The \pqasm semantics extends the \oqasm semantics (\Cref{fig:deno-sem}) with rules for \texttt{H} and \texttt{RZ} gates, as shown in the bottom of \Cref{fig:pqasm-sem}.

To support the new \texttt{Had} basis, we extend \Cref{def:well-formed} as follows:

\begin{definition}[Well-formed \oqasm state]\label{def:well-formed-1}\rm 
  A state $\varphi$ is \emph{well-formed}, written
  $\Sigma;\Omega \vdash \varphi$, iff:
\begin{itemize}
  \item For every $x \in \Omega$ such that $\Omegaty(x) = \texttt{Nor}$,
  for every $k <\Omegasz(x)$, $\varphi(x,k)$ has the form
  $\alpha(r)\ket{b}$.
\textcolor{red}{\item For every $x \in \Omega$ such that $\Omegaty(x) = \thad{n}$,
      for every $k <n$, $\varphi(x,k)$ has the form $\alpha(b)\qket{b'}$;
   and for every $n \le k <\Omegasz(x)$, $\varphi(x,k)$ has the form $\alpha(b)\ket{c}$.}
  \item For every $x \in \Omega$ such that $\Omegaty(x) = \tphi{n}$ and $n \le \Omegasz(x)$,
  there exists a value $\upsilon$ such that for
  every $k < \Omegasz(x)$, $\varphi(x,k)$ has the form
  $\alpha(r)\qket{\frac{\upsilon}{ 2^{n- k}}}$.
\end{itemize}
\end{definition}

We have re-proved type soundness (\Cref{thm:type-sound-oqasm}) for \pqasm, but the subgroupoid (\Cref{thm:subgroupoid}) and type reversibility (\Cref{thm:reversibility}) theorems \emph{do not} hold. We can re-state the quantum summation formula (\Cref{thm:sem-same}) as follows, allowing input in the \texttt{Nor} basis and output in the \texttt{Had} basis.

\begin{theorem}\label{thm:sem-same-1}\rm
  Let $\ket{y}$ be an abbreviation of $\bigotimes_{m=0}^{d-1} \alpha(r_m) \ket{b_m}$ for $b_m \in \{0,1\}$.
  If for every $i\in [0,2^d)$, $\llbracket \instr \rrbracket\ket{y_i}=\qket{r_i}$, then $\llbracket \instr \rrbracket (\sum_{i=0}^{2^d-1} \ket{y_i})=\sum_{i=0}^{2^d-1} \qket{r_i}$.
\end{theorem}

\ignore{
\subsection{Case Study: Compare and Contrast QFT and Approximate QFT Circuits}
\label{sec:qft-circuit}

\begin{figure*}[t]
\centering

\begin{coq}
Fixpoint many_CR (x:var) (b:nat) (n : nat) (i:nat) :=
  match n with
  | 0 | 1 => SKIP (x,n)
  | S m  => if b <=? m then (many_CR x b m i ; (CU (x,m+i) (RZ n (x,i)))) else SKIP (x,m)
  end.

(* approximate QFT for reducing b ending qubits. *)
Fixpoint appx_QFT (x:var) (b:nat) (n : nat) (size:nat) :=
  match n with
  | 0    => Oexp (SKIP (x,n))
  | S m => if b <=? m then ((appx_QFT x b m size)) ; (H (x,m))
                    ; (Oexp (many_CR x b (size-m) m)) else Oexp (SKIP (x,m))
  end.
\end{coq}
\caption{Approximate QFT Circuits}
\label{fig:qft-circuit-impl}
\end{figure*}

\oqasm assumes the AQFT circuit implementation from \Cref{fig:background-circuit-example}, but there are other possible ways to implement AQFT circuits, such as removing \texttt{RZ} gates that act on the middle of the qubit array, rather than those acting on the end.
These alternate implementations can be defined and analyzed in \pqasm. 
In analyzing the AQFT semantics, we use the random testing framework to test if a semantic pattern is correct based on the circuit in \Cref{fig:background-circuit-example}, with bitstring inputs representing \texttt{Nor} basis values. 
The procedure works as follows: we first guess a semantic property, and use the testing framework to test its correctness. 
If it fails, we then use the framework to see the distance between the property and the circuit implementation, then we repair the semantic property to approach the correct circuit semantics. 
This semantic property gives us the situation when inputs are in \texttt{Nor}. Once we have the property, we can then use the summation formula (\Cref{thm:sem-same-1}) to infer the behavior of the quantum circuit with superposition inputs.

There are many ways \pqasm can be useful.
First, QFT/AQFT circuits are designed to disentangle the entanglement \cite{ApproximateQFT,appox-qft2,appox-qft1} to achieve physical implementation benefits.
This makes it easy to analyze in \pqasm because we guarantees the non-entanglement so that states are separably analytical.
However, it is hard to analyze the reversed QFT/AQFT circuits $\texttt{QFT}^{-1}$/$\texttt{AQFT}^{-1}$, because their non-entanglement properties are not trivial.
A simple way of analyzing $\texttt{QFT}^{-1}$/$\texttt{AQFT}^{-1}$ circuits is to infer the reversed matrix value based on the $\texttt{QFT}$ circuit behavior.
By given the semantic output of applying QFT/AQFT devices on an input in \pqasm, we can infer the semantic output for the reversed circuit by computing the inverse value for each qubit, since we are sure that none of the qubits are entangled.

Second, we also often want to learn about the side-effects of these quantum patterns in a quantum algorithm. We use the \pqasm random testing framework to cooperate with \sqir verification framework to finish such analysis. Hietala \textsf{et al.} \cite{VOQC} previously proved the semantics of Quantum Phase Estimation (QPE) with the QFT circuit that uses a $\texttt{QFT}^{-1}$ device. We mimic a approximate QPE by replacing the $\texttt{QFT}^{-1}$ with our AQFT device ($\texttt{AQFT}^{-1}$).
By using random testing to estimate the maximum distance between circuits and their approximate circuits, we learn about the semantics for $\texttt{QFT}$ and $\texttt{AQFT}$. We then use the reversed circuit formula described above to infer the semantics of $\texttt{QFT}^{-1}$ and $\texttt{AQFT}^{-1}$.
After that we can plug the two semantic statements into the summation theorem in \Cref{thm:sem-same-1}. 
We then compile the circuits and proofs to \sqir as two \sqir semantic statements for $\texttt{QFT}^{-1}$ and ($\texttt{AQFT}^{-1}$. The final maximum distance theorem between QPE and approximate QPE uses the summation forms of the two semantic statements, and it is listed as:

 \begin{theorem}\label{thm:qpe-max-dist}\rm[QPE Maximum Distance]
    For a QPE circuit with a QFT device as $\texttt{QPE}(\texttt{QFT}^{-1})$, and an approximate QPE circuit with an $\texttt{AQFT}^{-1}$ device as $\texttt{QPE}(\texttt{AQFT}^{-1})$, if $n$ is the number of qubits for the input of $\texttt{QFT}^{-1}$ and $\texttt{AQFT}^{-1}$, and $i$ is the precision number of $\texttt{AQFT}$, if $O(n) \gg O(n-i)$, for any $\epsilon$ and eigenvalue input $\ket{u}$, there exists $k$, such that for $k<n$, $\dabs{\texttt{QPE}(\texttt{QFT}^{-1})\ket{u}-\texttt{QPE}(\texttt{AQFT}^{-1}\ket{u})} < \epsilon$.
 \end{theorem}

This theorem shows that the maximum distance of a QFT and an approximate QFT is insignificant in QPE circuits. The QFT substitution for an AQFT is a good approximation in the QPE circuit.
}

\section{\vqimp: Semantics, Typing, and Compilation}
\label{sec:appendix}

Though it is common practice, writing oracles in a host
metalanguage---like using Coq to write \vqir programs---is tedious and
error prone.  To make writing arithmetic-based quantum oracles easier,
we developed \vqimp, a source-level, imperative language. This section
describes \vqimp's design rationale and features, and sketches its
formal semantics, metatheory, and partially-verified compilation to
\vqir.

\subsection{Syntax and Overview}

\begin{figure}[t]
  \small
  \[\begin{array}{llcl}
      \text{Bitstring} & b       \\
      \text{Nat. Num} & m, n       \\
      \text{Variable} & x,y \\
      \text{Mode} & q & ::= & C \mid Q \\
      \text{Base type} & \omega & ::=  & \tbool \mid \tfixed \mid \tnat \\
      \text{Type} & \tau  & ::=  & \omega^q ~\mid~ \tarr{n}{\omega^q} \\
      \text{Value} & v & ::= &  l \mid \econst{\omega}{b} \\
      \text{LValue} & l & ::= & x \mid \eindex{x}{v} \\
      \text{Bool Expr} & e & ::= & v < v \mid v = v \mid \texttt{even}~v \mid ... \\
      \text{Operator} & op & ::= & + \mid - \mid \times \mid \otimes \mid ... \\
      \text{Statement} & s & ::= & \sassign{l}{op}{v}{v} \mid 
                                    \ssassign{l}{op}{v} \mid
                                    \ssassign{l}{}{v} \mid \scall{l}{f}{\bar{v}} \\
      & & \mid & \sinv{l} \mid
                                     \sfor{x}{v}{s} \mid
                                    \sif{e}{s}{s} \mid
                                    \sseq{s}{s} \\
      \text{FunDef} & d & ::= & \texttt{def}~f~(\overline{\omega^q~x})~\{~\overline{\tau~y}; s; \texttt{return}~v \} \\
      \text{Program} & P & ::= & \overline{\tau~x}; ~\overline{d} \\
    \end{array}
  \]
  \caption{Core \vqimp syntax}
  \label{fig:vqimp}
\end{figure}

The grammar of core \vqimp is given in \Cref{fig:vqimp}. An \vqimp
program $P$ is a sequence of global variable declarations
$\overline{\tau~x}$ followed by a sequence of function definitions
$\overline{d}$, where the last of these acts as the ``main''
function. A function definition $d$ declares parameters and local
variables; its body consists of a statement $s$; and it concludes
by returning a value $v$. All variables are initialized as $0$. 

Variables $x$ have types $\tau$, which are either primitive types
$\omega^q$ or arrays thereof, of size $n$. A primitive type pairs a
base type $\omega$ with a quantum mode $q$. There are three base
types: type $\tnat$ indicates non-negative (natural) numbers; type
$\tfixed$ indicates fixed-precision real numbers in the range $(-1,1)$;
and type $\tbool$ represents booleans. The programmer specifies the number of qubits to use to represent
$\tnat$ and $\tfixed$ numbers when invoking the \name compiler. We discuss modes $q$
and our rationale for the choice of primitive types, below.

\vqimp statements $s$ consist of assignments,
loops, conditionals, 
and sequences of statements, as is typical. Assignments are always
made to lvalues $l$, which are either variables or array locations,
and come in four forms. $\sassign{l}{op}{v}{v}$ assigns to $l$ the
result of applying binary operator $op$ on two value parameters
$v$. Values can be either lvalues or literals $\econst{\omega}{b}$,
where $b$ is a bitstring and $\omega$ is the base type that indicates
its interpretation. $\ssassign{l}{op}{v}$ is a unary assignment; it is
equivalent to $\sassign{l}{op}{l}{v}$. $\ssassign{l}{}{v}$ initializes
$l$ to a value. $\scall{l}{f}{\bar{v}}$ assigns to $l$ the result of
calling function $f$ with arguments $\bar{v}$.

\subsection{Semantics}
\label{sec:source-semantics}

\begin{figure*}[t]
{\small
\begin{center}
$
\begin{array}{c}
\begin{array}{l}
\AxiomC{$
\;
$}
\UnaryInfC{$ \Xi \vdash \sigma; \sassign{l}{op}{v_1}{v_2}\steps \sigma[l \mapsto \texttt{app}^{op}(v_1,v_2)] $}
\DisplayProof
\texttt{(bin)}
\end{array}
\;\;
\begin{array}{l}
\AxiomC{$
\sigma;v\steps n
\quad
\Xi \vdash \sigma[x\mapsto 0]; s;n\steps \sigma'
$}
\UnaryInfC{$ \Xi \vdash \sigma; \sfor{x}{v}{s}\steps \sigma' $}
\DisplayProof
\texttt{(for\_t)}
\end{array}
\\[1em]
\begin{array}{l}
\AxiomC{$
\;
$}
\UnaryInfC{$ \Xi \vdash \sigma; s;0\steps \sigma $}
\DisplayProof
\texttt{(for\_0)}
\end{array}
\quad
\begin{array}{l}
\AxiomC{$
\Xi\vdash \sigma ;s \steps \sigma'
\quad
\Xi\vdash \sigma';s;n \steps \sigma''
$}
\UnaryInfC{$ \Xi \vdash \sigma; s;\texttt{S}\;n\steps \sigma'' $}
\DisplayProof
\texttt{(for\_n)}
\end{array}
\end{array}
$
\end{center}
}
\caption{Select \vqimp semantics}\label{fig:vqimp-sem}
\end{figure*}

We define a big-step operational semantics for \vqimp; select rules are shown in \Cref{fig:vqimp-sem}.
The judgment $\Xi \vdash \sigma; s \steps r$ states
that under function environment $\Xi$ and input store $\sigma$,
statement $s$ evaluates to result $r$. Here, $\Xi$ is a partial map
from function variables $f$ to their definitions, and $\sigma$ is a
partial map from variables $x$ and array locations $\eindex{x}{n}$
to a \emph{history} of literal values
$\overline{\econst{\omega}{b}}$. A result $r$ is either an output store
$\sigma'$ or a run-time failure \texttt{Error}, which arises due to an
out-of-bounds index or a division by zero.
The semantics also defines the standard subsidiary judgments, e.g.,
$\Xi; \sigma \vdash e \steps \econst{\omega}{b}$ evaluates $e$ under
$\Xi$ and $\sigma$ to a literal result.

The \vqimp semantics is largely standard except for the treatment of
$\sigma$: An assignment of $v$ to a $Q$-mode variable $x$ pushes $v$ to the
history $\sigma(x)$, and likewise for assignments to
$\eindex{x}{n}$. Assignment to a $C$-mode variable $x$ replaces $x$'s
history with the singleton $[v]$. Lookup of $x$ returns the topmost literal of
$\sigma(x)$, while $\sinv{x}$ pops off the top element of
$\sigma(x)$. For example, in \Cref{fig:sine-impl}, executing the
statement $x_z\texttt{=}\texttt{pow}(x_{/8},n_4)$ updates the topmost
element of $\sigma(x_z)$ to $\texttt{pow}(x_{/8},n_4)$. After the
computation of $\sinv{x_z}$, this new value for $x_z$ is erased and the
store entry for $x_z$ reverts to what it was before. Evaluating
$\scall{l}{f}{\bar{v}}$ performs uncomputation automatically. It
amounts to evaluating $f$'s body $s$ (stored in $\Xi$) under
$\sigma$ extended with mappings for the function's parameters to
$\bar{v}$ and local variables to $0$. When execution
of $s$ concludes, the returned result will be copied to $l$, and the
added mappings will be dropped from the output $\sigma$ (effectively
uncomputing them). Moreover any
updates to global variables will be reverted, just as if
\texttt{inv} had been applied for each modification thereto.
To evaluate a program $P$, we populate
$\Xi$ with $P$'s function definitions, set the initial $\sigma$ to map
$P$'s global variables to $\econst{\omega}{0}$ (where $\omega$ is
extracted from the variable's declared type), and then evaluate
\texttt{main}'s body $s$ (which has no arguments);
\texttt{main}'s \texttt{return}~$v$ is evaluated using the $\sigma'$
that results.

\subsection{Typing}
\label{sec:source-typing}

\begin{figure*}[t]
{\small
\centering
$
\begin{array}{c}
\AxiomC{$
\;
$}
\UnaryInfC{$ \Gamma \vdash \econst{\omega}{b} : \omega^C$}
\DisplayProof
\quad
\AxiomC{$
\Gamma(x)=\tarr{n}{\omega^q}
\quad
\Gamma\vdash v : \omega^C
$}
\UnaryInfC{$ \Gamma \vdash \eindex{x}{v} : \omega^q$}
\DisplayProof
\quad
\AxiomC{$
\Gamma(x)= \omega^q
$}
\UnaryInfC{$ \Gamma \vdash x : \omega^q$}
\DisplayProof
\quad
\AxiomC{$
\Gamma \vdash v : \omega^q
\quad
q \sqsubset Q
$}
\UnaryInfC{$ \Gamma \vdash v : \omega^Q$}
\DisplayProof
\\[1.5em]
\AxiomC{$
\Gamma\vdash e : \tbool^q
\quad
\Xi;\Gamma; q \sqcup q' \vdash s_1 
\quad
\Xi;\Gamma; q \sqcup q' \vdash s_2 
$}
\UnaryInfC{$ \Xi;\Gamma; q' \vdash \sif{e}{s_1}{s_2}$}
\DisplayProof
\texttt{(if)}
\\[1em]
\AxiomC{$
\Gamma\vdash v_1 : \omega^C
\quad
\Gamma\vdash v_2 : \omega^C
\quad
\Gamma(l) = \omega^C
$}
\UnaryInfC{$ \Xi;\Gamma; C \vdash \sassign{l}{op}{v_1}{v_2} $}
\DisplayProof
\texttt{(binop\_c)}
\quad
\AxiomC{$
\begin{array}{c}
l \neq v_1 
  \quad l\neq v_2\\
  \Gamma\vdash v_1 : \omega^Q
\quad
\Gamma\vdash v_2 : \omega^Q
\quad
\Gamma(l) = \omega^Q
  \end{array}
$}
\UnaryInfC{$ \Xi;\Gamma; q \vdash \sassign{l}{op}{v_1}{v_2}$}
\DisplayProof
\texttt{(binop\_q)}
\\[2em]
\AxiomC{$
\Gamma(x) = \tnat^C
\quad
\Gamma\vdash v : \tnat^C
\quad
\Xi;\Gamma; q\vdash s 
$}
\UnaryInfC{$ \Xi;\Gamma; q \vdash \sfor{x}{v}{s} $}
\DisplayProof
\texttt{(for)}

\qquad 

\AxiomC{$
\Xi;\Gamma; q\vdash s_1 
\quad
\Xi;\Gamma; q\vdash s_2 
$}
\UnaryInfC{$ \Xi;\Gamma; q \vdash \sseq{s_1}{s_2} $}
\DisplayProof
\texttt{(seq)}
\\[1.5em]
\AxiomC{$
\begin{array}{c}
\Gamma\vdash l : \omega^{q}
\qquad
\Gamma'\vdash v : \omega^{q'}
\qquad
q' \sqsubset q
\\
\Xi(f)=((\tau_1~x_1...\tau_n~x_n),\overline{\tau~y},s,v,\Gamma')
\quad
\Gamma\vdash v_1:\omega_1^C ...
 \Gamma\vdash v_n:\omega_n^C
\quad
\Gamma'(x_1)=\omega_1^C ...
 \Gamma'(x_n)=\omega_n^C
\end{array}
$}
\UnaryInfC{$ \Xi;\Gamma;q \vdash \scall{l}{f}{v_1...v_n} $}
\DisplayProof
\texttt{(call)}
\\[2em]
\AxiomC{$
\Gamma'=\texttt{gen\_}\Gamma(\Gamma,\overline{\tau~x}@\overline{\tau~y})
\quad
\sseq{\Xi;\Gamma'}{C}\vdash s
\quad
\Gamma'\vdash v : \omega^q
$}
\UnaryInfC{$ \sseq{\Xi}{\Gamma} \vdash (\texttt{def}~f~(\overline{\tau~x})~\{~\overline{\tau~y}; s; \texttt{return}~v \}) \triangleright \Xi[f\mapsto (\overline{\tau~x},\overline{\tau~y},s,v,\Gamma')] $}
\DisplayProof
\texttt{(fun)}
\end{array}
$
}
\caption{Select \vqimp typing rules}
\label{fig:vqimp-type}
\end{figure*}

The semantics of \vqimp defines what a program $P$ would do
if we ran it directly, but our intention is not to do this,
but to compile $P$ to a quantum oracle. We can think of $P$ as a
computation whose inputs are the $Q$-mode global variables; all of the
$C$-mode globals---and function calls, loops, etc.---will be inlined away by the compiler,
so that what remains can be compiled directly to a quantum
circuit. When that quantum circuit is executed, it will produce the
result indicated by the \vqimp semantics.

To ensure that this is possible, the \vqimp type system restricts
how various constructs are used, based on when their variables are
available, as is standard for partial evaluation \cite{partialeval}.
The \vqimp type system in \Cref{fig:vqimp-type}
defines three judgments. The first is $\Gamma \vdash v : \omega^q $,
which states that under assumptions $\Gamma$, value $v$ has primitive
type $\omega^q$; a similar judgment (not shown) handles boolean
expressions $e$. As usual $\Gamma$ maps variables $x$ to types
$\tau$. Arrays are not first class, and always classical, so the array
index rule requires the index to have mode $C$. Modes $q$ are
organized as a lattice with $C \sqsubset Q$ for the purposes of
subtyping: a value known at compile-time ($C$) can have its use
deferred to run-time ($Q$) if need be. 

The second judgment has form $\Xi;\Gamma; q \vdash s$, which states that
in context $q$ and under assumptions $\Xi$ and $\Gamma$, statement $s$ is well
formed. The context $q$ indicates the mode of the data that
determines whether the current statement was reached, either classical
 $C$ (compile time) or quantum $Q$ (run time). At the outset,
the context $q$ is $C$---the program's execution depends on no prior
result---but the context can change at a conditional. If the guard
$e$'s type is $\tbool^Q$, then which branch executes depends on a
quantum result, so the branches should be checked in mode $Q$. If the
guard $e$'s type is $\tbool^C$, then the branches should be checked in
the current mode $q$. Both notions are captured in a single rule, with
the branches checked in mode $q' \sqcup q$, where $q'$ is the current
mode and $q$ is from the guard's.

Rule \texttt{(binop\_c)} types the assignment to a $C$-mode
(compile-time) lvalue. Here, both operands must be $C$ mode too, i.e.,
known at compile time. Such a statement can only be typed in context
$C$: It will have no run-time effect, so its execution must not be
conditional on quantum data. Rule \texttt{(binop\_q)} considers
assignments to $Q$-mode variables, which can occur in any context
$q$. Both operands are also in $Q$ mode (possibly made so through
subtyping), and moreover must be different from the output-variable
($l$), though the operands can be the same. This restriction is leveraged to
simplify compilation (the $\ssassign{l}{op}{v}$ form can be used to
update the left-hand side). If both operands are the same, one is
copied out to a temporary register, to avoid violating the no cloning
rule.

Rule \texttt{call} checks that the input arguments and the parameters 
for the function have the same size, the types are matched,
and the return value ($v$) of the function is a subtype of the
to-be-assigned variable ($l$).

A \texttt{for} loop may be evaluated in whatever context $q$ its
body $s$ may be evaluated in. During compilation it will be
unrolled, so we require the iterator $x$ and the bound value $v$ to be
type $C$.

The rule (\texttt{fun}), which type checks a function, has form
$\Xi;\Gamma \vdash d \triangleright \Xi'$ where $\Xi$ and $\Xi'$ are partial maps from function names
to a tuple of (i) a list of function arguments, (ii) a list of declarations
($\overline{\tau~y}$) in the function, (iii) a function body statement,  (iv) the
return value for the function, and (v) a map ($\Gamma$) recording the
types for variables in the function and global variables declared in the program. 
The judgment outputs $\Xi'$ containing the function information and acting as
the new function environment.
$\texttt{gen\_}\Gamma(\Gamma,\overline{\tau~x}@\overline{\tau~y})$
adds variable-type information to $\Gamma$ for the two lists
$\overline{\tau~x}$ and $\overline{\tau~y}$.  The list
$\overline{\tau~x}$ should contain only $C$-mode variables since we
require all function arguments to be $C$ (users can always use global
$Q$-mode variables), while the local declaration list
$\overline{\tau~y}$ contains $C$ or $Q$ mode variables.  The type
judgment for functions resembles the \texttt{(fun)} rule except that the input is a list of functions,
and the type judgment for the whole program resembles the \texttt{(call)} 
rule except that we also need to generate a type environment for
global variables.

Well-formedness of $\sinv{x}$ statements is checked in conjunction
with typing, as described in \Cref{sec:revcomp}.

\subsection{Soundness}
\label{app:source-soundness}

\begin{definition} (Store Consistency)
Let $s$ be an \vqimp statement and $\Gamma$ be a type environment such that $\Gamma \vdash s$. We say that store $S$ is consistent with $\Gamma$ if and only if for every variable $x$ and index $i$, if $\Gamma(x)$ is defined and $i < \texttt{sizeof}(x)$, then there exists a value $v$, such that $S(x,i) = v$.
\end{definition}

Type soundness is divided into two parts: progress and preservation. The progress theorem (for statements) says that every well typed \vqimp program can take a step:

\begin{theorem}\rm (Progress)
Let $s$ be a well-formed \vqimp statement, and $\Gamma$ and $\Gamma'$ be type environments such that $\Gamma \vdash s\triangleright \Gamma'$. Let $S$ be a store that is consistent with $\Gamma'$. Then there exists a $v$ such that  $\Xi \vdash_s S;s \steps v$.
\end{theorem}

The value $v$ above can be an $\verror$ state or a final store evaluated from the big-step semantics for statements ($\vdash_s$).
Because \vqimp uses a big-step operational semantics, it is trivial to prove that the final type environment after evaluating statement $s$ is the same as the input type environment. 
We thus choose to prove a different form of preservation related to store consistency:

\begin{theorem}\rm (Preservation)
Let $s$ be a well-formed \vqimp statement, and $\Gamma$ and $\Gamma'$ be type environments such that $\Gamma \vdash s~\triangleright \Gamma'$. Let $S$ and $S'$ be stores. If $S$ is consistent with $\Gamma$ and $\Xi \vdash_s S;s \steps S'$, then $S'$ is consistent with $\Gamma'$.
\end{theorem}

The preservation theorem says that if a statement evaluates to a non-$\verror$ state with resulting store $S'$, then $S'$ is consistent with the resulting type environment.
The proofs of the two theorems are done by induction on the statement $s$; both are mechanized in Coq.

\subsection{Compilation to \vqir}
\label{sec:source-compilation}

Compilation for \vqimp statements is a partial function of the form $\Xi;\Gamma;\Theta\vdash (n * \sigma * s) \rightsquigarrow \kappa$, where:

\begin{itemize}
  \item $\Xi$ is a map from function variables to their definitions;
  \item $\Gamma$ is the standard type environment;
  \item $\Theta$ is a map from \vqimp variables to \vqir variables;
  \item $n$ is the current amount of scratch space (i.e., number of ancilla qubits) needed;
  \item $\sigma$ is a dynamic store that records the values for all $C$-mode variables;
  \item $s$ is an \vqimp statement;
  \item and $\kappa$ is the result of compilation, which is either \texttt{Error} or a value $(n,\sigma,u)$, where $n$ is the final amount of scratch space, $\sigma$ is the resulting store for $C$-mode variables, and $u$ is the generated \oqasm circuit.
\end{itemize}
$\Xi$ and $\Gamma$ are generated during type checking and the amount of scratch space ($n$) is initialized to 0.




Compilation assumes (correct) \vqir implementations of primitive arithmetic operators. 
An addition operation $\ssassign{l}{+}{v}$ in \vqimp will be compiled to an \vqir program $\texttt{add}(l,v)$ that will either apply a constant or general adder from \Cref{sec:arith-oqasm}, depending on whether $v$ has mode $C$ or $Q$ (if both $l$ and $v$ have mode $C$, then the result is precomputed).
Compilation expects two global settings: flag $fl \in \{\texttt{Classical}, \texttt{QFT}\}$ indicates whether \vqir operators should use Toffoli-based or QFT-based arithmetic, and size $sz \in \mathbb{N}$ fixes the bit size of operations.


\begin{figure*}[t]
{\small\centering
$
\begin{array}{c}
\AxiomC{$ 
\Gamma\vdash l : \omega^Q
\quad
\Gamma\vdash v : \omega^Q
\quad
u=\texttt{get\_op}(op)(fl,\Theta(l),\Theta(v),sz)
$}
\UnaryInfC{$\Xi;\Gamma;\Theta\vdash (n,\sigma,\ssassign{l}{op}{v})\rightsquigarrow (n,\sigma,u) $}
\DisplayProof
\texttt{(bin\_q)}
\\[2em]
\AxiomC{$ 
\Gamma\vdash l : \omega^C
\quad
\Gamma\vdash v : \omega^C
\quad
\Xi \vdash_s \sigma;\ssassign{l}{op}{v} \steps \sigma'
$}
\UnaryInfC{$\Xi;\Gamma;\Theta\vdash (n,\sigma,\ssassign{l}{op}{v}) \rightsquigarrow (n,\sigma',\iskip{(\Theta(l),0)}) $}
\DisplayProof
\texttt{(bin\_c)}
\quad
\AxiomC{$ 
\begin{array}{c}
\Gamma\vdash e : \tbool^C
\quad
\Xi\vdash \sigma;e \steps (\tbool)\texttt{true}
\\
\Xi;\Gamma;\Theta\vdash (n,\sigma,s_1) \rightsquigarrow (n',\sigma',u)
\end{array}
$}
\UnaryInfC{$\Xi;\Gamma;\Theta;\vdash (n,\sigma,\sif{e}{s_1}{s_2}) \rightsquigarrow (n',\sigma',u) $}
\DisplayProof
\texttt{(if\_c)}
\\[2em]
\AxiomC{$ 
\begin{array}{c}
\Gamma\vdash e : \tbool^Q
\quad
\Xi;\Gamma;\Theta\vdash (n,\sigma,e) \rightsquigarrow (n_e,\sigma_e,u_e)
\quad
\Xi;\Gamma;\Theta\vdash (n_e,\sigma_e,s_1) \rightsquigarrow (n_1,\sigma_1,u_1)
\\
\Xi;\Gamma;\Theta\vdash (n_1,\sigma_e,s_2) \rightsquigarrow (n_2,\sigma_2,u_2)
\quad
u'=\sseq{u_e}{\sseq{\ictrl{(\chi,n)}{u_1}}{\sseq{\inot{(\chi,n)}}{\ictrl{(\chi,n)}{u_2}}}}
\end{array}
$}
\UnaryInfC{$
\Xi;\Gamma;\Theta\vdash (n,\sigma,\sif{e}{s_1}{s_2}) \rightsquigarrow (n_2,\sigma_e,u') $}
\DisplayProof
\texttt{(if\_q)}
\\[2.5em]
\AxiomC{$
\begin{array}{c}
\Xi(f)=(\overline{\tau\;x},\overline{\tau\;y},s,l',\Gamma')
\quad
\Theta'=\texttt{add\_}\Theta^Q(\Theta,\Gamma',\overline{\tau\;y})
\quad
\Gamma' \vdash l' :\omega^{Q}
\quad
\Gamma \vdash l :\omega^{Q}
\\[0.5em]
\sigma'=\texttt{init\_}\sigma^C(\overline{\tau\;x},\overline{v},\overline{\tau\;y})
\quad
\Xi;\Gamma';\Theta'\vdash (n,\sigma',s) \rightsquigarrow (n',\sigma'',u)
\quad
u'=\sseq{u}{\sseq{\texttt{copy}(\Theta'(l'),\Theta(l))}{\texttt{qinv}(u)}}
\end{array}
$}
\UnaryInfC{$
\Xi;\Gamma;\Theta \vdash (n,\sigma,\scall{l}{f}{\overline{v}}) \rightsquigarrow (n,\sigma,u') $}
\DisplayProof
\texttt{(call)}
\end{array}
$
}
\caption{Select \vqimp to \vqir compilation rules}
\label{fig:compile-vqimp}
\end{figure*}

\Cref{fig:compile-vqimp} provides a select set of compilation rules from \vqimp to \vqir.
The first two rules compile an assignment operation. 
If $l$ and $v$ are both typed as $Q$-mode, we compile the assignment to a pre-defined \vqir program, which we look up using \texttt{get\_op}. 
For example, if $op$ is an addition, $\texttt{get\_op}(op)$ produces the \vqir program \texttt{add}, described above.
The rule for the case where one variable has mode $Q$ and the other variables has mode $C$ is similar.
On the other hand, if $l$ and $v$ are both typed as $C$-mode, we update the store $\sigma$ by computing the addition directly using the \vqimp semantics (\Cref{sec:source-semantics}) and generate the trivial \oqasm program \texttt{ID}.

The rules \texttt{(if\_c)} and \texttt{(if\_q)} are for branching operations. If the Boolean guard has mode $C$, we evaluate the expression to its value and choose one of the branches for further compilation; \texttt{(if\_c)} shows the case where $e$ evaluates to \texttt{true}. 
If the Boolean guard has mode $Q$, we generate an \vqir expression to compute the guard and store the result in $(\chi,n)$, where $\chi$ is the scratch space variable and $n$ is the current scratch space index.
We then compile the two branches conditioned on $(\chi,n)$, as shown in rule \texttt{(if\_q)}. 

Rule \texttt{(call)} compiles a function call.
$\texttt{add\_}\Theta^Q$ extends the \vqimp-to-\vqir variable map with the new $Q$-mode variables in the local declaration list $\overline{\tau~y}$.
The \vqimp type system requires that all function arguments ($\overline{\tau~x}$) have mode $C$. 
$\texttt{init\_}\sigma^C$ initializes the values of all variables in $\overline{\tau~x}$ to their corresponding values in $\overline{v}$ and all $C$-mode variables in $\overline{\tau~y}$ to 0.
$\texttt{copy}(x,y)$ is an \vqir program that copies the states of all qubits in $x$ to $y$ using a series of $\ictrl{p_x}{\inot{p_y}}$ operations. 
In the case where $l'$ (the return value of $f$) has mode $C$, we do not need to generate a circuit for the function $f$; instead, we just generate a circuit to set $l$ to the value of $l'$.
If $l$ also has mode $C$, we just update $l$'s in $\sigma$ without generating any circuit.

\ignore{
Here are some details about $\Theta$ generation in function calls.
A function call might contain other functions that have local variables having same names in \vqimp, we alpha-convert these variables to new names, and assign them different variables in \vqir. In a function, it is also possible to call functions consecutively, where these functions have same-name local variables. In generating $\Theta$, we reuse variables in \vqir as much as possible in order to reduce qubit usage size. In the mechanization, variables in \vqir are implemented as natural numbers, so we keep a variable counter in generating $\Theta$ for statements. If we compile two consecutive function calls, the counter inputs for both of them are the same so that we can reuse variables generated for the first function call in the second one. This is possible because all functions are reversible and have no side-effects in \vqimp.
}


We prove that compilation is correct: Given an \vqimp program $P$ that compiles to an \oqasm circuit $C$, evaluating $P$ according to the \vqimp semantics will produce a value consistent with evaluating $C$ according to the \oqasm semantics. 
The proof is mechanized in Coq and proceeds by
induction on the compilation judgment, relying on proofs of
correctness for \vqir arithmetic operators, as discussed in
\Cref{sec:vqir-compilation}.

\ignore{
\subsection{Compilation Correctness}
\label{sec:source-compilation-correctness}

Before describing the main theorem proved, we discuss the equivalence relation on the final values. For an \vqimp program $P$, 
the equivalence of final values for the \vqimp semantics of the compilation is defined as the evaluation result of the program matches the execution result of the circuit generated from the $P$ compilation.
For an \vqimp statement $s$, the store is a little complicated. 
Given an \vqimp statment $s$ and an initial state $\sigma$, we can evaluate $s$ to get a final store $\sigma_q$ by executing the \vqimp semantics, or we can compile $s$ to a tuple $(\sigma_c,u)$, and execute $u$ in \vqir on an initialized \vqir state $\varphi$ to get a final state $\varphi_q$. Given a type environment $\Gamma$ for variables in $\sigma_q$, an \vqimp to \vqir variable map $\Theta$, and a bit-size map $\Sigma$, we construct the following equivalence relation between $\sigma_q$ and $(\sigma_c,\varphi_q)$:

\begin{definition} (Statement Final Value Equivalence)
Let $s$ be an \vqimp statement, $\Gamma$ the type environment, $\Theta$ an \vqimp to \vqir variable map, $\Sigma$ a bit-size map such that for all $(x,i)\in \texttt{dom}(\Theta)$, $\Sigma(\Theta(x,i))$ is defined and it is equal to $\tsizeof(\Gamma,l)$. 
$\sigma_q$ is an \vqimp state, and $(\sigma_c,\varphi_q)$ a pair of \vqimp and \vqir states.
Then
$\Sigma;\Theta;\Gamma \vdash \sigma_q \simeq (\sigma_c,\varphi_q)$, if and only if, for all indexed-variables $(x,i)$ having $\Gamma(x)=\omega^Q$, we have $[\sigma_q(x,i)]_{\Sigma(\Theta(x,i))} = \texttt{cval}^{\Theta(x,i)}(\Sigma(\Theta(x,i)))~\varphi_q$, and for all indexed-variables $(x,i)$ having $\Gamma(x)=\omega^C$, we have $\sigma_q(x,i)=\sigma_c(x,i)$.
\end{definition}

{\small
\begin{center}
$
\begin{array}{ll}
\texttt{cval}^x~0~\varphi = \lambda i.0 & \texttt{cval}^x(n+1)~\varphi = (\texttt{cval}^x~n~\varphi)[n\mapsto \downarrow(\varphi(x,n))]\\
&\qquad\texttt{where}\;\;
\downarrow(\inval{c}{b})=c
\end{array}
$
\end{center}
}

\texttt{cval} is for getting bit-values from $n$ qubit state described above, while $[b]_{n}$ gets the first $n$ bits of the bit-string based on the bit-size number of the type of a given \vqimp value. We show the main theorem to prove below:

\begin{theorem}\rm
Let $p=\sseq{(\overline{\tau}@[\tau\;x])}{\overline{d}}$ be a well-formed \vqimp program, $\Xi$ is generated by the \vqimp function type rule in \Cref{fig:vqimp-type}, $\Theta$ an \vqimp to \vqir variable map for $P$, $\Sigma$ a record of the bit-string sizes for the mappings in $\Theta$ for the global variables in $P$. Then for every $\sigma_q$ $\sigma_c$ $r$ $r'$ $\varphi$ $\varphi_q$, such that $\Sigma;\Theta;\Gamma\vdash\sigma_q\simeq (\sigma_c,\varphi)$, $p\rightsquigarrow \kappa$, $\Xi\vdash \sigma_q; p \steps r$,
if $r$ is \texttt{Error}, then $\kappa$ is also \texttt{Error}; if $r$ is a value $v$, then $\kappa$ is $(n,u)$, such that 
$\Sigma;\varphi \vdash u \steps \varphi_q$, thus, $\texttt{cval}^{\Theta(x)}(\Sigma(\Theta(x)))~\varphi_q=v$.

\end{theorem}

The above theorem only discusses what happens at the top level. It is best to question the compilation correctness of individual \vqimp statements. Here is the theorem for a statement:

\begin{theorem}\rm
Let $s$ be a well-formed \vqimp statement, $\Xi$ the proper function type environment, $q$ a statement mode, $\Gamma$ and $\Gamma'$ type environments such that $\Xi;\Gamma;q \vdash s \triangleright \Gamma'$,  $\Theta$ a proper \vqimp to \vqir variable map for $s$, $\Sigma$ a record of the bit-string size for the mappings in $\Theta$ for $Q$-mode variables in $s$, $n$ the current scratch space number, $\sigma_c$ an initial state containing the values for the \texttt{C}-mode variables in $s$, $\sigma_q$ an initial state containing the values for the variables in $s$, $\varphi$ the \vqir state such that $\Sigma;\Theta;\Gamma\vdash\sigma_q \simeq (\sigma_c,\varphi)$. 
Then, for every $r$ $r'$, such that $\Xi;\Gamma;\Theta\vdash (n,\sigma_c,s) \rightsquigarrow \kappa$ and $\Xi \vdash_s \sseq{\sigma_q}{s} \steps r$, if $r$ is \texttt{Error}, so do $\kappa$, if $r$ is a store $\sigma'_q$, then $\kappa=(n',\sigma'_c,u)$ and $\Sigma;\varphi \vdash u \steps \varphi_q$ and $\Sigma;\Theta;\Gamma\vdash\sigma'_q \simeq (\sigma'_c,\varphi_q)$.

\end{theorem}

All the theorem proofs have been mechanized in Coq. The proof is by
induction on the compilation judgment, and relies on proofs of
correctness of the functions we define to compile primitive \vqimp
operations to \vqir, as discussed in
\Cref{sec:vqir-compilation}.
}

\ignore{

\section{Constant Addition Circuit}
\label{app:const-add}

\begin{figure}[h]
{\footnotesize
\begin{tabular}{c}
\begin{minipage}{\textwidth}
\begin{coq}
Fixpoint rz_adder_c' (a) (b:nat -> bool) (n:nat) (size:nat) :=
  match n with 
  | 0 => ID (a,0)
  | S m => rz_adder_c' a b m size
           ; if b m then SR (size - n) a else ID (a,m)
  end.
\end{coq}
\end{minipage} 
\\
\begin{minipage}{\textwidth}
\begin{coq}
Definition rz_adder_c (a:var) (b:nat -> bool) (n:nat) := 
  Rev a ; $QFT$ a ;
  rz_adder_c' a b n n;
  $QFT^{-1}$ a; Rev a.
\end{coq}
\end{minipage}
\end{tabular}
}
\caption{Adding an \vqimp variable to a constant, in \vqir}
\label{fig:circuit-example2}
\end{figure}

\Cref{fig:circuit-example2} shows \coqe{rz_adder_c},
which adds a variable to a constant. 

}

\ignore{

\section{Sine Oracle in \vqimp}
\label{sec:eval-sine}

With limited resources on near-term quantum computers, the most anticipated application is Hamiltonian simulation. Many sophisticated algorithms are developed \cite{Childs_2009,Berry_2006,Berry_2015,Low_2017,Low_2019} for near optimal quantum simulation. These algorithms rely heavily on implementations of different oracles. Among them, a quantum walk based algorithm proposed by Childs \cite{Childs_2009} has an essential subroutine $F_t$ realizing a phase manipulation $F_s\ket{x}=e^{-2\pi i s\mathrm{sin}(2\pi x)}\ket{x}.$ This subroutine requires an approach to encode the sine function into the phase. This is easy to implement in \vqimp. 
\begin{figure}[h]
  \centering
  \begin{minipage}{0.6\textwidth}
    \[\small
    \begin{array}{l}
      \texttt{\textcolor{red}{fixedp} phase\_calc}(C\;\texttt{nat }n,\; C\;\texttt{fixedp }s,\; \textcolor{red}{Q\;\texttt{fixedp} \;x},\;\textcolor{red}{Q\;\texttt{fixedp } z})\{ \\
      \quad\textcolor{red}{a}\texttt{ = }2\pi * x;\\
      \quad\textcolor{red}{b}\texttt{ = sin}(a);\\
      \quad\textcolor{red}{z}\texttt{ = }s*b;\\
      \quad\texttt{return }\textcolor{red}{z};\\
      \}
    \end{array}
    \]
    \subcaption{\vqimp code for the phase calculation oracle.}
    \label{fig:hs1}
  \end{minipage}
  \quad
  \begin{minipage}{0.6\textwidth}
\begin{coq}
Fixpoint VarRz (b:var) (n:nat) :=
  match n with 
  | 0 => Rz 1 (b,0)
  | S m => VarRz a m; Rz n (b,m)
  end.
Definition Fs a b n s :=
  phase\_calc n s a b; VarRz b n; phase\_calc$^{-1}$ n s a b
\end{coq}
\subcaption{\vqir code for the phase manipulation.}
    \label{fig:hs2}
  \end{minipage}
  \caption{Implementation of the phase subroutine in quantum walk-based Hamiltonian simulation.}
\end{figure}

To implement this subroutine, we construct an oracle in \vqimp (\Cref{fig:hs1}), which calculates the phase and stores it into an ancillary register, resulting in $\ket{x}\ket{0}\rightarrow \ket{x}\ket{\sin(2\pi x)}$. Then we compile it into \vqir as a function \texttt{PhaseCalc}, which takes in the input and output variables. The \texttt{VarRz} function in \Cref{fig:hs2} applies a series of z-axis rotations realizing transformation $\ket{y}\rightarrow e^{-2\pi i y}\ket{y}.$ Since the oracle outputs to an ancillary variable for $F_s$ subrouting, we apply the inverse of phase calculation oracle to uncompute the ancillae.

\paragraph*{Correctness}
We have manually tested our \vqimp sine program. Some of its components (addition/subtraction of quantum variables) are full verified, while the rest (multiplication of quantum variables, division by a constant) are randomly tested (\Cref{fig:op-table}).

}

\end{document}